\documentclass[letterpaper,onecolumn,superscriptaddress,nofootinbib,floatfix,showpacs]{revtex4}
\usepackage{graphicx}
\usepackage{graphicx,amsfonts,amssymb,amsmath,hyperref,hypcap,enumerate}
\usepackage{latexsym}
\usepackage{amssymb}
\usepackage{amsmath}
\usepackage{amsfonts}
\usepackage{bm}
\usepackage{multirow}
\usepackage{color}
\usepackage{slashed}
\usepackage{comment}
\usepackage{ulem}
\newcommand{\ii}{\mathrm{i}}

\renewcommand{\Re}{\mathop{\mathrm{Re}}}
\renewcommand{\Im}{\mathop{\mathrm{Im}}}

\newcommand{\beq}{\begin{equation}}
\newcommand{\eeq}{\end{equation}}
\newcommand{\beqn}{\begin{eqnarray}}
\newcommand{\eeqn}{\end{eqnarray}}
\newcommand{\nn}{\nonumber}

\DeclareMathAlphabet{\mathbbold}{U}{bbold}{m}{n}

\def\sgn{{\rm sgn}}

\makeatletter
\newcommand\xleftrightarrow[2][]{%
\ext@arrow 9999{\longleftrightarrowfill@}{#1}{#2}}
\newcommand\longleftrightarrowfill@{%
\arrowfill@\leftarrow\relbar\rightarrow} \makeatother

\newcommand{\T}{\mathcal{T}}
\newcommand{\CT}{\mathcal{CT}}

\newcommand{\be}{\begin{equation}}
\newcommand{\ba}{\begin{align}}
\newcommand{\ee}{\end{equation}}
\newcommand{\bea}{\begin{eqnarray}}
\newcommand{\eea}{\end{eqnarray}}

\newcommand{\ra}{\rightarrow}

\newcommand{\Z}{\mathbb{Z}}
\newcommand{\LL}{\mathcal{L}}

\renewcommand{\vec}[1]{{\bf #1}}
\renewcommand{\hat}[1]{{\widehat #1}}
\renewcommand{\Re}{{\rm \, Re\,}}
\renewcommand{\Im}{{\rm \, Im\,}}
\def\nn{\nonumber\\}

\begin{document}
\title{Duality between $(2+1)d$ Quantum Critical Points}

\author{T. Senthil}
\affiliation{Department of Physics, Massachusetts Institute of
Technology, Cambridge, MA 02139, USA}

\author{Dam Thanh Son}
\affiliation{Kadanoff Center for Theoretical Physics, University
of Chicago, Chicago, Illinois 60637, USA}

\author{Chong Wang}
\affiliation{Department of Physics,  Harvard University,
Cambridge, MA 02138, USA}
\affiliation{Perimeter Institute for Theoretical Physics, Waterloo, ON N2L 2Y5, Canada}

\author{Cenke Xu}
\affiliation{Department of Physics, University of California,
Santa Barbara, CA 93106, USA}

\date{\today}

\begin{abstract}

Duality refers to two equivalent descriptions of the same theory
from different points of view. Recently there has been tremendous
progress in formulating and understanding possible dualities of
quantum many body theories in $2+1$-spacetime dimensions. Of
particular interest are dualities that describe conformally
invariant quantum field theories in $(2+1)d$. These arise as
descriptions of quantum critical points in condensed matter
physics. The appreciation of the possible dual descriptions of
such theories has greatly enhanced our understanding of some
challenging questions about such quantum critical points. Perhaps
surprisingly the same dualities also underlie recent progress in
our understanding of other problems
such as the half-filled Landau level and correlated surface states
of topological insulators. Here we provide a pedagogical review of
these recent developments from a point of view geared toward
condensed matter physics.

\end{abstract}

\maketitle

\tableofcontents


\section{Introduction}
\label{intro}

A crowning achievement of the 19th century is the discovery that
electric and magnetic fields are two sides of the same coin, and
they are described by one set of elegant unified equations. In
Maxwell's equations, the electric field and magnetic field enjoy
an intriguing ``duality" transformation: the form of the equations
remain unchanged after interchanging  electric and magnetic
fields, as long as one introduces magnetic charges and magnetic
currents. At the quantum level, such an electric-magnetic duality
(and the more general ``S-duality" form) was established in
certain supersymmetric Yang-Mills
theories~\cite{Montonen1977,seiberg1995,seibergwitten1,seibergwitten2}
and Abelian gauge theory without
supersymmetry~\cite{cardy1,cardy2,wittenu1}.

Quite generally, duality refers to two equivalent descriptions of
the same theory but from different points of view. In condensed
matter physics, duality methods originated in the work of Kramers
and Wannier on the statistical mechanics of the $2d$ classical
Ising model~\cite{kramers41}. Duality transformations have since
been used in subsequent decades on a variety of problems to obtain
powerful non-perturbative insights into the phase diagram and
universal properties of strongly interacting many particle systems
in both classical and quantum many body physics. In the context of
supersymmetric quantum field theories many powerful dualities have
been discovered in diverse dimensions that demonstrate the
equivalence of seemingly distinct theories.

In quantum many body physics duality methods are perhaps most
familiar in $d = 1$ space dimension. A well known example is the
low energy description of a quantum Bose fluid at finite density
in $1d$.  This system may be described as a `fluctuating
superfluid' through a quadratic Lagrangian written in terms of the
phase of the superfluid order parameter.  Correlation functions
(such as the boson Green's function) decay as a power law.  It is
well known that there is an alternate but equivalent description
of the same phase as a `fluctuating crystal'. In this `dual' point
of view the low energy physics is again described in terms of the
phonon mode of the crystal order parameter.  Thus the $1d$ Bose
fluid can be viewed either as a  fluctuating superfluid or as a
fluctuating crystal.  As is well known, in $1d$, this duality
generalizes to interacting fermions and leads to the familiar
`bosonized'
description~\cite{coleman,mandelstam,luther,wittenbosonization,Giamarchi2004}.
Unlike in higher dimension even for weak interaction strength
Landau's celebrated Fermi liquid theory breaks down, and it is not
possible to describe the physics using standard Landau
quasiparticles. The bosonized theory gives a powerful framework to
understand universal aspects of the resulting  $1d$ Luttinger
liquid using a completely different set of variables.

Moving to $d = 2$ space dimensions, for systems of interacting
bosons, there is a famous duality that interchanges a description
in terms of the bosonic particles with a description in terms of
vortices in the phase of the boson field. This is known as
charge-vortex duality. Specifically let us consider a bose Hubbard
model at integer filling on a $2d$ lattice. This model has two
obvious phases: a superfluid phase and a Mott insulating phase.
These two phases are separated by a quantum critical point in the
$(2+1)d$ $XY$ universality class. Both phases and the phase
transition can be understood simply in terms of the bosonic
particles (the `charge' picture). Equivalently there is a dual
description of the same system in terms of vortices coupled to a
dynamical $U(1)$ gauge field~\cite{Peskin,halperindual,leedual}.
This dual description captures the universal long wavelength low
energy physics of both phases and the phase transition. However it
is not an exact re-write of the original bose Hubbard model, and
cannot be used to calculate non-universal properties.

The dual `vortex' point of view of the interacting boson system in
$2d$ has been extremely useful in thinking about correlated boson
systems. In the condensed matter context it is an example of a
`weak duality'. It tells us that both the charge and vortex
descriptions describe the same physical system, {\em i.e}, have
the same local operators, and the same global symmetries.  The
weak duality  opens up  the possibility of  a different notion of
duality (known naturally as a `strong duality'): that of the
continuum quantum field theories obtained from either the charge
or vortex descriptions.  The quantum critical point describing the
superfluid-Mott transition is described by a $(2+1)d$ Conformal
Field Theory (CFT). In the charge picture, a natural continuum
Lagrangian that flows to this CFT is that of a $|\phi|^4$ theory
of a complex scalar $\phi$ tuned to its critical point. A
different continuum field theory is obtained from the vortex
picture, namely a theory of a (different) complex scalar coupled
to a dynamical $U(1)$ gauge field. Strong duality is the
statement that both continuum Lagrangians flow to the same
Infra-Red (IR) CFT.  If the strong duality holds for the CFT,
then by deforming by adding a relevant perturbation we get dual
descriptions of the two phases on either side of the phase
transition. For the bosonic charge-vortex duality, the weak
duality is a rigorously correct statement while the strong duality
is a conjecture that is supported by existing numerical
calculations~\cite{sudbo99,Kajantie04}.

In the last few years tremendous
progress~\cite{son2015,wangsenthil1,tsymmu1,maxashvin,xudual,mrossduality,mengxu,seiberg1,karchtong,seiberg2,xumajorana,seiberg3,kachru1,kachru2,potterdual,SO5}
has been made in unearthing many other dualities in $(2+1)d$.
These include dualities involving  theories written in terms of
fermions, and relate them either to theories written in terms of
other fermions, or in terms of bosons. These fermion-fermion and
fermion-boson dualities are expected to have many powerful
applications to condensed matter physics, and indeed they
originated from modern work on diverse problems in condensed
matter physics and quantum field theory. Some of these
applications have already been explored.   Many
previously unrelated problems in condensed matter physics have
been connected through the duality program, including the theory of the
half-filled Landau level   in quantum Hall regime, strongly
correlated topological insulators, $U(1)$ spin liquids in three
dimensions and quantum phase transitions beyond the Landau
paradigm.

In parallel to these developments in condensed matter physics,
very similar quantum field theories and their dualities have been
studied in the high energy literature in recent years.  One
starting point is the well-known level-rank duality of
Chern-Simons gauge theories \cite{Naculich:1990pa,Mlawer:1990uv}.
Mirror symmetry, a supersymmetric version of particle-vortex
duality, has been known since mid-1990s~\cite{mirror1}.  Through a
rather circuitous route, dualities between nonsupersymmetric
Chern-Simons theories were arrived at~\cite{aharony2} and verified
diagrammically at large $N$. By synthesising various bits of
information, conjectures on dualities of Chern-Simons matter
theories away from the large-$N$ limit has been
formulated~\cite{aharony3}. At very small values of $N$, these
dualities reduce to ones discussed in the condensed matter
literature.

Our goal here is to review these recent developments with a
viewpoint geared heavily towards condensed matter physics. We will
describe the new dualities, and several ways of thinking about
them. We will demonstrate their utility in condensed matter
physics by reviewing a few examples where these dualities have had
direct and significant impact. It is important to emphasize that
all the recent dualities (except in special large-$N$ limits) are
(like the bosonic charge-vortex duality) well established as
`weak' dualities but are conjectural as `strong' dualities. It is
important to check the strong dualities through numerical
calculations. We show how some of the dualities lend themselves to
numerical tests and discuss the current state of evidence in their
support.

 We should emphasize that both ``weak" and ``strong"
dualities can be of great use. If our goal is to explore
interesting phases of matter, ``weak" dualities will typically
suffice -- this is the situation we will encounter in the problems
of half-filled Landau level and correlated topological insulators.
In those cases the dualities provide alternative pictures to
formulate the problems, and certain mean-field ansatz (plus
appropriate fluctuations) can be motivated to construct
interesting phases. This approach should be familiar from
composite boson/fermion theories in quantum Hall effects and
parton (slave particle) theories in quantum spin liquids.  where
an exact re-writing of the problem becomes useful to motivate
interesting low energy effective theories capable of accessing
non-trivial phases/phase transitions that are hard to otherwise
describe.

Strong dualities, on the other hand, are needed if we are
interested in a critical point, typically described by an
interacting CFT. Similar to the bosonic charge-vortex duality many
of the new dualities also map one interacting problem to another
interacting problem. Both sides are difficult to solve by
themselves. Nevertheless knowing the duality between two difficult
problems can still be very useful. For instance it may reveal some
hidden symmetries of the system, which become more obvious in one
side  of the duality than the other. A more subtle situation
arises when the full symmetry of the system is not manifest in any
single formulation  but becomes apparent when different dual
formulations are viewed together. We will see examples of this
phenomenon. Nontrivial (and surprising) predictions can be made
for low energy correlation functions based on these hidden
symmetries. A possibly familiar field theoretic example of hidden
symmetries is the SO(4) symmetry of a single compact boson in
$(1+1)d$ with certain compactification radius (or certain
Luttinger parameter depending on the choice of convention). The
SO(4) symmetry is  not explicit in the compact boson formalism,
but becomes obvious when we reformulate this system as an O(4)
nonlinear Sigma model with a Wess-Zumino-Witten
term~\cite{wittenbosonization,affleck1985}, or the $(1+1)d$
$N_f=2$ quantum electrodynamics (QED). Finally, a duality mapping
may map a problem to another which is much easier to study
numerically for technical reasons. We will also see these examples
in this review.

The rest of the paper is organized as follows. In
Sec.~\ref{1dDualities} we review some familiar dualities in
$(1+1)d$ including the Kramers-Wannier self-duality of Ising model
and the Jordan-Wigner duality between Ising model and free
Majorana fermions. We will see that many structures of the
higher-dimensional dualities are already revealed in these basic
$(1+1)d$ dualities. In Sec.~\ref{bosonvortex} we review the
$(2+1)d$ Peskin-Dasgupta-Halperin boson-vortex duality, and its
connection to electric-magnetic duality (S-duality) of $(3+1)d$
electrodynamics. In Sec.~\ref{DFermions} we review some basic
aspects of Dirac fermions in $(2+1)d$ and summarize the statements
of some basic dualities proposed recently. In
Sec.~\ref{DualityWeb} we discuss in detail a ``seed" duality
sometimes known as $(2+1)d$ bosonization, and relate it to other
dualities involving Dirac fermions (dubbed a web of dualities). We
also discuss how time-reversal symmetry is realized nontrivially
in these dualities. In Sec.~\ref{EMDuality} we relate those
$(2+1)d$ dualities (and their nontrivial symmetry realizations) to
electric-magnetic dualities of $U(1)$ gauge theories in $(3+1)d$.
In Sec.~\ref{Evidences} we discuss other theoretical evidence
supporting (but not necessarily proving) these $(2+1)d$ dualities,
either by defining the field theories on lattices and coupled
wires, or by generalizing the theories to some large-$N$ limit
that are theoretically more controlled. In Sec.~\ref{CMApp} we
discuss applications of the  fermion-fermion duality to the problem of
half-filled Landau level, focusing on the intriguing realization of particle-hole
symmetry, and the problem of correlated surface states of
topological insulators. In Sec.~\ref{miniweb} we discuss the
application of dualities to deconfined quantum criticality -- a
class of exotic quantum phase transitions beyond the traditional
Landau-Ginzburg paradigm. Sec.~\ref{Numerics} discusses some
recent numerical simulations testing some of the dualities,
especially those related to deconfined criticality. In
Sec.~\ref{MajoranaBosonization} we discuss an example of
non-abelian duality, between a free gapless Majorana fermion in
$(2+1)d$ and an $SO(3)$ vector Higgs model with a Chern-Simons
term at level one. We conclude with some discussions on some other
related developments and possible future directions in
Sec.~\ref{discussion}.


\section{Lightning review of some familiar dualities}
\label{1dDualities}

We begin with a review of the physical basis of some dualities
familiar in condensed matter physics.  Readers interested in
details are urged to consult treatments in textbooks.

We start with the Kramers-Wannier duality of the Ising model. This
can be formulated either in terms of a classical $2d$ Ising model
or in terms of the quantum transverse field Ising model in spatial
dimension $d = 1$.  We describe the duality in the latter context
below.

The Hamiltonian for the quantum transverse field Ising model in
spatial dimension $d = 1$ is
\begin{equation}
\label{tfim} H = -J \sum_i \sigma^z_i \sigma^z_{i+1} - h \sum_i
\sigma^x_i
\end{equation}
where $\vec \sigma_i$ are Pauli matrices on a $d = 1$ spatial
lattice with sites labelled by $i$.  The model has a global $Z_2$
symmetry which takes $\sigma^{z,y}_i \rightarrow -\sigma^{z,y}_i$.
For $J > h$ the ground state spontaneously breaks the global $Z_2$
symmetry (ferromagnetically ordered) while for $h > J$ (the
paramagnetic state) the global symmetry is unbroken.  The critical
point is at $h = J$. The duality transformation is obtained by
introducing new spin variables $\vec \tau_{i + \frac{1}{2}}$ that
live at the midpoints of the bonds of the original $1d$ lattice.
We write
\begin{eqnarray}
\tau^x_{i + \frac{1}{2}} & = & \sigma^z_i \sigma^z_{i+1} \\
\tau^z_{i - \frac{1}{2}}\tau^z_{i + \frac{1}{2}} & = & \sigma^x_i
\end{eqnarray}
The second equation is readily solved to obtain
\begin{equation}
\label{tauz} \tau^z_{i + \frac{1}{2}} = \prod_{j \leq  i}
\sigma^x_i
\end{equation}
Clearly the Hamiltonian re-expressed in terms of the $\tau$-spins
has the form also of the transverse field Ising model but with an
interchange of the couplings $h$ and $J$, and hence the
ferromagnetic and  paramagnetic phases.

The critical point is left invariant by the duality
transformation, and hence is ``self-dual".  Physically the duality
reflects an equivalence between two alternate descriptions of the
model: we can  use  either  the spin degrees of freedom
($\sigma^z$) themselves or   domain wall configurations starting
with a reference ferromagnetically ordered state. To see this note
that the $\tau^z$ operator in Eqn.~\ref{tauz} flips the $\sigma^z$
values of all spins to the left of site $i+1$ and hence creates a
domain wall at $i + \frac{1}{2}$. The domain walls are topological
defects of the order parameter of the Ising ferromagnet. In the
ordered phase the spins are ordered but the domain walls are
costly. In the paramagnetic phase the spins are disordered but the
domain walls have proliferated.

It is also well known that the Ising model can be solved by a
mapping to a free fermion model. In the quantum context this is
accomplished by a Jordan-Wigner transformation.  We can view this
as another duality of the Ising model. Not only is the Ising model
self-dual it is also dual to a free fermion model. For the
transverse field Ising model in Eqn. \ref{tfim} the corresponding
free fermion model is a chain of Majorana fermions $\eta_r$
defined at sites $r$ of another $1d$ lattice:
\begin{equation}
\label{kitc} H_f = iJ\sum_{r=2i} \eta_r \eta_{r+1} + ih
\sum_{r=2i+1} \eta_r \eta_{r+1},
\end{equation}
where
\begin{equation}
\eta_{2i}=i\sigma^z_i\tau^z_{i+1/2}, \hspace{10pt}
\eta_{2i+1}=\tau^z_{i+1/2}\sigma^z_{i+1}.
\end{equation}
Clearly the Majorana fermions $\eta_r$ can be interpreted as
spin-kink composites.

Though the mapping to free fermions has been known for decades, a
proper interpretation of the free fermion model and its relation
to the original Ising model was fully clarified only much later.
From a modern perspective  the Hamiltonian in Eqn. \ref{kitc} is
known as the Kitaev Majorana chain. It has two phases which are
topologically distinct from each other. A physical statement of
the distinction is that with open boundary conditions, in one
phase (the topological phase) there are a pair of Majorana zero
modes localised at the two edges while in the other phase (the
trivial phase) there are no such edge modes. The topological phase
thus has a two-fold degenerate ground state corresponding to the
two dimensional Hilbert space of the pair of Majorana zero modes.
Carefully tracking the Jordan-Wigner mapping with open boundary
conditions shows that the topological phase maps to the ordered
phase of the Ising model. The two-fold ground state degeneracy of
the topological phase then corresponds to the two ferromagnetic
ground states of the Ising model.

The critical point of the Ising model (at $J = h$) maps to the
trivial-topological phase transition of the Kitaev chain. Note
that at the critical point the Kitaev chain has a symmetry under
translations by one lattice spacing $r \rightarrow r+1$. What does
this symmetry correspond to in the Ising Hamiltonian? Though not
apparent in  Eqn. \ref{tfim}, the translation symmetry of the
Kitaev chain maps to the self-duality symmetry of the critical
Ising model.

It is interesting to extract from these well known facts some
statements about dualities of continuum field theories that
describe the vicinity of the Ising critical point.  The critical
Ising model is described by the Infra-Red (IR) CFT fixed point of
the theory of an {\em interacting} real scalar $\phi$. We
schematically denote the (Minkowski) Lagrangian of this theory by
\beq \label{isingdr} {\cal L} = (\partial \phi)^2 - \phi^4 \eeq
The $\phi$ field should be viewed as the long wavelength version
of the lattice Ising order parameter. The relevant perturbation of
this theory will be denoted $r \phi^2$ and drives the system away
from the critical point. $r > 0$ corresponds to the paramagnetic
phase and $r < 0$ to the ordered phase. The Ising self-duality
means that there is a different field theory that flows to the
same IR CFT which may be written \beq \label{isingdl} \hat{{\cal
L}} = (\partial \hat{\phi})^2 - \hat{\phi}^4 \eeq The $\hat{\phi}$
is the order parameter of the dual Ising model. The relevant
perturbation is now $- \hat{r} \hat{\phi}^2$ so that the duality
interchanges the ordered and disordered phases. Finally both these
theories are equivalent in the IR to the free massless Majorana
fermion CFT: \beq \label{isingm} {\cal L}_m =
\bar{\chi}i\slashed{\partial}\chi \eeq with $\slashed{\partial} =
\gamma^\mu
\partial_\mu$. (In $(1+1)d$ the $\gamma$ matrices are $2 \times
2$ matrices.) A Majorana mass term $m\bar{\chi}\chi$ corresponds
to the relevant perturbation $r \phi^2 \sim - \hat{r}
\hat{\phi}^2$.

Here we should emphasize the two logically distinct, though
physically closely related, notions of dualities. The lattice
duality was derived exactly and holds for all length/energy scale.
The continuum field theory duality, motivated by the lattice
duality, is based on the belief that the $\phi^4$ theory (a
renormalizable continuum field theory) describes the same IR
physics as the critical Ising model -- in the specific context of
$(1+1)d$ Ising model there is little room to doubt this belief,
but later we will see more complicated examples in which the
relations between lattice models and continuum field theories are
essentially conjectural. For this reason the continuum dualities
are sometimes called ``strong dualities".  Another distinction
between the two notions of dualities is that the continuum one
often only holds in the IR limit. For example, in the
Ising-Majorana duality, the $\phi^4$ theory (as a
super-renormalizable continuum theory) is free in the UV, which
makes it clearly different from a free fermion theory.

Thus far we have been somewhat cavalier about global issues (such
as boundary conditions) related to the duality, though given the
explicit transformations on the lattice it is always possible to
keep track of these subtleties.  Further we have also not
carefully specified how the global Ising symmetry which is
manifest as $\phi \ra - \phi$ in the theory of Eqn.~\ref{isingdr}
acts on the other dual theories. As a pedagogical example, the
precise form of the (continuum) Kramers-Wannier duality can be
written as \beq
(D_B\phi)^2-\phi^4\hspace{10pt}\Longleftrightarrow\hspace{10pt}(D_b\hat{\phi})^2-\hat{\phi}^4+\pi
b\wedge B, \eeq where $B$ is a background (probe) $\Z_2$ gauge
field that couples to the Ising charge, and $b$ is a dynamical
$\Z_2$ gauge field. A nontrivial $\Z_2$ gauge flux over a loop in
the space-time manifold essentially corresponds to an
anti-periodic boundary condition over this loop. The last term
$\pi b\wedge B$ assigns a nontrivial global $\Z_2$ charge to each
$\Z_2$ instanton of $b$ (a tunneling event that flips the boundary
condition in $\hat{\phi}$), thereby identifying the instanton with
$\phi$ on the left side. The $\Z_2$ gauge field $b$, unlike
continuous gauge fields, is flat and has no dynamics of the
Maxwell type (instantons are also suppressed because of the global
$\Z_2$ symmetry). Therefore $b$ has no nontrivial dynamics and
only imposes a global constraint, and is often dropped when global
issues such as boundary conditions are neglected, making the
duality a ``self-duality". In Appendix \ref{1dnotes} we carefully
state the other $1+1$-D dualities (such as the Jordan-Wigner
duality) paying special attention to these subtleties.

For now we point out one aspect of symmetry realization of the
continuum dualities which we already alluded to at the lattice
level. The microscopic lattice translation symmetry of the
critical Majorana chain is realized in the continuum field theory
as an internal symmetry under which  $\chi \ra \gamma^5 \chi$
(which flips the sign of the left moving fermion alone). How is
this symmetry realized in the dual $\phi^4$ theory? From the
lattice discussion we know that it  is realized as the duality
transformation $\phi \leftrightarrow \hat{\phi}$. This is an
example of a ``quantum symmetry":  it is not a symmetry of the
Lagrangian but is a symmetry of the partition function. Later we
will see other examples of this phenomenon in higher dimensions
where an ordinary-looking symmetry on one member of the duality
web is realized as a duality transformation on other members of
the web.

To better understand the unconventional realization of the chiral
symmetry $\mathcal{S}: \chi\to\gamma^5\chi$ in the dual theories,
it is helpful to view these theories as the boundary of a $\Z_2$
topological order (a deconfined $\Z_2$ gauge theory) in $(2+1)d$.
The $\Z_2$ topological order has three nontrivial particle
excitations (superselection sectors) in the bulk, often labeled as
$(e,m,\epsilon)$ where $e$ (charge) and $m$ (vison) are bosonic
and $\epsilon\sim e\times m$ is fermionic. The boundary of this
bulk topological order is naturally an Ising theory, where $\phi$
can be interpreted as the boundary descendent of $e$, $\hat{\phi}$
as that of $m$, and $\chi\sim \phi\hat{\phi}$ as that of
$\epsilon$. The boundary gapless Majorana fermion $\chi$ with
on-site chiral symmetry $\mathcal{S}$ can be realized when the
$\epsilon$ fermion in the bulk forms a topological
superconductor\footnote{When realized as a lattice translation
symmetry (not on-site), a bulk is not needed. Instead structures
similar to the Lieb-Schultz-Mattis constraint\cite{HHG} become the
field theory anomaly in the continuum limit.}. In this topological
superconductor the $\mathcal{S}$-even fermions form a $p+ip$
chiral superconductor, while the $\mathcal{S}$-odd fermions form a
$p-ip$ superconductor. A $\pi$-vortex of this superconductor will
then trap two Majorana zero modes $\gamma_+$ and $\gamma_-$, from
the $\mathcal{S}$-even and $\mathcal{S}$-odd fermions,
respectively. The $e$ and $m$ particles correspond to vortices
with opposite fermion parity $(-1)^F=i\gamma_+\gamma_-$, which
flips sign under $\mathcal{S}$. This means that $e$ and $m$ are
exchanged under $\mathcal{S}$. On the boundary this implies that
$\phi$ and $\hat{\phi}$ are exchanged under $\mathcal{S}$ --
precisely what we expected from the lattice argument.

We also show in Appendix~\ref{1dnotes} the more formal structure
of the $(1+1)d$ web of dualities and its interpretation from a
$(2+1)d$ point of view. As we shall see later, very similar
structures appear in one dimension higher, where the role of the
$\Z_2$ symmetry is played by a $U(1)$ symmetry.

Next we remind the reader of  another famous model where duality
transformations play a crucial role in describing the physics:
this is the classical $XY$ model in $2d$. It is well known that
this model has two phases as a function of temperature: a low
temperature phase with power law correlations of the $XY$ spins
and a high temperature phase with exponentially decaying
correlations. The phase transition is driven by the proliferation
of topological defects, {\em i.e} the vortices of the $XY$ order
parameter.  In the low-$T$ phase the vortices cost an energy
logarithmically large in the system size (equivalently a
vortex-antivortex pair has an energy that grows logarithmically
with their separation).  Note that the logarithmic interaction is
also what is expected from a Coulomb potential in two spatial
dimensions.  In the high-$T$ phase the vortex-antivortex pairs
unbind from each other.  The duality of the $XY$ model
reformulates it as a gas of $\pm$ point charges interacting with
each other through the $2d$ Coulomb potential. These charges have
the interpretation as the vortex/antivortex topological defects of
the XY order parameter.

In contrast to the Kramers-Wannier duality for the nearest
neighbor Ising model described above, the $XY$- Coulomb gas
duality is not exact microscopically for the $XY$ model. However
it describes an equivalence of the  {\em universal} long
wavelength properties of the two models.

The physics of the classical $2d$ $XY$ model can readily be
re-interpreted to yield the physics of the $O(2)$ quantum rotor
model\footnote{This model can be viewed as describing a
superfluid-insulator transition of bosons (with a global $U(1)$
symmetry) at integer filling on a $1d$ lattice. The superfluid  in
$d = 1$ has power law order while the Mott insulator has
exponentially decaying correlations.}. In the context of quantum
many body physics in $1d$, dualities such as these are
tremendously powerful and are part of the standard theoretical
toolbox. As the Ising example shows they include as a subset the
well known bosonization methods for $1+1$-D continuum field
theories.

What about quantum matter in $2d$? An old and important duality of
strongly correlated {\em boson} systems in $2d$ was described a
long time ago~\cite{Peskin,halperindual,leedual}. and is known as
the charge-vortex duality of bosons. We turn to this next.

\section{Charge-vortex duality for bosons}

\label{bosonvortex} A simple and paradigmatic model of a strongly
correlated boson system on a lattice is the Bose Hubbard model
with Hamiltonian
 \beq
 H = - t \sum_{rr'} b^\dagger_r b_{r'} + h.c + \frac{U}{2} \sum_r n_r (n_r - 1)
 \eeq
Here $b_r$ are boson destruction operators at sites $r$ of a $2d$
square lattice. We specialize to a situation where the boson
density is such that, on average, there is one boson per site. If
$t \gg U$, the ground state (the superfluid phase) spontaneously
breaks the global $U(1)$ symmetry (associated with letting $b_r
\ra b_r e^{i\alpha}$ for all $r$), and there is a corresponding
gapless Goldstone mode.  In the opposite limit $U \gg t$ this
symmetry is preserved, and there is a gap to all excitations (the
Mott insulating phase). The quantum phase transition between these
two phases is second order, and is described by the Wilson-Fisher
fixed point of the theory of a single complex scalar (which also
describes the critical point of the $3D$ classical XY model).  In
the vicinity of this critical point we may describe the system by
a coarse-grained continuum field theory with the Minkowski action
  \beq
  \label{xycont}
 {\cal L} = |D_A \phi|^2 - r|\phi|^2 - u  |\phi|^4
 \eeq
The covariant derivative $D_A$ includes a minimal coupling to a
{\em background }\footnote{We do not integrate over these fields
in the path integral.} external $U(1)$ gauge field $A$. Including
this enables us to keep track of the global $U(1)$ symmetry of the
model.  By tuning $r,$ the theory can be placed at its critical
point - at that point the IR physics is described by the 3D-XY
Wilson-Fisher CFT.  We will schematically write the Lagrangian for
this CFT as
 \beq
 \label{XYcft}
 {\cal L}_{WF} = |D_A \phi|^2 -  |\phi|^4
 \eeq

The phases and phase transition of the boson Hubbard model have an
alternate dual description in terms of vortices of the superfluid
order parameter.  This duality can be established at the lattice
level along the same lines as the duality of the $2d$ classical
$XY$ model to the Coulomb gas mentioned above. Here we give a
physical description.

Let us start with the low energy theory of the Goldstone mode in
the superfluid phase. This has the $(2+1)d$ (Euclidean)
Lagrangian:
 \beqn
 \label{SFgstn}
\mathcal{L} = \frac{K}{2} (\partial_\mu \theta)^2, \eeqn where
$\theta$ is the phase   of the superfluid order parameter.   At
low energies we can ignore the fact that $\theta$ is defined
periodically ($\theta$ is identified as $\theta + 2\pi$).  This
theory is exactly dual to the theory of a free massless photon
(also in $(2+1)d$) described by the Maxwell action \beq
\label{FreeM} {\cal L}_M  = \frac{1}{2e^2}  \left( \epsilon^{\mu
\nu \lambda}
\partial_\nu a_\lambda \right)^2 \eeq with $e^2 = 4\pi^2 K$.  To
see this equivalence we first rewrite the path integral
corresponding to Eqn. \ref{SFgstn} as \beq Z = \int [{\cal D}j_\mu
{\cal D}\theta] e^{- \int d^3x \frac{j_\mu^2}{2K} + ij_\mu
\partial_\mu \theta} \eeq The $j_\mu$ can be identified with the
3-current associated with the global $U(1)$ symmetry. The
$\theta$-integral can now be performed and leads to the continuity
equation \beq
\partial_\mu j_\mu = 0
\eeq This is readily solved by writing \beq j_\mu = \frac{1}{2\pi}
\epsilon_{\mu \nu \lambda} \partial_\nu a_\lambda \eeq
Substituting this into the remaining path integral over $j_\mu$
immediately leads to Eqn. \ref{FreeM}.

In two space dimensions the photon has only one polarization and
hence the free Maxwell theory describes a single linear dispersing
massless mode just as the Goldstone theory of Eqn. \ref{SFgstn}.
Note that the particle density $j_0$ is identified with the dual
magnetic flux (in units of $2\pi$) while the particle 2-currents
$j_{x,y}$ are identified with the dual electric field rotated by
90 degrees (again in units of $2\pi$).

Including the periodicity of $\theta$ into the theory of the
superfluid phase leads to the existence of vortex defects which
cost logarithmically large energy. In the dual Maxwell theory
these have the interpretation of electrically charged matter
fields coupled minimally to the dynamical $U(1)$ gauge field
$a_\mu$.  Thus a full theory of the superfluid phase consists of a
gapped complex boson $\hat{\phi}$ coupled minimally to a dynamical
$U(1)$ gauge field.  This then motivates a dual description   in
terms of a Minkowski Lagrangian \beq \label{xydualcont} {\cal L}_d
= |D_a \hat{\phi} |^2 - \hat{r}|\hat {\phi} |^2 - \hat{u}
|\hat{\phi} |^4 +  \frac{1}{2e^2}  f^{\mu\nu }f_{\mu\nu}  +
\frac{1}{2\pi} \epsilon_{\mu\nu\lambda} A_\mu \partial_\nu
a_\lambda
 \eeq
where $\hat{\phi}$ is gapped in the superfluid phase.   The  field
strength $f_{\mu\nu} =  \partial_\mu a_\nu - \partial_\nu a_\mu$.
As in the $2D$ classical $XY$ model, we can hope to recover the
symmetry preserving (Mott insulating) phase with exponentially
decaying correlations by proliferating the vortices. Formally this
corresponds to the Higgs phase of Eqn. \ref{xydualcont} (where
$\langle \hat{\phi} \rangle \neq 0$). In the Higgs phase the flux
of $a_\mu$ is quantized in units of $2\pi$: these correspond
precisely to the particle excitations with quantized $U_A(1)$
charge above the Mott gap in the insulator. Note that in the
superfluid phase the charges are condensed but the vortices are
costly. In the insulating phase on the other hand the charge is
gapped but the vortices have ``condensed".

The superfluid-insulator critical point will be described in this
dual description as the critical point of Eqn. \ref{xydualcont}
obtained by tuning the parameters $\hat{r}$. Note that the dual
Lagrangian Eqn. \ref{xydualcont} has the structure of a gauged
version of the original one in eqn. \ref{xycont}.  Thus we may try
to access the critical fixed point of the dual theory by starting
with the Wilson-Fisher fixed point  and coupling in a dynamical
$U(1)$ gauge field. Schematically we therefore write the dual
Lagrangian of the CFT as \beq \label{xydualcft} \hat{ {\cal
L}}_{WF} = |D_a \hat{\phi} |^2 -  |\hat{\phi} |^4 + \frac{1}{2\pi}
Ada
 \eeq
A strong version of the bosonic charge-vortex duality is the
assertion that $\hat{ {\cal L}}_{WF}$ and ${\cal L}_{WF}$ describe
the same CFT. Note that as before the relevant perturbation $r
|\phi|^2$ is mapped to $-\hat{r} |\hat{\phi}|^2$ under the
duality. Further the boson operator $\phi$ is mapped to the
monopole operator ${\cal M}_a$ (which destroys a $2\pi$ magnetic
flux of $a$) of the dual theory. It follows that these operators
will have the same scaling dimensions:
 \beqn
 \Delta[\phi] = \Delta[\mathcal{M}_a], \ \ \ \
\Delta[|\phi|^2] = \Delta[|\hat{\phi}|^2], \label{pvduality2}\
\eeqn

In principle the predictions in Eq.~\ref{pvduality2} can be
verified by calculating the critical exponents of both theories
through the standard renormalization group methods. However, due
to the nonintegrability of either theory, and the lack of a
controlled perturbative method, it is  not possible to do such a
computation. Fortunately, both phase transitions (the MI-SF
transition and the Higgs transition with one flavor of bosonic
matter field) can be realized as lattice models, and  simulated
with numerical methods. Indeed, it can be shown explicitly that
the partition function of a $3d$ lattice O(2) model is dual to
that of a $3d$ boson coupled to a lattice U(1) gauge
field~\cite{Peskin,halperindual}.  If we further {\em assume} that
the continuum limit of both models land us in  the same second
order phase boundary then we might reasonably guess that they are
both controlled by the same fixed point\footnote{This is  commonly
assumed in studies of models in statistical mechanics.}.

Thus the strong version of the duality is  strongly supported (but
not proven) by the lattice derivations of the charge-vortex
duality. It is also supported by numerical simulations which take
the continuum limit of both sides of the
duality\cite{sudbo99,Kajantie04}. Conversely if we assume this
strong version of the duality we can then perturb the CFT by its
relevant operator and obtain an equivalence of the two field
theories away from the critical point as well.

The charge-vortex duality of bosons provides a powerful conceptual
framework to think about many novel phenomena in correlated
bosonic systems in two spatial dimensions.  For instance it
provides a useful point of view\cite{mpafsit} to think about the
destruction of supercondiuctivity in thin films as either the
thickness or a magnetic field is  tuned. An interesting
application is to the hierarchy of fractional quantum Hall states
of electrons\cite{leedual}. The electrons  are first converted to
bosons through flux attachment and then the resulting bosonic
field theories are dualized to obtain useful effective field
theories for the hieirarchy states.  Another
application\cite{z2long} is to understand bosonic Mott  insulator
phases\footnote{These may also be useful interpreted as quantum
$XY$ magnets.} with fractional charge excitations and the
associated topological order.  Finally the bosonic charge-vortex
duality plays a crucial role in the theory of non-Landau quantum
critical points\cite{deconfine1,deconfine2} of spins/bosons in two
space dimensions.

\subsection{Relation to $(3+1)d$ electromagnetic duality}
\label{bcvem}

There is an interesting relationship between the charge-vortex
duality of the $(2+1)d$ system just described and the
electric-magnetic duality of $(3+1)d$ Maxwell theory.
Relationships of this kind  will be very insightful in
understanding the other dualities we will describe later in the
paper, and we therefore review it here for the bosonic
charge-vortex duality.

Consider the boson system of interest (for instance the boson
Hubbard model) as living at the boundary of a bosonic insulator in
$(3+1)d$. While the bulk remains gapped and insulating we can
imagine tuning parameters such that the surface theory undergoes a
superfluid-insulator transition. The boson $\phi$ is a good
excitation in the bulk irrespective of the fate of the surface.

It is extremely useful now to modify the theory by coupling the
bosons to a dynamical $U(1)$ gauge field that lives inside the
$(3+1)d$ sample. How should we view this bulk $U(1)$ gauge theory?
We will regard it as a $(3+1)d$ quantum liquid with an emergent
$U(1)$ gauge field of a UV systems of spins/bosons with a tensor
product Hilbert space.  All local operators in this theory are
gauge invariant bosons.  The bulk $U(1)$ gauge field will have
Maxwell dynamics and hence a propagating massless photon at low
energies. Specific microscopic models of quantum phases with
emergent photons were constructed some time ago in Ref.
\cite{bosfrc3d,wen03,hfb04,3ddmr,lesikts05,kdybk,shannon} in
diverse systems. Importantly such phases  will also have magnetic
monopole excitations. The strength of these monopoles will be
quantized by the usual Dirac quantization conditions. In general
by considering bound states (known as dyons) of electric charges
$q_e$ and magnetic charges $q_m$ we can build up a full set of
allowed (massive) particles labelled by $(q_e, q_m)$.  Clearly
these can be represented as points on a two dimensional lattice
(which we denote  the charge-monopole lattice).  We call the
particle with $(q_e = 1, q_m = 0)$   the $E$ particle and that
with $(q_e = 0, q_m = 1)$  the $M$ particle.

Now consider the boundary. The $E$ particles of the $U(1)$ gauge
theory of course are the bosonic particles $\phi$ we originally
had. The $M$ particles  correspond to vortices $\hat \phi$. This
is readily seen by going to a surface superfluid state of the
original ungauged theory. After introducing the gauge field, the
superfluid vortices will trap quantized gauge flux and will
precisely be the surface avatars of the bulk magnetic monopoles.

Thus by studying the magnetic monopoles $M$  in the gauged bulk we
can infer the properties of the vortices of the surface theory.
Alternately if we understand the surface vortices we can describe
the properties of the bulk monopoles.

The bulk $U(1)$ gauge theory has a duality transformation that
interchanges electric and magnetic fields, and correspondingly the
electric and magnetic charges.  Thus we can describe the same
$U(1)$ gauge theory either from the electric point of view (as a
gauged insulator of the $E$ particles) or from the magnetic point
of view (as a gauged insulator of the $M$ particles).

Let us now think about the surface. From the electric point of
view, there is a `Higgs' phase where the  $E$-particle is
condensed.  Let us call this the E-Higgs phase. This descends from
the surface superfluid of the original ungauged boson system. From
the magnetic point of view, in this surface phase,   $M$ is
gapped. The insulating surface phase of the original bosons goes
over to a distinct surface state after gauging in which the $E$
particle is gapped.    In the magnetic point of view this
corresponds to a condensation of the $M$ particle at the surface
({\em i.e} a descendent of the vortex condensate). Thus this is a
magnetic Higgs phase, or M-Higgs in short.

Clearly in the presence of the boundary the electric-magnetic
duality of the $U(1)$ gauge theory induces a duality between the
E-Higgs and M-Higgs phases.  Right at  the phase transition
between the E-Higgs and M-Higgs phases, it is natural then that
the combined bulk + boundary theory is self-dual. As we show below
this assumption directly leads - in the ungauged theory - to the
charge-vortex duality of the Wilson-Fisher CFT describing the
superfluid-insulator transition.

It is  sufficient and extremely convenient to consider the bulk
theory at energy scales below the gap of all charged matter so
that the only relevant excitation is the photon. Consider
therefore the partition function of free Maxwell electrodynamics
on a closed manifold in $(3+1)d$.
 \beq
Z = \int{\cal D}A_\mu {\cal D}F_{\mu\nu}  \delta( F_{\mu\nu} -
(\partial_\mu A_\nu - \partial_\nu A_\mu) ) e^{- \int d^4x
\frac{1}{4e^2} F^{\mu\nu} F_{\mu\nu}}
 \eeq
To avoid notational clutter we henceforth simply write $F = dA$
(where $A$ is the one-form corresponding to $A_\mu$ and $F$ is the
two-form corresponding to the field strength). We implement the
delta function constraint by integrating over an auxiliary
two-form $F'$:
 \beq
Z = \int {\cal D}A {\cal D}F {\cal D}F' e^{- \int d^4x
\frac{1}{4e^2} F^2 + \frac{i}{2\pi} F' \wedge (F - dA)}
 \eeq
Now we can freely do the integral over $F$ to find
 \beq
Z = \int {\cal D}A {\cal D}F {\cal D}F' e^{- \int d^4x
\frac{e^2}{4(2\pi)^2} F'^2 - \frac{i}{2\pi} F' \wedge (dA)}
 \eeq
Doing the integral over $A$ now tells us that $F' = dA'$ locally
and the resulting path integral is identical to that of the
original Maxwell theory but at a different coupling ${4\pi^2 \over
e^2}$. This is the famous electric-magnetic duality of free
Maxwell theory on a closed space-time manifold.

Now consider the same theory in the presence of a boundary where
there is a $\phi$ field that couples minimally to the boundary
value of $A$. The full theory -   surface + bulk - then has the
path integral
 \beq
Z_{SB} =  \int [{\cal D}\phi{\cal D}A {\cal D}F] \delta(F - dA)
e^{- S_{3D}[\phi, A]} e^{-\int d^4x \frac{1}{4e^2} F^2}
 \eeq
where $S_{3D}[\phi, A]$ is the action of the surface   degrees of
freedom.   Let us now repeat the steps of the  bulk duality
transformation in the presence of the boundary.  The  only change
is that we do not immediately integrate over boundary values of
$A$. In the bulk doing the $A$-integral again gives us $F' = dA'$
locally. The last term $ -{i \over 2\pi} {dA' \wedge dA} $ then
leads to an extra boundary contribution
 \beq
- {i \over 2\pi} A'dA
 \eeq
Thus the duality of the bulk has induced a change in the boundary
action to
 \beq
 S_{3D}[\phi, A, A'] = S_{3D}[\phi, A]  - \int d^3x {i \over 2\pi} A'dA
 \eeq
This is identical to the charge-vortex duality transform of the
boundary $(2+1)d$ theory exactly as expected on physical grounds.
Now let us tune the boundary theory to the phase transition
between the $E$-Higgs and $M$-Higgs phases. As explained above a
natural assumption is that then the full theory - surface + bulk -
is self-dual so long as we make the replacement $e^2
\leftrightarrow {4\pi^2 \over e^2}$. In other words consider the
partition function
 \beq
 \label{zsb}
Z_{SB}[e^2] = \int [{\cal D}A {\cal D}F] Z_{WF}[A] e^{-\int d^4x
\frac{1}{4e^2} F^2}
 \eeq
where $Z_{WF}[A]$ is the partition function of the Wilson-Fisher
CFT in the presence of a {\em background} $U(1)$ gauge field $A$.
We then include the bulk dynamics for $A$ and integrate over its
values both at the surface and bulk. The assumption made above on
the self-duality of the theory then becomes the statement
 \beq
 \label{sbselfd}
Z_{SB}[e^2] = Z_{SB}\left[ \frac{4\pi^2}{e^2} \right]
 \eeq
Applying now the duality transformation to Eqn. \ref{zsb} we find
 \beq
 \label{sbdual}
Z_{SB}[e^2]  =  \int [{\cal D} A' {\cal D}F']\left( \int [{\cal
D}A] Z_{WF}[A]  e^ {\int d^3x {i \over 2\pi} A'dA}\right) e^{-\int
d^4x \frac{e^2}{4(2\pi)^2} F'^2}~~ =
\tilde{Z}_{SB}\left[\frac{4\pi^2}{e^2}\right]
 \eeq
Now the integral over $A$ inside the $\left( \right)$ is only over
the boundary while the integral over $A'$ is over both bulk and
boundary.  Indeed the term inside the $\left( \right)$ defines the
partition function of the dual vortex CFT that corresponds to  the
$(2+1)d$ Wilson-Fisher CFT. From Eqns. \ref{sbselfd} and
\ref{sbdual} we therefore find
 \beq
  Z_{SB}[e^2] = \tilde{Z}_{SB}[e^2]
  \eeq
Since this equality holds for any $e^2$ we are free to take the
limit $e^2 \ra 0$. In this case we can ignore the bulk dynamics of
$A$ in $Z$ and of  $A'$ in $Z'$. For the boundary theory they
simply become background fields. We thus have the equality (after
a trivial renaming of $A$ and $A'$ in $\tilde{Z}$):
 \beq
Z_{WF}[A]  =  \left( \int [{\cal D}A'  ] Z_{WF}[A']  e^ {\int d^3x
{i \over 2\pi} A'dA}\right)
 \eeq
which is precisely the statement of charge-vortex duality of the
$(2+1)d$ Wilson-Fisher CFT.

The charge-vortex duality of the surface thus ties in nicely with
the electric-magnetic duality of the bulk gauge theory. There is
also a nice correspondence between the fields describing the
$(2+1)d$ theory of interest (after the global $U(1)$ is gauged)
and that of particles in a bulk $(3+1)d$ $U(1)$ gauge theory
obtained by extending the gauge fields to the bulk.

\section{Charge-vortex duality of fermions in $(2+1)d$}

\label{DFermions}

Given the success and utility of  charge-vortex duality  in
thinking about correlated boson systems in 2 space dimensions, it
is natural to wonder if there is an analogous duality for
fermions. In 2015 precisely such a duality was found, partly
inspired by a stimulating proposal by Son\cite{son2015} for a
theory of the half-filled Landau level with particle-hole
symmetry, by the theory of $3+1$-dimensional quantum spin
liquids\cite{tsymmu1}, and by questions in the theory of
correlated surfaces of topological
insulators\cite{wangsenthil1,maxashvin}. Specifically the simplest
fermion theory - that of a free massless Dirac fermion - was
proposed to have a dual description in terms of a theory of other
Dirac fermions coupled to $U(1)$ gauge fields.   In this section
we describe some preliminaries that will set the stage to
describing this duality. We will focus on the physics here; some
of the more formal concepts are reviewed in
Appendix~\ref{formalstuff}.

\subsection{The free massless Dirac fermion in $(2+1)d$}

Consider a single massless $2$-component Dirac fermion in $(2+1)d$
space-time dimensions\footnote{Though we will primarily only be
interested in theories in flat space-time it will be convenient to
demand, as a non-trivial consistency check,  that the theory can
be formulated on an arbitrary space-time manifold. We will
therefore often assume that there is a background metric $g$ that
is potentially different from the standard flat space Euclidean
metric. We will however restrict ourselves to orientable
manifolds, {\em i.e}, we do not ``gauge" time-reversal.}.   This
has the Lagrangian
\begin{equation}
\label{freeD} {\cal L}_D = \bar{\psi} i\slashed{D}_A \psi
\end{equation}
Here $\slashed{D}_A = \gamma^\mu (\partial_\mu - i A_\mu )$ is the
covariant derivative.  In $(2+1)d$ the $\gamma$ matrices are $2
\times 2$ Pauli matrices which we take to be $\gamma_0 =
i\sigma_2, \gamma_1 = \sigma_1, \gamma_2 = \sigma_3$. $A_\mu$ is a
background $U(1)$ gauge field\footnote{Strictly speaking we should
take $A$ to be what is known as a ``spin$_c$ connection"  (see
Appendix \ref{formalstuff} for a brief review and references)
rather than a $U(1)$ gauge field. This may be viewed as a
book-keeping device that ensures that in the free Dirac theory
operators with odd electric charge have half-integer spin while
those with even electric charge have integer spin.  A spin$_c$
connection is locally the same as a $U(1)$ gauge field but its
Dirac quantization condition is altered. Specifically
\begin{equation}
\int_C \frac{dA}{2\pi} = \int_C \frac{w_2}{2} ~~(mod~ Z)
\end{equation}
for every oriented 2-cycle $C$ and $w_2$ is the second
Stiefel-Whitney class\cite{nakaharabook} of the space-time tangent
bundle.  The Lagrangian in Eqn. \ref{freeD}, with $A$ taken to be
such a spin$_c$ connection can be formulated on an arbitrary
orientable space-time 3-manifold.  Recall that any such
three-dimensional manifold is a spin manifold, meaning it can be
assigned  a spin structure. In general there can be multiple
inequivalent spin structures.  Eqn. \ref{freeD} is defined without
a choice of a specific spin structure. Physically this means that
there is no charge neutral fermion in the theory.}.  Note that
with this choice the $\gamma$ matrices are real and satisfy
$\{\gamma_\mu, \gamma_\nu\} = 2 \eta_{\mu\nu}$ with $\eta =
diag(-1,1,1)$.

It is important, in defining the free Dirac theory, to have some
regularization in mind. As is well-known this theory cannot be
regularized in a time-reversal invariant manner in a strictly
$(2+1)d$ system.  This is known as the parity
anomaly\cite{semenoff,redlich}. It is convenient to choose a
regularization  where we assume that there is another  Dirac
fermion with a heavy mass $M < 0$:
 \beq {\cal L}_H =
\bar{\psi}_H\left(i \slashed{D}_A + M \right) \psi_H \eeq We leave
this implicit in the definition of Eqn. \ref{freeD}.  Formally
with this definition the partition function of the free Dirac
fermion field includes a contribution from the massive Dirac
fermion. This may be loosely written as a level-$1/2$ Chern-Simons
term for the background gauge field\footnote{A precise way to
define this theory is to use the procedure in
Ref.~\cite{wittenreview,seiberg1} where the partition function of
a massless Dirac fermion is written as $ Z_\psi = |Z_\psi|
e^{-i\pi \eta[A,g]/2}$, where $A$ is the $U(1)$ gauge field (more
correctly a spin$_c$ connection), either dynamical or background,
and $g$ is the space-time metric. $\eta$ is defined in terms of
eigenvalues of the Dirac operator \cite{wittenreview}. We review
this briefly in Appendix \ref{formalstuff}.}.

This free massless Dirac theory  is familiar in condensed matter
physics as a theory of the surface of the standard three
dimensional topological insulator of electrons. An interesting and
useful perspective on the theory is to regard it as a quantum
critical point between two phases. Specifically consider adding a
mass term $m\bar{\psi}\psi$. For either sign of $m$, the fermions
are gapped. However the theories with the two signs differ in
their Hall conductances. When the signs of $m$ and $M$ match, {\em
i.e} $m <  0$  there is a net Hall conductance $\sigma_{xy} = -1$
while when $m > 0$, the contributions from the light fermion
$\psi$ and the heavy fermion $\psi_H$ cancel and thus $\sigma_{xy}
= 0$. Both phases have zero longitudinal conductivity. Thus the
free massless Dirac theory can be viewed as a theory of an integer
quantum Hall transition of a system of electrons.  Indeed it is
easy to construct lattice tightbinding models for the integer
quantum Hall transition where the continuum theory of the critical
point is the free Dirac fermion.

From a formal point of view, the partition function of the free
Dirac fermion perturbed by a mass $m$ has a phase that depends on
the sign of the mass. The ratio of the partition function for the
two signs of $m$ is readily seen to be \beq \frac{Z[m; A,g]}{Z[-m;
A,g]} = e^{i \int d^3x \left( \frac{1}{4\pi} AdA + 2 CS_g \right)}
\eeq The level-$1$ Chern-Simons term for the background gauge
field $A$ signifies the difference of the electrical Hall
conductivity of $1$ between the two signs of $m$. The term $CS_g$
is a gravitational Chern-Simons term~\footnote{For the interested
reader we provide the explicit definition of this term in Appendix
\ref{formalstuff}.} that physically corresponds to a thermal Hall
conductivity, which is given by the chiral central charge on the
edge if the $(2+1)d$ theory is gapped in the bulk\cite{KaneFisher}
-- the normalization is chosen so that an integer quantum Hall
state with one complex chiral fermion on the edge corresponds to
$2CS_g$. The above equation then simply means that the two gapped
phases obtained by turning on opposite masses in a free Dirac
fermion differ by an integer quantum Hall state.

Time reversal acts in a simple way on the free massless Dirac
theory.  It takes \beqn \label{Tdirac}
\psi & \rightarrow & i\sigma_2 \psi \nn \\
\psi_H & \ra & i\sigma_2 \psi_H \nn \\
A_i & \ra & - A_i \eeqn Here in the last equation we specify the
transformation of the spatial components $A_i$ of the 3-vector
$A_\mu$. The $A_0$ will then transform with the opposite sign from
$A_i$. With these transformations time reversal is not a symmetry
of the theory. However it is a symmetry up to an additive
Chern-Simons term that depends on the background gauge field and
metric but not on the dynamical fields: \beq \label{Tdirac} \T:
{\cal L}_D \ra {\cal L}_D + {1 \over 4\pi} AdA + 2CS_g \eeq The
extra background contributions come from the reversed sign of the
heavy mass $M$. This is the parity anomaly. As is well known if
the theory arises at the boundary of a three dimensional
topological insulator then the background contributions combine
with those of the bulk response of the topological insulator to
give a time reversal invariant answer (for a clear review see Ref.
\cite{wittenreview}). We will say that the free Dirac theory
in Eqn. \ref{freeD} is time reversal invariant up to an anomaly.

It is also useful to consider a different anti-unitary discrete
charge-conjugation symmetry $\CT$ under which \beqn
\psi & \ra & \psi^\dagger \nn \\
\psi_H & \ra & \psi_H^\dagger \nn \\
A_i & \ra & A_i \eeqn Again $A_0$ transforms with the opposite
sign from $A_i$. $\CT$ is also a symmetry only upto an anomaly:
\beq \CT:  {\cal L}_D \ra {\cal L}_D + {1 \over 4\pi} AdA + 2CS_g
\eeq

\subsection{Dualities of the Dirac fermion}
The proposed\cite{wangsenthil1,maxashvin}  fermionic dual theory
of the free massless Dirac fermion may loosely be written
\begin{equation}
\label{dualD1} {\cal L} = \bar{\chi}i \slashed{D}_a \chi + {1
\over 8\pi} ada +  \frac{1}{4\pi} A da + {1 \over 8\pi} AdA
\end{equation}
Again in our definition of the massless Dirac Lagrangian we have
left implicit a massive Dirac fermion $\chi_H$ that also couples
minimally to $a$. Eqn. \ref{dualD1} is, as written, not strictly
well-defined. For instance, the coupling to  $A$  is not gauge
invariant.  Later we will see how to refine the dual Dirac theory
to make it well-defined.  But for now notice the similarity of
this fermion-fermion duality to the charge-vortex duality of
bosons. In both cases the dual Lagrangian is a gauged version of
the original theory.  The field $\chi$ may loosely be interpreted
as a ``vortex" in the electron field $\psi$. Specifically it
corresponds to a $4\pi$ vortex - thus $\chi$ sees the density
$\rho_\psi$ of the original electrons as a magnetic flux $b = 4\pi
\rho_\chi$.  As we will see there is a close correspondence
between the $\chi$ field and the composite fermions that appear in
discussions of quantum Hall phenomena.  Indeed the
proposal\cite{son2015} that the composite fermion in that context
may be a Dirac fermion partly motivated this duality.

In addition to this proposed duality (known as a fermion-fermion
duality) other `bosonization' dualities\cite{seiberg1,karchtong}
which relate the free Dirac fermion to theories written in terms
of bosonic fields can be written down. The simplest is the WF
theory coupled to a dynamical $U(1)$ gauge field $b$ with a
Chern-Simons term at level-$1$. \beq \label{basicduality} i \bar
\Psi \slashed{D}_A \Psi \qquad \longleftrightarrow  \qquad
|D_b\phi|^2 - |\phi|^4 + {1 \over 4\pi} bdb +{1 \over 2\pi} bdA
\eeq

A closely related bosonization duality\cite{seiberg1,karchtong} takes the
form \beq
 \label{Tbasicduality}
i \bar \Psi \slashed{D}_A \Psi\qquad \longleftrightarrow  \qquad
|D_{ -\hat b} \hat \phi|^2 -|\hat \phi|^4 -{1 \over 4\pi} \hat
bd\hat b - {1 \over 2\pi} \hat bdA -{1 \over 4\pi} AdA-2CS_g. \eeq

We will explain the reasoning behind these dualities and their
relationship in subsequent sections.

\section{Fermion-boson duality, and fermion-fermion duality}
\label{DualityWeb}

We begin with the fermion-boson duality.  A crucial  physical
insight is provided by the flux attachment procedure  developed to
transmute statistics in two space dimensions. Flux attachment has
been successfully used in theories of quantum Hall phenomena for
many decades. Here we will provide a modern treatment that is well
suited to applying the flux attachment idea to the  CFTs of
primary interest to us.

Consider an electronic system in a translation invariant lattice
that is undergoing an integer quantum Hall transition (or
equivalently a transition from a Chern insulator to a trivial
insulator). As already described, in the absence of interactions a
continuum low energy description of this transition realizes the
free massless Dirac fermion. Now let us consider the same
electronic system but we allow for arbitrary short ranged
interactions that preserve both phases and admit a direct phase
transition between them. A different description of this system is
to use a parton (or slave boson) representation by writing the
electron operator as \beq \label{parton} \psi_r = \hat \phi_r  f_r
\eeq Here $r$ are the sites of the spatial lattice.   $\hat{\phi}_r$ is
a boson operator that carries the global $U(1)$ charge of the
electron while $f_r$ is a fermion that is neutral under the global
$U(1)$ symmetry. This representation comes with a $U(1)$ gauge
redundancy associated with the transformations \beq \hat \phi_r
\rightarrow \hat \phi_r e^{i\alpha_r}; ~~ f_r \rightarrow f_r e^{
-i\alpha_r} \eeq at each site $r$.  Correspondingly there is a
constraint $\hat \phi^\dagger_r \hat \phi_r =   f^\dagger_r f_r$
at each site $r$.

Given any particular microscopic Hamiltonian in terms  of $\psi_r$
we can clearly trade it for a description in terms of $(\hat
\phi_r, f_r)$. Here we are not interested in any specific
microscopic Hamiltonian but rather in the structure of any theory
of the underlying  electronic system when expressed in terms of
$\hat \phi, f$. This structure is largely determined by general
considerations. Clearly any effective theory in terms of $(\hat
\phi,f)$ must include a dynamical $U(1)$ gauge field which we
denote $\hat b_\mu$ under which $\hat \phi, f$ carry charges $- 1,
1$ respectively.  We also assign global $U_A(1)$ charges   of
$(1,0)$ to $(\hat \phi, f)$ to reproduce the global charge of the
electron. Thus we schematically write \beq {\cal L} = {\cal
L}[\hat \phi, A -  \hat b] + {\cal L}[f, \hat b] \eeq Given a
Lagrangian of this sort it is clear that local, {\em i.e} gauge
invariant (under $\hat{b}$) operators are precisely the same as in
the original electronic system.  To reproduce the integer quantum
hall phases and their transitions we now consider a specific
example of this Lagrangian\footnote{This may be motivated by
considering a parton `mean field ansatz'  where the $f$ fermions
form a band insulator (possibly with a Chern number $C = 0, -1$)
and the $\hat \phi$ bosons are described by a boson Hubbard model.
Replacing the theory of  both $f$ and $\phi$ by their continuum
versions and including the $U(1)$ gauge field $\hat b$ leads to
the Lagrangians below.}: \beqn
{\cal L}[\hat \phi, A - \hat b] & = & |D_{A - \hat b} \hat \phi|^2 - r|\hat \phi|^2 - u  |\hat \phi|^4 \nn \\
{\cal L}[f, \hat b] & = &  \bar{f} \left( i\slashed{D}_{\hat b} +
m \right) f \eeqn As before we assume that in defining ${\cal
L}[f, \hat b]$ that there is also a heavy fermion $\psi_H$ that
has a negative mass $M$.

Consider the phases of the theory when $m <  0$ and $\hat \phi$ is
uncondensed (i.e $\hat \phi$ is in a Mott insulator phase). As
$\hat \phi, f$ are both gapped they can be integrated out. The
resulting induced long wavelength action for $\hat b$ takes the
form \beq \label{Leff1} {\cal L}_{eff} = - {1 \over 4\pi} \hat
bd\hat b - 2CS_g ..... \eeq The ellipses include in particular a
Maxwell term for $A - \hat b$. The dynamics of $\hat b$ is
described by a $U(1)_1$ Chern-Simons action. The $\hat b$ can now
be integrated out and it is readily seen that it yields a trivial
confined gapped phase ({\em i.e} where $\hat{\phi}, f$  are
confined and physical excitations are just the original electrons
and their composites). Further there is no induced Chern-Simons
term for $A$; in other words when $m <  0$, and $\hat \phi$ is
uncondensed, we get a completely trivial insulator.  As $r, u$ are
changed and $\phi$ condenses, the $\hat b$ will be locked to the
external gauge field $A$.  Replacing $\hat b$ by $A$ in Eqn.
\ref{Leff1} we see that  we recover the integer quantum Hall
insulator with $\sigma_{xy} = -1$ (and a thermal Hall conductivity
$-1$).  As $m <  0$ on  both sides and at the transition itself we
can describe the vicinity of the transition by integrating out $f$
(but not $\hat \phi$).

A continuum theory for the transition  thus takes the form (after
a shift $\hat b \ra \hat b - A$) \beq \label{cbeff} {\cal L}_{cb}
= {\cal L}[\hat \phi, -\hat b ] - {1\over 4\pi} \hat bd \hat b  -
{1 \over 2\pi} \hat bdA -{1 \over 4\pi} AdA- 2 CS_g \eeq Following
standard terminology we will refer to $\hat \phi$ as composite
bosons. Thus in this parton description the integer quantum hall
transition of electrons is mapped to the superfluid-insulator
transition of the composite boson $\hat \phi$ in the presence of a
Chern-Simons gauge field.

To see the connection with the familiar flux attachment ideas,
consider the equation of motion of $\hat b$: \beq j^\hat{\phi} =
{1 \over 2\pi} (db + dA) \eeq The left hand side is the 3-current
of $\hat \phi$. In the absence of the background gauge field $A$
this means that a $2\pi$ flux of $\hat b$ is attached to each
$\hat \phi$ particle. Thus - as usual - we can think of the
composite boson $\hat \phi$ as being obtained from the original
electrons by attaching $2\pi$ flux.

It is powerful to recast this intuition in terms that are suitable
even at a putative quantum critical point where there are gapless
excitations and the standard flux attachment procedure is a bit
subtle.  To that end consider the monopole operator
$\mathcal{M}_{\hat b}$.  This destroys a $2\pi$ flux of the gauge
field $\hat b$. Due to the Chern-Simons terms for $\hat b$,  a
$2\pi$ flux carries a gauge charge $q_{\hat b} = 1$ and a global
$U_A(1)$ charge $1$. Thus  $\mathcal{M}_{\hat b}$ is by itself not
gauge invariant under the internal $U_\hat b(1)$. However the
bound state $\hat{\phi} \mathcal{M}_{\hat b}$ is gauge invariant,
and carries global $U_A(1)$ charge $1$. Further the $\hat b$
Chern-Simons term implies that this operator has spin-$1/2$ under
spatial rotations ({\em i.e} it is a fermion operator) . Thus we
identify this  with the physical electron $\psi$: \beq \psi =
\hat{\phi} \mathcal{M}_{\hat b} \eeq

Now consider the composite boson theory in Eqn.  \ref{cbeff} when
$r, u$ are tuned to the phase transition associated with the Higgs
condensation of $\hat \phi$.  We {\em assume} that this transition
is second order. Following the same logic as in the discussion of
the bosonic charge-vortex duality we may try to access this
critical point by starting with the Wilson-Fisher fixed point of
the $\hat \phi$ theory, and then coupling in the gauge field $\hat
b$ with dynamics given by the Chern-Simons terms of Eqn.
\ref{cbeff}.  We write the resulting theory as \beq \label{cbcft}
|D_{- \hat b} \hat \phi|^2 - |\hat \phi|^4 -{1 \over 4\pi} \hat
bd\hat b - {1 \over 2\pi} \hat bdA -{1 \over 4\pi} AdA-2CS_g \eeq

We have thus far argued that this Lagrangian describes a theory of
electrons with global $U_A(1)$ charge $1$, ({\em i.e} all local
operators are electrons or their composites),  and that it
describes an integer quantum Hall phase transition where the Hall
conductivity jumps from $-1$ to $0$.  We know that a different
theory for this same phase transition is just the free massless
Dirac fermion. This then leads us to {\em conjecture} that the
theory of Eqn. \ref{cbcft} is {\em equivalent} in the IR to the
free massless Dirac fermion. This is precisely the bosonization
duality statement of Eqn. \ref{Tbasicduality}.

Now we perform the bosonic charge-vortex duality on Eqn.
\ref{cbcft} to get a Lagrangian \beq |D_b \phi|^2 - |\phi|^4 - {1
\over 2\pi} \hat b db -{1 \over 4\pi} \hat bd\hat b - {1 \over
2\pi} \hat bdA -{1 \over 4\pi} AdA-2CS_g \eeq The only terms
involving $\hat b$ now are the Chern-Simons terms.  The $U(1)_1$
Chern-Simons term for $\hat b$ leads to a trivial theory (with
fermions). The $\hat b$ integral can now be readily done and leads
exactly to the Lagrangian \beq |D_b\phi|^2 - |\phi|^4 + {1 \over
4\pi} bdb +{1 \over 2\pi} bdA \eeq which is precisely the right
side of the other proposed bosonization duality of Eqn.
\ref{basicduality}.

We have thus motivated both bosonization dualities. They however
pose a crucial puzzle. We know that the free massless Dirac
fermion is time reversal invariant upto an anomaly. But how can
the bosonized theories possibly be time reversal invariant  even
allowing for the possibility of an anomaly? From a condensed
matter perspective  the  flux attachment procedure seems to
manifestly break time reversal. How then do we  know that the
bosonized theories are properly time reversal invariant as they
need to be if the dualities are correct? We will see below that
(much like in the $(1+1)d$ Ising/Majorana dualities) time reversal
is a `quantum symmetry' of the bosonized theories.  In the
bosonized versions it is implemented as the duality that
interchanges the two distinct bosonized theories.

\subsection{The duality web}

The methods we have described so far can be extended to build a
full web of dualities.  We now provide a compact way to summarize
and think about the duality web in terms of two elementary
operations $S$ and $T$ defined in the space of quantum field
theories with a global $U(1)$ symmetry.

For any $(2+1)d$ CFT with a global $U(1)$ symmetry, the   $S$
operation is  defined in terms of its action on the path integral
as follows:
\begin{equation}
\label{Sdef} Z_S[B] = \int {\cal D}A ~ Z_{CFT_1}[A]
e^{\frac{i}{2\pi} \int d^3x AdB}
\end{equation}
Here $Z_{CFT_1}$[A] is the partition function of the $(2+1)d$ CFT
in the presence of a {\em background} $U(1)$ gauge field $A$. The
$S$ operation converts this  background gauge field  into a
dynamical one but without including a kinetic term. $B$ is a new
background $U(1)$ gauge field that couples to $\frac{1}{2\pi} dA$
which is conserved. This operation was defined and used by
Kapustin and Strassler\cite{kapstr}, and by Witten\cite{witten03}.
The different operation $T$ was also introduced by Witten - it
simply shifts the level of the Chern-Simons term of the background
gauge field by $1$.\footnote{Strictly speaking, if we want to
avoid introducing charge-neutral fermions  (spin-structure
dependence) into the theory, we should not add $\frac{1}{4\pi}AdA$
alone. For a spin$_c$ structure we should add
$\frac{1}{4\pi}AdA+2CS_g$, while for an ordinary gauge field $B$
we should add $\frac{1}{4\pi} BdB+\frac{1}{2\pi}BdA$ where $A$ is
a spin$_c$ structure. These rules are familiar in the quantum Hall
literature where the probe gauge field coupling to the physical
electrons is a spin$_c$ structure, while the dynamical gauge
fields in a $K$-matrix Chern-Simons theory are typically ordinary
$U(1)$ gauge fields. }

From a formal point of view  $Z_S[B]$ is the partition function of
a new theory with a new global $U(1)$ symmetry (to whose currents
the background gauge field $B$ couples). Further formally the
path-integral over $A$ is conformally invariant - it is to ensure
this that no kinetic term for $A$ is introduced in the definition
of $S$. Thus $Z_S$ will then define a new conformal field
theory\footnote{Caution is needed here; this assumes that the
path-integral on the right hand side of Eqn. \ref{Sdef} is
well-defined. In principle we need to define it as the limit of a
regularized theory - for instance we could add a Maxwell term for
$A$ with a coupling $e^2$ and take the limit $e^2 \ra \infty$. It
is not a priori clear that the limit exists. Physically this means
that turning on a  coupling to a dynamical gauge field may lead to
first order transition. In common with much of the literature we
will simply {\em assume} that this does not happen and that
$Z_S[B]$ and the other formal manipulations below are well defined
for the theories we will consider here.  For more discussion of
these and other concerns, see Appendix C of Ref. \cite{SO5}.
} which we denote CFT$_2$. Note that the theories obtained by
either $S$ or $T$ are, in general, inequivalent to the original
theory.

Schematically we write the $S$ operation  as  $S[CFT_1] = CFT_2$
where both CFTs have a global $U(1)$ symmetry.  The combination of
$S$ and $T$ then leads to an $SL(2,Z)$ action in the space of
$(2+1)d$ CFTs with a global $U(1)$ symmetry\cite{witten03}.
Specifically the $S$ and $T$ can be formally shown to satisfy the
equations\footnote{$S^2 = -1$ means that the $S^2$ theory has the
sign of the gauge coupling reversed compared to the original
theory, {\em i.e}, it is the charge-conjugated version.} $S^2 = -
1, \left(S T \right)^3 = 1$ which together generate $SL(2, Z)$.

Let us first recast some of the $(2+1)d$ dualities we have
discussed so far in the language of these $S$ and $T$ operations.
We denote the Wilson-Fisher fixed point of the $3D$ XY model as
$WF$ and the free masless Dirac fermion as $D$. Then the classic
bosonic charge-vortex duality of this CFT may be compactly written
as the equality \beq \label{bcvst} [WF] = S[WF] \eeq The
bosonization duality of the Dirac fermion in Eqn.
\ref{basicduality} becomes the equality \beq \label{basicdst} [D]
= ST[WF] \eeq The second bosonization duality in Eqn.
\ref{Tbasicduality} is written \beq \label{tbasicdst} [D] = T^{-1}
S^{-1} T^{-1}[WF] \eeq

The equivalence of the two bosonization dualities finds  compact
expression in this notation. Indeed using Eqns. \ref{bcvst} and
\ref{basicdst} we may write $[D] = STS[WF]$. Now the equation
$(ST)^3 = 1$ then implies that $ STS = T^{-1}S^{-1}T^{-1}$ and we
immediately get Eqn. \ref{tbasicdst}.

Given these equalities we can apply any combination of $S$ and $T$
to generate other dualities.  Inverting each of Eqns.
\ref{basicdst} and \ref{tbasicdst} we find fermionized versions of
the $3D$ XY Wilson-Fisher fixed point: \beqn \label{wffermion1}
[WF] &  = & T^{-1} S^{-1}[D]  \\
\label{wffermion2} [WF] & = & TST[D] \eeqn The first of these was
conjectured many years ago\cite{wufisher,BarkeshliMcGreevy}.  In
Appendix~\ref{fparton} we provide a flux-attachment/parton
understanding of this duality, similar to what we did  in the
previous subsection. As with the bosonization duality of the Dirac
fermion,  a long-standing concern about this conjecture was about
how the fermionized version could possibly be time reversal
symmetric (which the $[WF]$ theory manifestly is). The
resolution\cite{seiberg1}  once again is that time reversal is
realized as a quantum symmetry as we will discuss later.

For practice let us explicitly write out the Lagrangians for these
two dualities: Eqn. \ref{wffermion1} becomes \beq \label{wffex1}
|D_A \phi|^2 -  |\phi|^4 \leftrightarrow i\bar{\psi}
\slashed{D}_a\psi - \frac{1}{2\pi} Ada - \frac{1}{4\pi} AdA
 \eeq
 and Eqn. \ref{wffermion2} becomes
 \beq
 \label{wffex2}
|D_A \phi|^2 -  |\phi|^4 \leftrightarrow  i\bar{\psi}
\slashed{D}_a\psi  + \frac{1}{4\pi} ada + \frac{1}{2\pi} Ada +
\frac{1}{4\pi} AdA + 2CS_g
  \eeq
We emphasize that in our definition a massless Dirac fermion
always comes with a regulator  in the form of a heavy Dirac
fermion that couples to the same gauge field. Thus  the theory in
the right side of Eqn. \ref{wffex1} is often referred to as as ``a
fermion coupled to $U(1)_{-1/2}$" . This loosely means that the
Dirac fermion is coupled to a $U(1)$ gauge field with a
level-$-1/2$ Chern-Simons term (which is a short hand for
remembering the presence of the massive Dirac fermion).

Notice that we have obtained two different fermionic duals for
$[WF]$ which must therefore be equivalent to each other: \beq
 TST[D] = T^{-1} S^{-1}[D]
\eeq

Acting on both sides with $T^{-1}S^{-1}T^{-1}$ then gives us
fermionic duals of the free massless Dirac fermion: \beq
\label{dualDst} [D]  =  T^{-1} S^{-1} T^{-2} S^{-1}[D] \eeq

Let us write down the right side of the last line. The Lagrangian
of this theory is \beq \label{dualD1} \bar{\chi} i\slashed{D}_a
\chi - {1 \over 2\pi} adb - {2 \over 4\pi} bdb - {1 \over 2\pi} b
dA - {1 \over 4\pi} AdA - 2CS_g \eeq This then is the precise
version of the proposed fermionic dual of the free massless Dirac
fermion. Note that if we naively integrate out $b$ we recover the
loose form of the fermionic dual theory described earlier.
Alternately since $S^{-1}$ is the same as $S$ after charge
conjugation we can write the dual Dirac theory as \beq
\label{dualD2} {\cal L} =  \bar{\chi} i\slashed{D}_a \chi + {1
\over 2\pi} adb - {2\over 4\pi} bdb + {1 \over 2\pi} b dA - {1
\over 4\pi} AdA - 2CS_g \eeq

It is interesting to take the fermion-fermion duality as given and
ask what happens if we apply it to the Dirac fermion theory
appearing in the right of Eqn. \ref{wffermion1}. Clearly we  then
end up with the second duality Eqn. \ref{wffermion2}.  Thus we may
regard the Dirac fermion of Eqn. \ref{wffermion2} as the dual
fermionic vortex of the fermion in Eqn. \ref{wffermion1}. Note
again the similarity with our discussion of the two bosonization
dualities of the free Dirac theory.

\subsection{Symmetry realization}

The global $U(1)$ symmetry of all theories involved in the various
dualities is explicit and determines the coupling to the
background $U(1)$ gauge field.  As we have indicated on several
occasions the fate of time reversal is less clear. We now address
the realization of time reversal symmetry. We use  the symbolic
description of the previous subsection in terms of $S$ and $T$.
Let us denote the time reverse transform of a theory by $\T$. The
Wilson-Fisher theory is explicitly time reversal invariant. We
therefore write \beq \T([WF]) = [WF] \eeq The time reversal
transform of the massless Dirac theory is given by Eqn.
\ref{Tdirac} and reflects the parity anomaly.  We see that time
reversal is equivalent to performing a $T$ operation. We write
this as \beq \T([D] )= T[D] \eeq

From the definitions of $S$ and $T$ we notice that \beqn
\T(S[CFT]) & = &  S^{-1}\T[CFT] \\
\T(T[CFT]) & = & T^{-1}\T[CFT] \eeqn

Now let us examine the time reversal properties of the bosonized
duals of the free Dirac theory. Acting with $\T$ on both sides of
Eqn. \ref{basicdst}, we find \beq T[D] = S^{-1}T^{-1}[WF]~~= T [
T^{-1}S^{-1} T^{-1}[WF]] \eeq Thus time reversal takes the first
bosonic dual to the second one (upto the same anomaly as the Dirac
theory). Since the two bosonized versions are themselves related
by the standard charge-vortex duality, {\em i.e} the $S$
transformation, we see that time reversal is implemented as a
duality transformation on the bosonized theories. Note the close
similarity to the realization of the chiral $\Z_2$ symmetry in the
Ising/Majorana duality web in $(1+1)d$.

A similar phenomenon happens for the two fermionized dualities of
the Wilson-Fisher theory (Eqns. \ref{wffermion1} and
\ref{wffermion2}).  Applying $\T$ to both sides of Eqn.
\ref{wffermion1} gives \beq [WF] = TST[D] \eeq which is the second
duality.  Thus  time reversal is implemented in the fermion side
as a quantum symmetry and  acts as a duality that interchanges the
two fermionic versions.

One can also physically visualize the unconventional time-reversal
transforms as follows. A Dirac fermion can be viewed as a
composite of a boson and its vortex -- a structure revealed by the
bosonization duality or even the traditional composite-boson
theory. One should view the boson and the vortex to be displaced
from each other by a distance $\vec{d}$, which gives an
interpretation of the emergent Dirac spinor structure. A similar
picture for the Dirac charge-vortex duality was discussed in the
context of a half-filled Landau level\cite{wangsenthil2}. The
notion can also be defined precisely at the operator level when
the theory is appropriately UV completed, for example on
coupled-wire systems\cite{mrossduality, mrossduality2}, which we
review in Sec.~\ref{wire}. Under the fermionic time-reversal
transform, the Dirac fermion keeps its integrity but flips its
Dirac spin. The only way to make the picture consistent is for the
boson and vortex to be exchanged under time-reversal, giving rise
to the non-local time-reversal action in the bosonization duality.
Now under a bosonic time-reversal transform, the boson keeps its
integrity while the vortex becomes an anti-vortex. The fermion
(boson $+$ vortex) now becomes another fermion (boson $+$
anti-vortex), which is a relative $4\pi$-vortex of the original
fermion (since the vortex and anti-vortex differ by a two-fold
vortex), this is nothing but the vortex dual of the Dirac fermion.
This gives the non-local time-reversal action in the
fermionization duality.

\section{Relation to $(3+1)d$ electromagnetic duality}
\label{EMDuality}

An intuitive and physical  understanding of the $(2+1)d$ dualities
we have discussed thus far comes from viewing them as boundary
theories of a $(3+1)d$ system. We discussed this for the bosonic
charge-vortex duality in Sec. \ref{bcvem}. We now sketch the basic
ideas of this perspective.

As in Sec. \ref{bcvem} we will regard the $(2+1)d$ theory as
living at the boundary of a $(3+1)d$ system with a gap to all
excitations. We also extend the background $U(1)$ gauge field to
the inside of the $(3+1)d$ system. Next we modify the theory  by
making this $U(1)$ gauge field dynamical. The bulk should then be
viewed as a $(3+1)d$ quantum liquid of a UV spin/boson system with
an emergent photon. The photon will be gapless but there will be
electrically and/or magnetically charged quasiparticles as gapped
excitations described by a charge-monopole lattice labelled by
charges $(q_e, q_m)$.

\begin{figure}[h]
\begin{center}
\includegraphics[width=1.9in]{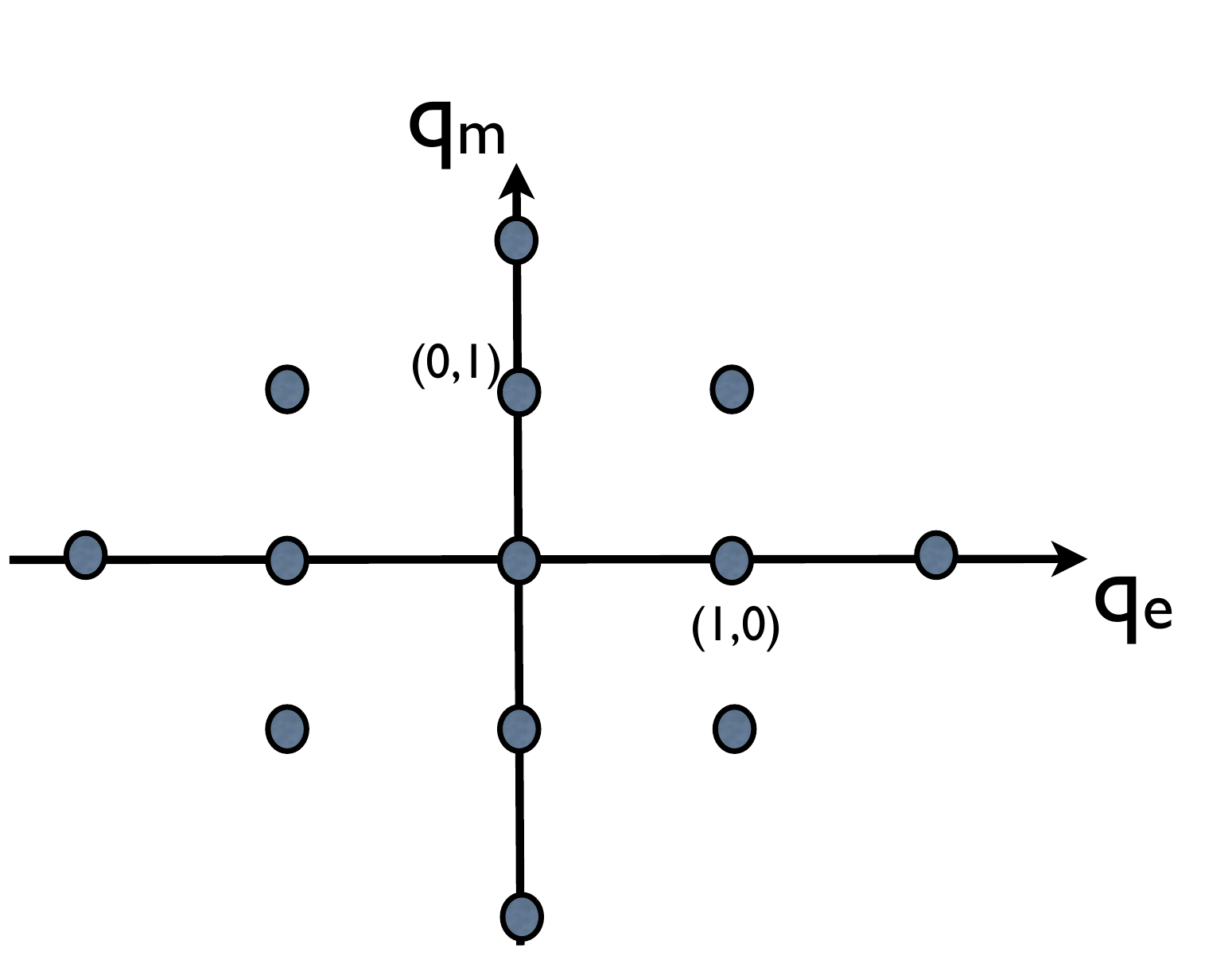}
\end{center}

\caption{Charge-monopole lattice at $\theta = n\pi$ with $n$ even.
} \label{cmlat1}

\end{figure}

\begin{figure}[h]
\begin{center}
\includegraphics[width=1.9in]{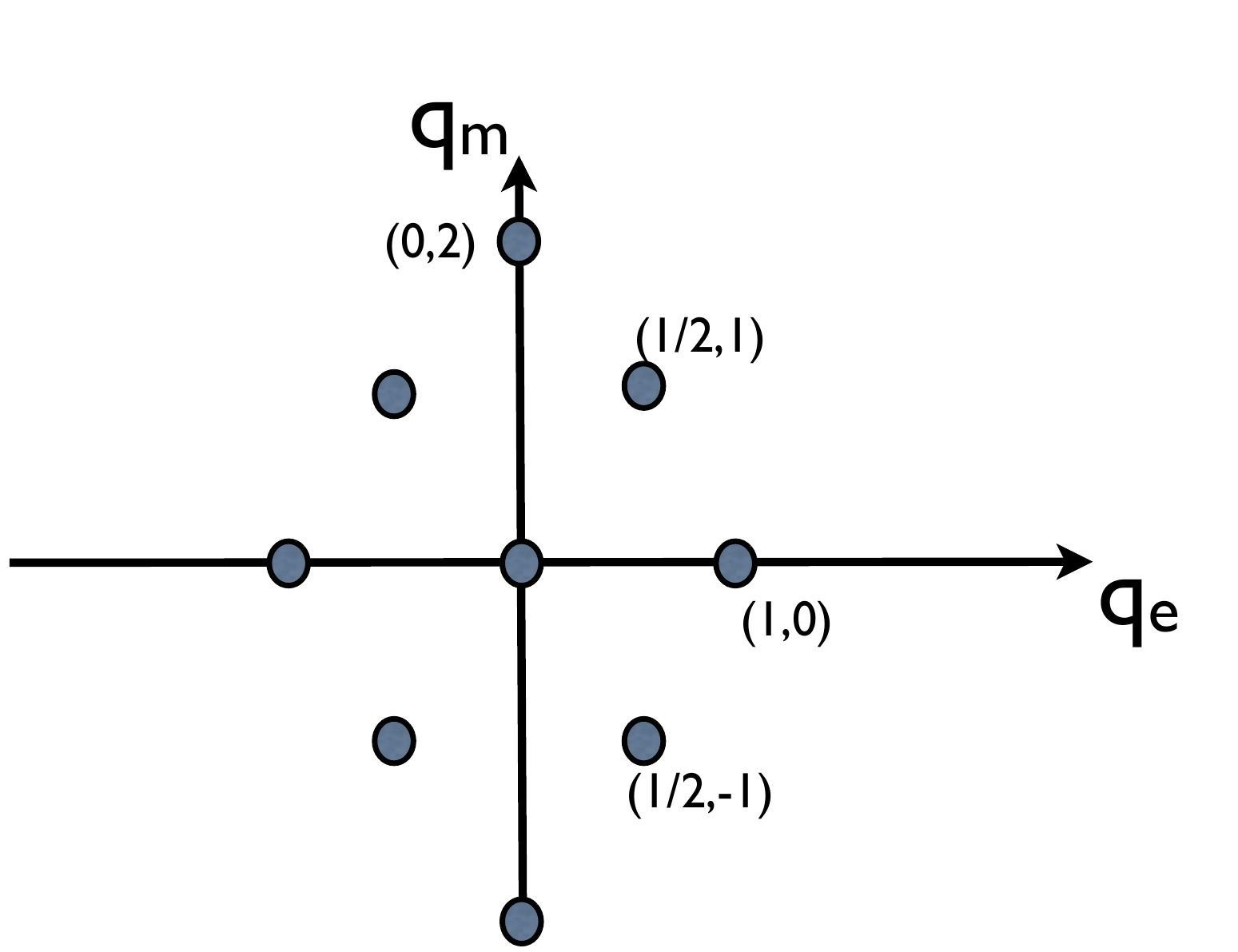}
\end{center}

\caption{ Charge-monopole lattice at $\theta = n\pi$ with $n$ odd.
} \label{cmlat2}

\end{figure}

Let us begin with the Lagrangian of the free Maxwell theory. To
fully discuss this theory we must allow for a $\theta$
term\footnote{Strictly speaking to fully make contact with the
earlier discussion we should also placing the theory on a
non-trivial manifold with a metric $g$, and distinguish between
$U(1)$ gauge fields and spin$_c$ connections. In particular we
should allow for gravitational theta terms which will  yield the
gravitational Chern-Simons terms of the boundary theories. In our
discussion below we will assume we are in flat space-time $R^4$ to
understand the essential idea. The generalization to arbitrary
oriented space-time manifolds is straightforward.}  : \beq S_M =
\int d^4x \frac{1}{4e^2} F^{\mu \nu} F_{\mu \nu} +
\frac{\theta}{32 \pi^2} \epsilon_{\mu \nu \lambda \kappa} F^{\mu
\nu} F^{\lambda \kappa} \eeq As is well known the presence of the
theta term manifests itself on the spectrum through the Witten
effect: there is an induced electric charge ${\theta \over 2\pi}$
on the monopole. The theta term has the effect of tilting the
charge-monopole lattice.  In Fig. \ref{cmlat1} we show the
charge-monopole lattice for $\theta = n\pi$ ($n$ even) and in Fig.
\ref{cmlat2} for $\theta = n\pi$ ($n$ odd).

\begin{figure}[hh]
\begin{center}
\includegraphics[width=5in]{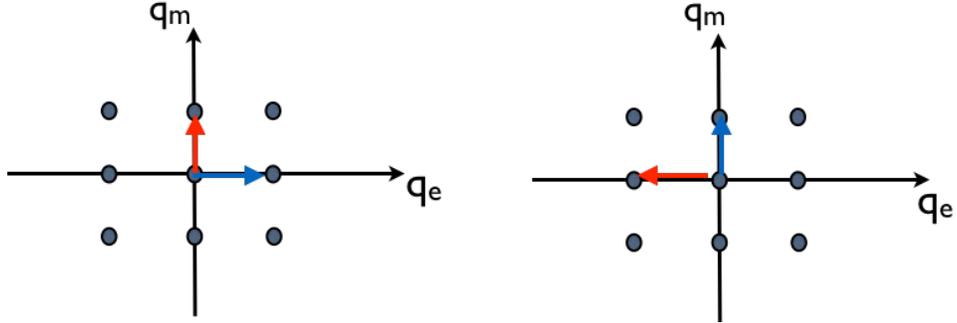}
\end{center}

\caption{Two different basis choices for the charge-monopole
lattice at $\theta = n\pi$ with $n$ even.   The blue arrow points
to what - in that basis - is the electric charge and the red arrow
points to the corresponding magnetic charge. The basis to the left
is the standard one while that to the right is obtained by a 90
degree rotation, {\em i.e} by an $S$-transformation. }
\label{cmlatbsis1}

\end{figure}

\begin{figure}[hh]
\begin{center}
\includegraphics[width=5in]{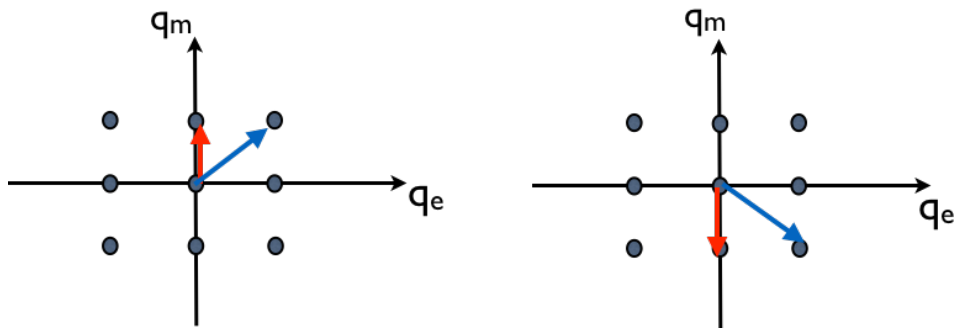}
\end{center}

\caption{Two other basis choices for the charge-monopole lattice
at $\theta = n\pi$ with $n$ even.  Note that time reversal does
not keep the basis vectors fixed. Rather the basis in the left
figure is transformed to the one in the right figure and vice
versa. } \label{cmlatbsis2}

\end{figure}

We note that Maxwell electrodynamics has a large set of dualities
that can be summarized by different basis choices of the
charge-monopole lattice. For some examples, see Figs.
\ref{cmlatbsis1} and \ref{cmlatbsis2}.  In the bulk we can define
two elementary operations $S$ and $T$ which act on the lattice as
follows: \beqn \label{bulkS}
S: (q_e, q_m) &  \ra & (q_m, -q_e)  \\
\label{bulkT} T: (q_e, q_m) & \ra & (q_e + q_m, q_e) \eeqn
Together these generate  an $SL(2,Z)$ group of transformations
which leaves the charge-monopole lattice invariant. We thus have
many equivalent points of view on the $U(1)$ spin liquid. We pick
any basis we want for the charge-monopole lattice and couple one
of the basic particles to a dynamical $U(1)$ gauge field.

In the presence of the boundary this bulk $SL(2, Z)$
transformation generates a corresponding $SL(2,Z)$ transformation
of the boundary theories. We can describe the boundary from the
point of view of any particle in the lattice. The same phase/phase
transition at the boundary will then have multiple equivalent
descriptions, {\em i.e} dualities, depending on the bulk particle
chosen. In  Sec. \ref{bcvem} we  compared the boundary
descriptions from the $(1,0)$ (electric) and $(0,1)$ magnetic
points of view, and showed how they were related to the bosonic
charge-vortex duality in $(2+1)d$. Here we generalize this bulk
description.

Now it is well known that the bound state of a bosonic charge and
a bosonic monopole in three space dimensions yields a
fermion\cite{asgdyon,statwitt}. Thus if in Fig. \ref{cmlat1} the
$(1,0)$ and $(0,1)$ are both bosons then the $(1, 1)$ and $(1,
-1)$ dyons are both fermions. The dual theory Eqn.
\ref{wffermion1} corresponds to describing the surface from the
point of view of the excitation that corresponds, in the bulk, to
the $(1,1)$ dyon.  Likewise we can identify the other dual theory
Eqn. \ref{wffermion2} as describing the surface from the point of
view of the $(1, -1)$ dyon.

This relation to the bulk gives an appealingly simple
understanding of the non-trivial action of time reversal on the
$(2+1)d$ web of dualities. Though there are many equivalent basis
choices for the  charge-monopole lattice,  the action of time
reversal symmetry $\T$ (and similarly for $\CT$) may be
non-trivial depending on the basis choice. This is illustrated in
Fig. \ref{cmlatbsis1} and Fig. \ref{cmlatbsis2}.  The standard
basis choice (left of Fig. \ref{cmlatbsis1}) leaves the $E$
particle invariant under time reversal $\T$ while the $M$ goes to
its antiparticle.  In the S-transformed theory the new electric
particle $E'$ is precisely the $M$ particle of the original theory
(see Fig. \ref{cmlatbsis2}) while the new magnetic charge $M'$ is
$E^{-1}$.  Clearly under $\T$, $E'$ goes to its antiparticle while
$M'$ goes to itself.  Thus an S-transformation of the bulk theory
interchanges $\T$ and $\CT$ symmetry\footnote{Equivalently we may
say that in the S-transformed theory, time reversal is implemented
as $S^2\T$ as $S^2$ is precisely the charge conjugation
operation.}. This is clearly related to the interchange between
$\T$ and $\CT$ in the bosonic charge-vortex duality.    With other
basis choices, such as in Fig. \ref{cmlatbsis2}   time reversal
acts in a more drastic manner and  is implemented as a combination
of $\T$ and  a non-trivial further $SL(2,Z)$ transformation.

All of the non-trivial $\T$ actions on the $(2+1)d$ duality web
can be given simple pictorial descriptions in terms of actions on
different basis choices of the charge-monopole lattice of the
$(3+1)d$ $U(1)$ gauge theory.

Below we will flesh this out more formally and precisely. As in
Sec. \ref{bcvem} consider the bulk Maxwell theory obtained by
integrating out all matter fields, and in the absence of any
boundaries, {\em i.e} on a closed orientable 4-manifold.   As it
will in general be necessary we will  keep both the Maxwell and
the theta term.   It is convenient to combine $e^2$ and $\theta$
into a single complex coupling constant $\tau$ defined in the
complex upper-half plane; \beq \label{tau} \tau =
\frac{\theta}{2\pi} + \frac{2\pi i}{e^2} \eeq

Repeating the same steps as in Sec. \ref{bcvem} it is readily seen
that the duality transformation maps the theory at a coupling
$\tau$ to the same theory at a different coupling $\tau'$ where
\beq \label{Stau} \tau' \equiv S(\tau) = - {1 \over \tau} \eeq We
have denoted this transformation of the coupling constant $S$ as
it   affects the matter fields by precisely the $S$ transformation
of Eqn. \ref{bulkS}. The $T$ operation of Eqn. \ref{bulkT} is
reproduced by shifting $\theta \ra \theta + 2\pi$. By the Witten
effect this changes the electric charge of a strength $q_m$
monopole by precisely $q_m$ in agreement with Eqn. \ref{bulkT}.
The effect of this shift on the complex coupling constant $\tau$
is \beq \label{Ttau} T(\tau) = \tau + 1 \eeq

As expected the $S$ and $T$ operations on $\tau$ generate an
$SL(2,Z)$ transformation\footnote{ Note that an element of
$SL(2,\Z)$ is a $2\times 2$ integer-valued matrix of determinant
1: \beq
\begin{pmatrix}
a& b \\
c& d
\end{pmatrix} , ~~ad-bc=1,
\eeq Note that the element $-1\in SL(2,\Z)$ acts trivially on
$\tau$ (so the group that acts faithfully on $\tau$ is actually
the quotient group $SL(2,\Z)/\{\pm 1\}=PSL(2,\Z)$).  Accordingly,
$-1$ is a symmetry for any $\tau$ and can be shown to simply be
charge conjugation. }
  $\tau \ra M(\tau)$
\beq M(\tau)={a\tau+b\over c\tau+d}. ~~~ad-bc = 1 \eeq

In Sec. \ref{bcvem} we saw the effect of the $S$-operation on the
bulk in the presence of a boundary $(2+1)d$ theory. Specifically
if we define
 \beq
 \label{zsbgen}
 Z_{SB}[\tau] = \int [{\cal D}A ] Z_{CFT}[A] e^{-\int d^4x  S_M[A; \tau]}
 \eeq
then after the $S$-transformation, we get
 \beq
Z_{SB}[\tau] =  \int [{\cal D}A']  \left( [{\cal D}A] Z_{CFT}[A]
e^{{i \over 2\pi} \int d^3x AdA'}\right)e^{-\int d^4x  S_M[A'; -{1
\over \tau}]}
 \eeq
As before inside the $\left( \right)$ the integration is only over
the boundary values of $A$.

Since the bulk $T$-operation corresponds to a shift $\theta \ra
\theta + 2\pi$, it follows that its effect on the surface is to
shift the  level  of the Chern-Simons term at the surface by $1$.
Clearly the bulk $S$ and $T$ operations are closely related to the
$S$ and $T$ operations introduced earlier in $(2+1)d$.

As explained above describing the surface of the same theory in
terms of different bulk excitations gives rise to $(2+1)d$
dualities. The corresponding basis change in the bulk is
implemented - in the low energy free Maxwell theory - by the
$SL(2, Z)$ transformation. Let us start with the fermionic duals
of the Wilson-Fisher theory. As explained above we can think of
these as a description of the surface in terms of what in the bulk
is either the $(1,1)$ or $(1,-1)$ dyon.  To go from a
representation of the bulk in which the basic electric charge is
the bosonic $E$ particle ($(q_e = 1, q_m = 0)$) to one where it is
the fermionic $(1,1)$ particle, we transform by $S^{-1}T^{-1}$.
Consider therefore the partition function of Eqn. \ref{zsbgen}
when the $CFT$ is the $WF$ theory and a general coupling constant
$\tau$. \beq
 \label{zsbgenwf}
 Z_{SB}^{WF}[\tau] = \int [{\cal D}A ] Z_{WF}[A] e^{-\int d^4x  S_M[A; \tau]}
 \eeq

The ST transformation  changes the bulk coupling constant to \beq
\tau' = -{1 \over \tau - 1} \eeq In the resulting theory the role
of $E$ is now played by the $(1,1)$ fermion.  Consider therefore
the partition function of Eqn. \ref{zsbgen}  where the surface is
tuned so that this is a massless Dirac fermion: \beq
\label{zsbgenD}
 Z_{SB}^D[\tau'] = \int [{\cal D}A ] Z_{D}[A] e^{-\int d^4x  S_M[A; \tau']}
\eeq We now assume that \beq \label{sbwfd} Z_{SB}^{WF}[\tau] =
Z_{SB}^D[\tau'] \eeq This is similar to the assumption in Eqn.
\ref{sbselfd} which was used to find a bulk interpretation of the
standard bosonic charge-vortex duality.  Note that this equality
is certainly true in the absence of the boundary. So the real
assumption is that with the particular boundary theories of both
sides the equality continues to be true. Eqn. \ref{sbwfd} relates
theories at two different couplings $\tau$ and $\tau'$. We can
however transform Eqn. \ref{zsbgenD} to a theory defined at
coupling $\tau$ by doing the inverse of $S^{-1}T^{-1}$, {\em i.e}
$TS$ in the bulk.  This yields \beqn
Z_{SB}^D[\tau'] & \equiv & \tilde{Z}_{SB}^D[\tau] \\
\tilde{Z}_{SB}^D[\tau] & =  &  \int [{\cal D}A ']\left(  \int
[{\cal D}A] Z_{D}[A]  e^{- \int d^3x {i \over 2\pi}AdA' - {i \over
4\pi} A'dA' } \right) e^{-\int d^4x  S_M[A'; \tau]} \eeqn We
therefore find
 \beq
   Z_{SB}^{WF}[\tau] = \tilde{Z}_{SB}^D[\tau]
   \eeq
Now we can specialize to the limit $\tau =  i \infty$ in which
case the bulk gauge field becomes a background gauge field thereby
finding
  \beq
Z_{WF}[A] =  \int [{\cal D}A']  Z_{D}[A']  e^{ -  \int d^3x {i
\over 2\pi} A'dA -  {i \over 4\pi} AdA}
  \eeq
This is precisely one of the fermionic duals of the Wilson-Fisher
fixed point.

All of the members of the duality web can be given similar bulk
interpretations.  As another example consider the fermion-fermion
duality from the bulk point of view.

We start with the free massless Dirac electrons in $(2+1)d$ and
view it as living at the boundary of a $(3+1)d$ system with gapped
electrons. A concrete physical realization of such a system is a
topological insulator of electrons with $U(1) \rtimes \T$
symmetry. It is assumed that $\T^2 = -1$ when acting on a single
electron. Now consider gauging the global $U(1)$ symmetry of this
system (surface + bulk).  Let us consider the fate of the
charge-monopole lattice.  As is well known in such a topological
insulator a magnetic monopole of strength $1$ has a fractional
electric charge $\frac{1}{2} ~(mod ~Z)$.  Correspondingly the
charge-monopole lattice takes the form shown in Fig. \ref{cmlat2}.
Equivalently in the absence of any boundaries the induced Maxwell
action of the topological insulator includes a $\theta$ term at
$\theta = \pi$.  The $(q_e = \frac{1}{2}, q_m = \pm 1 )$ dyons are
both bosons which are time reversal partners of each other. The
$(q_e = 0, q_m = 2)$ monopole is however a fermion.

The discussion presented thus far described a $(3+1)d$ $U(1)$
quantum liquid  obtained by gauging a fermionic topological
insulator, {\em i.e} we view it as a topological insulator of a
fermionic $(1,0)$ particle. Interestingly this same $U(1)$ gauge
theory can also be viewed as a time reversal invariant topological
insulator of the $(0,2)$ particle. This is readily seen from the
structure of the charge-monopole lattice. Note however that time
reversal takes $(0,2)$ to $(0,-2)$ but $\CT$ takes $(0,2)$ to
itself. Thus we may view the same phase as a gauged fermionic
topological insulator with $U(1) \times \CT$
symmetry\footnote{This is denoted class $AIII$ in the condensed
matter literature.}.

The dual description of this time reversal symmetric gauged
topological insulator suggests that, in the presence of a
boundary, the original ungauged free massless Dirac fermion will
have a dual description in terms of fields that are the surface
descendents of the $(0,2)$ monopole.  These will be fermionic
fields as the bulk $(0,2)$ particle is a fermion, and following
the discussion of the bosonic charge-vortex duality, will be
coupled to $U(1)$ gauge fields . Further as the bulk is a
topological insulator in terms of the $(0,2)$ particle it is
natural that these dual fermions are also massless Dirac fermions.

These considerations directly motivate - from this bulk point of
view - the fermion-fermion duality. Here we sketch how to obtain
the more precise form along the lines of our formal discussion
above.  Consider the theory in Eqn. \ref{zsbgenD} at some general
coupling $\tau$. It is convenient to label the points  of the
charge-monopole lattice by their coordinates $(n_e, n_m)$ for real
$\tau = \frac{2\pi i}{e^2}$. When $\theta \neq 0$, these evolve
into the lattice points $(q_e = n_e + \frac{\theta}{2\pi} n_m, q_m
= n_m)$. Thus what is the $(0,2)$ monopole at $\theta = \pi$ will
be labeled by $(n_e = -1, n_m = 2)$.  To go from a representation
in terms of the $E$-fermion to one  in terms of  this $(n_e = -1,
n_m = 2)$ particle  we do an $ST^2ST$ transformation\footnote{It
is readily checked that the action of $ST^2ST$ takes $(n_e = -1,
n_m = 2)$ to $(n_e = 1, n_m = 0)$.}. This takes the coupling
$\tau$ to $\tau'$ where \beq \label{tau'ff} \tau' =
-\frac{\tau+1}{2\tau + 1} \eeq The bulk free Maxwell theory is
invariant under this transformation so long as $\tau$ is replaced
by $\tau'$. As in previous examples we assume this continues to be
true in the presence of boundary fermion fields tuned to their
massless points and which couple minimally to the bulk $U(1)$
gauge field, {\em i.e} \beq Z_{SB}^D[\tau]  =  Z_{SB}^{D}[\tau']
\eeq for the $\tau'$ given in Eqn. \ref{tau'ff}.  Once again this
is an equality of the partition function for two different
coupling constants. To get a useful equality we now do a
$T^{-1}S^{-1}T^{-2}S^{-1}$ transformation on the right side. This
transforms it to \beqn
Z_{SB}^{dD}[\tau] & = &  \int [{\cal D}A ] Z_{dD}[A] e^{-\int d^4x  S_M[A; \tau']} \\
Z_{dD}[A] & = & \int [{\cal D}A']  Z_{D}[A']  e^{  \int d^3x {i
\over 2\pi} A'dA -   \frac{2i bdb}{4\pi} +\frac{i}{2\pi} bdA - {i
\over 4\pi} AdA} \eeqn We then have the equality \beq
Z_{SB}^D[\tau] = Z_{SB}^{dD}[\tau] \eeq Now both theories are at
the same coupling $\tau$.  Taking the limit $\tau = i\infty$ again
converts $A$ to a background gauge field and we get the $(2+1)d$
fermion-fermion duality.

There is, however, a nontrivial subtlety in time-reversal symmetry
in the above discussion. We interpreted the $\theta=\pi$ Maxwell
theory from the electric point of view as a gauged standard
topological insulator. In the magnetic picture we interpreted the
gauge theory as a gauged $U(1)\times \mathcal{CT}$ topological
insulator (sometimes also called topological superconductors with
$S_z$ symmetry). However, there are multiple distinct $U(1)\times
\mathcal{CT}$ topological superconductors that can give
$\theta=\pi$ (hence identical charge-monopole lattice) when
gauged: the classification of interacting topological
superconductors (or SPT phases) in this symmetry class is
given\cite{3dfSPT2} by $\mathbb{Z}_8\times\mathbb{Z}_2$, and any
state with an odd index in the $\mathbb{Z}_8$ group will give
$\theta=\pi$ when gauged. The question is which one is dual to the
gauged standard topological insulator? To thoroughly answer this
question, one should gauge not only the $U(1)$ symmetry but also
time-reversal symmetry, by defining the theories on non-oriented
manifolds. This was done in Ref.~\cite{Metlitski2015}, where the
topological part of the partition functions of both theories
(gauged TI and gauged TSC) were calculated on non-orientable
manifolds. Specifically, it was found that on $\mathbb{RP}^4$ the
two agree only if the $U(1)\times \mathcal{CT}$ topological
superconductor corresponded to the ``minimal" state, with only one
Dirac fermion on the surface\footnote{There is another
possibility, having to do with the extra $\mathbb{Z}_2$ in the SPT
classification of this symmetry class, which can be ruled out
easily by computing the surface thermal Hall conductance.}. This
gives  strong support for the above discussion.

\section{Further evidence for the duality }
\label{Evidences}

So far we have only justified the $(2+1)d$ dualities at the
kinematic level, namely by matching the qualitative aspects of the
two sides such as global symmetries, anomalies and phase diagrams.
The dynamical aspects of the duality, such as the critical
exponents, are much harder to justify since the theories typically
flow to strong coupling in the IR. There are roughly two ways to
extend the field theories to enable analytical progress.

The first approach, which we review in Sec.~\ref{wire} and
\ref{lattice} below, is to define the field theories with an
explicit UV cutoff (for example on a lattice), typically at strong
coupling, so that the two theories in the duality have identical
partition function -- this makes the duality true in both UV and
IR. This kind of duality is sometimes called ``weak" in the
literature. In order to obtain the strong dualities -- dualities
of the (super-)renormalizable continuum field theories that are
free in the UV -- one has to further assume that the strongly
coupled UV theory can be smoothly deformed to the weakly coupled
continuum limit, say, without encountering another multi-critical
point. There is typically no analytical tool to justify such an
assumption. A virtue of the weak dualities, besides giving more
support to the strong dualities, is to offer a clearer physical
picture on how the operators (local or non-local) are mapped to
each other under the duality.

The second approach is to work directly with field theories
defined in the continuum, but generalized in a way that allows
analytical control. In Sec.~\ref{largeN} we briefly review the
large-$N$ generalizations of the dualities, which have been
justified in the field theory literature through many explicit
calculations. Another approach, which we will not review in detail
here, is to start from certain supersymmetric (SUSY) dualities,
deform the theories with some SUSY-breaking perturbations, and
reach non-SUSY dualities like the ones discussed in this review.
Analytically the SUSY dualities are better justified as strong
dualities (e.g. by matching free energies on a sphere on both
sides), although the SUSY-breaking deformations are typically less
controlled, making the deformed (non-SUSY) dualities ``weak". More
details can be found in Refs.~\cite{kachru1,kachru2}.

\subsubsection{Coupled wire constructions}
\label{wire}

Most of the discussions in this part will follow the logic of
Refs~\cite{mrossduality,mrossduality2}, in which more detail can
also be found. For concreteness, let us consider a Dirac fermion
defined on a coupled-wire system. Consider a set of parallel wires
aligned in the $\hat{x}$ direction and stacked in the $\hat{y}$
direction, with a left-moving Weyl fermion defined on each $y$-odd
wire and a right-moving Weyl fermion on each $y$-even wire. The
action is simply \beq \mathcal{S}[\psi]=\int
dtdx\sum_y\left(i\psi^{\dagger}_y[\partial_t-(-1)^y\partial_x]\psi_y-g(-1)^y(\psi^{\dagger}_y\psi_{y+1}+c.c.)
\right), \eeq which describes a two-component Dirac fermion
$\Psi(x,y)=(\psi_{2y}(x),\psi_{2y+1}(x))^T$ when the continuum
limit in $\hat{y}$ direction is taken. The massless nature is
protected by the time-reversal-like symmetry $\mathcal{T}$, which
is an ordinary time-reversal transform followed by a translation
in $\hat{y}$ direction by one unit. The collection of chiral
fermions on the wires can be bosonized: \beq \label{wireDirac}
\mathcal{S}[\phi]=\int
dtdx\sum_y\left(\frac{(-1)^y}{4\pi}\partial_x\phi_y\partial_t\phi_y+(\partial_x\phi_y)^2-g(-1)^y\cos(\phi_y-\phi_{y+1})
\right). \eeq The first term enforces the commutation relation
$[\phi_y(x),\phi_{y'}(x)]=i\pi\delta_{y,y'}(-1)^y\sgn(x-x')$,
which leads to the definition of the chiral fermion on each wire
as $\psi_y(x)=\eta_ye^{i\phi_y(x)}$ where $\eta_y$ are Klein
factors that make fermions on different wires anti-commute. In
general one can also add other terms when the fermions are
interacting, which respect (1) global symmetries such as
time-reversal $\mathcal{T}: \phi_y\to
-\phi_{y+1}+\frac{1+(-1)^y}{2}\pi$ and (2) locality of the fermion
field $\sim e^{i\phi_y}$.

Dualities on this system can be viewed as non-local changes of
variables. For example, to obtain the fermionic charge-vortex
duality, we define a set of new fields \beq
\label{dualchiralboson} \tilde{\phi}_y=\sum_{y'\neq
y}\sgn(y-y')(-1)^{y'}\phi_{y'}, \eeq which obey
$[\tilde{\phi}_y(x),\tilde{\phi}_{y'}(x')]=-[\phi_y(x),\phi_{y'}(x)]$.
Therefore $\tilde{\phi}_y$ represents another chiral fermion
$\tilde{\psi}_y\sim\eta_y e^{i\tilde{\phi}_y}$ with chirality
opposite to that of $\phi_y$. One can also show, from the
commutation relations, that when the original fermion $e^{i\phi}$
encircles the dual fermion $e^{i\tilde{\phi}}$, a phase of $4\pi$
is picked up, so the dual fermion is indeed interpreted as a
$4\pi$-vortex as anticipated.

One can then re-write the theory Eq.~\eqref{wireDirac} in terms of
the dual fields $\tilde{\phi}$, which is also a theory coupled
chiral fermions but with non-local coupling. The non-locality can
be seen from terms like $(\partial_x\phi_y)^2$: when expressed in
$\tilde{\phi}_y$, this term contains couplings between two
$\tilde{\phi}$ fields that are arbitrarily far away in $\hat{y}$
direction. We can then introduce a $U(1)$ gauge field defined on
the wires $a_{y}(x)$ that couples to $\tilde{\psi}$, and view the
non-local coupling of $\tilde{\phi}$ as a result of integrating
out the gapless photon field $a_y(x)$. The local action looks like
(see Ref.~\cite{mrossduality} for more details) \beq
\mathcal{S}[\tilde{\phi},a]=\int
dtdx\sum_y\left(-\frac{(-1)^y}{4\pi}\partial_x\tilde{\phi}_y\partial_t\tilde{\phi}_y-\frac{(-1)^y}{2\pi}a_{0,y}\partial_x\tilde{\phi}_y+\frac{1}{4\pi}(\partial_x\tilde{\phi}_y-a_{1,y})^2-g(-1)^y\cos(\tilde{\phi}_y-\tilde{\phi}_{y+1})
+ \mathcal{S}_{Maxwell}[a_{\mu}]\right). \eeq Importantly,
time-reversal symmetry is present and acts on $\tilde{\phi}$
simply as $\mathcal{T}: \tilde{\phi}_y\to \tilde{\phi}_{y+1}$ up
to an unimportant overall phase. The means that the dynamical
gauge field $a_{\mu}$ has a lattice Maxwell term
$\mathcal{S}_{Maxwell}[a_{\mu}]$ in the action but no overall
Chern-Simons term. This theory can then be viewed as a lattice
definition of Eq.~\eqref{dualD1}, namely a Dirac fermion
$\tilde{\Psi}\sim (\tilde{\psi}_{2y},\tilde{\psi}_{2y+1})$ coupled
to a dynamical Maxwell $U(1)$ gauge field $a_{\mu}$. In this
definition the duality to a free Dirac fermion holds at the level
of partition function (UV and IR).

The emergence of the $U(1)$ gauge structure in the dual variables
Eq.~\eqref{dualchiralboson} can also be understood at the
kinematic level. Consider putting the coupled wires on a cylinder
with a fixed boundary condition for $\phi$ in the $\hat{y}$
direction, i.e. $\phi_1=\phi_{2N+1}+\Phi(x)$ and
$\phi_2=\phi_{2N+2}+\Phi(x)$ with an $x$-dependent flux $\Phi(x)$.
In this case $(\Phi(x_1)-\Phi(x_2))$ is simply the total magnetic
flux normal to the cylinder surface between the two cycles at
$x_1$ and $x_2$. The ability to fix such a boundary condition is
another way to declare the locality of $\psi\sim e^{i\phi}$. It is
easy to see now that the boundary condition for $\tilde{\phi}$
becomes dynamical and can no longer be fixed. One can define
$\tilde{\Phi}(x)=\tilde{\phi}_{2N+2}(x)-\tilde{\phi}_2(x)$ as the
total flux seen by the dual fermion threading the cylinder at $x$,
and realize that \beq \label{wiredualflux}
\tilde{\Phi}(x_1)-\tilde{\Phi}(x_2)=2\sum_y\int_{x_2}^{x_1}dx(-1)^y\partial_x\phi_y=4\pi
Q(x_1,x_2), \eeq where $Q$ is the total electric charge enclosed
by the two cycles at $x_1$ and $x_2$ (recall that
$\rho=\frac{(-1)^y}{2\pi}\partial_x\phi$ is the charge density on
the wires). Similarly the boundary condition on $\phi$ enforces
the total charge of the dual fermions to vanish between any two
cycles: \beq \label{wiredualcharge}
\tilde{Q}(x_1,x_2)=-\frac{1}{2\pi}\sum_y\int_{x_2}^{x_1}dx(-1)^y\partial_x\tilde{\phi}_y=-\frac{1}{4\pi}(\Phi(x_1)-\Phi(x_2)).
\eeq The dynamical boundary condition Eq.~\eqref{wiredualflux} and
the constraint on total charge Eq.~\eqref{wiredualcharge} are both
hallmarks of a dynamical gauge theory. In fact these conditions
are exactly what one anticipates from the dual Dirac theory
Eq.~\eqref{dualD1}.  This emergence of gauge structure manifested
from boundary conditions is very similar to the Ising/Majorana
dualities in $(1+1)d$, where the $\mathbb{Z}_2$ gauge structure in
the non-local description can be detected when a periodic boundary
condition is imposed.

We can also consider a different non-local change of variable, by
defining a non-chiral boson (living only on wires at even $y$)
\beq
\varphi_{2y}=\frac{\phi_{2y}+\tilde{\phi}_{2y}}{2}=\frac{\phi_{2y+1}+\tilde{\phi}_{2y+1}}{2},
\eeq where the second identity follows from the definition of
$\tilde{\phi}$ in Eq.~\eqref{dualchiralboson}. It is easy to show
that $e^{i\varphi}$ is a boson. One can repeat either of the
previous arguments to conclude that the theory (which was local in
$e^{i\phi}$), when written in terms of $\varphi$, becomes a theory
of bosons coupled to a $U(1)$ gauge field with a Chern-Simons term
at level $k=1$, which can be viewed as a UV definition (at strong
coupling) of the right hand side of the bosonization duality
Eq.~\eqref{basicduality}.

Under time-reversal transform,
$\tilde{\phi}_y\to\tilde{\phi}_{y+1}$ and $\phi_y\to-\phi_y$ (up
to a shift which is not important here). Therefore the boson field
$\varphi_{2y}\to \tilde{\varphi}_{2y+1}$, defined (on wires at odd
$y$) as \beq
\tilde{\varphi}_{2y+1}=\frac{-\phi_{2y+1}+\tilde{\phi}_{2y+1}}{2}=\frac{-\phi_{2y+2}+\tilde{\phi}_{2y+2}}{2}.
\eeq

Using the commutation relations, one can see that
$e^{i\tilde{\varphi}}$ is again a boson, and when encircled by
$e^{i\varphi}$, a phase of $2\pi$ is produced. This means that
$e^{i\varphi}$ and $e^{i\tilde{\varphi}}$ are mutual vortices.
This offers an explicit picture, at the operator level, of the
non-local time-reversal transform in the bosonization duality,
which takes the composite fermion to its vortex dual. The chiral
bosons representing the original and dual fermions can be written
as \beqn \label{bosonfermionwire}
\phi_{2y}=\varphi_{2y}-\tilde{\varphi}_{2y-1}, && \phi_{2y+1}=\varphi_{2y}-\tilde{\varphi}_{2y+1}, \nonumber \\
\tilde{\phi}_{2y}=\varphi_{2y}+\tilde{\varphi}_{2y-1}, &&
\tilde{\phi}_{2y+1}=\varphi_{2y}+\tilde{\varphi}_{2y+1}. \eeqn
This offers a vivid picture of ``flux attachment" at the operator
level: the fermions are the composites of bosons ($\varphi$) and
vortices ($\tilde{\varphi}$), slightly displaced from each other
(in this model in the $\hat{y}$ direction). The ``spinor"
structure of the Dirac fermion comes from the relative direction
of the boson and vortex -- in particular, under time-reversal the
boson and vortex are exchanged, which flips the Dirac spin as
expected.

One can define different theories on the coupled wire system, by
demanding the locality of different operators. For example, we can
demand the boson field $e^{i\varphi}$ to be local, in the sense
that (1) terms in the action must be local in $e^{i\varphi}$ and
(2) periodic boundary conditions in $\hat{y}$ direction (possibly
with a flux) can be fixed. In this case we have defined an
interacting bosonic theory on the coupled-wire system. One can
invert the above non-local transforms to get various non-local
degrees of freedoms (relative to $e^{i\varphi}$), and produce
coupled-wire definitions of either the boson-vortex duality (using
$\tilde{\varphi}$) or the fermionization duality in
Eq.~\eqref{wffex1} or \eqref{wffex2} (using $\phi$ or
$\tilde{\phi}$). Since the boson phase $\varphi$ and vortex phase
$\tilde{\varphi}$ acquire opposite signs under time-reversal
transform, it is immediately obvious from
Eq.~\eqref{bosonfermionwire} that time-reversal would take the
``composite fermion" fields $e^{i\phi}$ to there vortex duals
$e^{i\tilde{\phi}}$, consistent with our previous discussion.

\subsubsection{Bosonization duality on the lattice}
\label{lattice}

The bosonization duality Eq.~\eqref{basicduality}, which plays an
important role in the duality web, can also be defined exactly on
a space-time lattice, as was done in Ref.~\cite{Chenlattice}
which we now review. Consider a Euclidean $3D$ cubic lattice. On
each cite $n$ we define a bosonic (rotor) variable
$\theta_n\in(-\pi,\pi]$ and a two-component Grassmann variable
$\chi_n=(\chi_{1,n},\chi_{2,n})^T$ and its conjugate
$\bar{\chi}_n$. One each link $(n,\hat{\mu})$
($\hat{\mu}\in\{\pm\hat{x},\hat{y},\hat{\tau}\}$) we define a
dynamical compact $U(1)$ gauge field $e^{ib_{n,\hat{\mu}}}$. The
bosonic fields carry gauge charge $1$ under $b_{\mu}$ while the
Grassmann fields carry charge $(-1)$, so the gauge invariant
object is the composite $\psi\sim e^{-i\theta}\chi$ -- this can be
viewed as a lattice implementation of the parton construction in
Eq.~\eqref{parton}.

Now we define the dynamics of the model. The bosonic sector is
described by a $3D$ XY model \beq
Z_{T}[b]=\int_{-\pi}^{\pi}\prod_n\frac{d\theta_n}{2\pi}\exp\left[-\frac{1}{T}\sum_{n,\hat{\mu}}\cos\left(\theta_{n+\hat{\mu}}-\theta_n-b_{n,\hat{\mu}}
\right)\right]. \eeq The fermionic sector is described by a Wilson
fermion -- a lattice regularization of Dirac fermions: \beqn
Z_{M,U}[-b]&=&\int \prod_{n}d^2\bar{\chi}_nd^2\chi_n\exp\left(-H_M-H_U \right), \nonumber \\
-H_M&=&\sum_{n,\hat{\mu}}\left(\bar{\chi}_n\frac{\sigma^{\mu}-1}{2}e^{ib_{n,\hat{\mu}}}\chi_{n+\hat{\mu}}+\bar{\chi}_{n+\hat{\mu}}\frac{-\sigma^{\mu}-1}{2}e^{-ib_{n,\hat{\mu}}}\chi_{n} \right)+\sum_nM\bar{\chi}_n\chi_n, \nonumber \\
-H_U&=&U\sum_{n,\hat{\mu}}\left(\bar{\chi}_n\frac{\sigma^{\mu}-1}{2}\chi_{n+\hat{\mu}}\right)\left(\bar{\chi}_{n+\hat{\mu}}\frac{-\sigma^{\mu}-1}{2}\chi_{n}
\right), \eeqn where the $H_M$ terms gives the fermions a Dirac
dispersion in the continuum limit, gapless when $|M|=1,3$. When
the Dirac fermions are gapped, they give a Hall conductance $C$ to
the gauge field $b$, given by \beq \label{Chernnumber} C=\left\{
\begin{array}{cc}
               0, & |M|>3 \\
               1, & 1<|M|<3 \\
               -2, & |M|<1. \end{array} \right.
\eeq The $H_U$ term is an on-site interaction. The form of the
action is chosen to make the theory particularly simple when
expanded in $\{\bar{\chi},\chi\}$ (notice that
$(\sigma^{\mu}\pm1)/2$ is a projector): \beq
\label{grassmannintegralsimplified}
e^{(-H_M-H_U)}=e^{M\bar{\chi}_n\chi_n}\left[1+\bar{\chi}_n\frac{\sigma^{\mu}-1}{2}e^{ib_{n,\hat{\mu}}}\chi_{n+\hat{\mu}}+\bar{\chi}_{n+\hat{\mu}}\frac{-\sigma^{\mu}-1}{2}e^{-ib_{n,\hat{\mu}}}\chi_{n}+(1+U)\left(\bar{\chi}_n\frac{\sigma^{\mu}-1}{2}\chi_{n+\hat{\mu}}\right)\left(\bar{\chi}_{n+\hat{\mu}}\frac{-\sigma^{\mu}-1}{2}\chi_{n}
\right) \right]. \eeq

The total partition function is given by \beq Z=\int_{-\pi}^{\pi}
\prod_{n\mu}\frac{db_{n,\hat{\mu}}}{2\pi}Z_{T}[b]Z_{M,U}[A-b],
\eeq where $A$ is the external $U(1)$ gauge field. The particular
form of the theory was chosen to drastically simplify subsequent
manipulations and even allow certain exact statements to be made.
The theory can be made more generic by adding other
symmetry-allowed local terms such as modified fermion hopping or
lattice Maxwell term for $b_{\mu}$ (the $U(1)$ gauge theory
without Maxwell term is formally at infinite coupling strength at
the lattice scale) -- the universal physics will not change much
so long as the modifications are  small.

The bosonic path integral can be re-written through a Fourier
transform \beq
e^{\frac{1}{T}\cos\left(\theta_{n+\hat{\mu}}-\theta_n-b_{n,\hat{\mu}}
\right)}=\sum_{j_{n,\hat{\mu}}=-\infty}^{+\infty}I_{j_{n,\hat{\mu}}}\left(\frac{1}{T}\right)e^{i\left(\theta_{n+\hat{\mu}}-\theta_n-b_{n,\hat{\mu}}
\right)j_{n,\hat{\mu}}}, \eeq where $I_k$ is the modified Bessel
function of the first kind, and $j_{n,\hat{\mu}}$ is an
integer-valued variable defined on the links that can be
interpreted as the boson current. Now integrating over $\theta_n$
simply puts a ``Gauss law" constraint on each site demanding
vanishing of the lattice divergence of $j$: \beq
Z_T[b]=\sum_{\{j_{n,\hat{\mu}}\}}
\prod_{n,\hat{\mu}}\delta_{\Delta_{\hat{\mu}}j_{n,\hat{\mu}}}I_{j_{n,\hat{\mu}}}e^{-ib_{n,\hat{\mu}}j_{n,\hat{\mu}}}.
\eeq

The fermion action Eq.~\eqref{grassmannintegralsimplified}
contains only terms with fermion current $0, \pm1$ (i.e. terms
either independent of $b$ or proportional to $e^{\pm b}$). Now
integrating out the gauge field $b$ will simply force the bosonic
and fermionic currents to be identical on each link. The Gauss law
constraint for the bosonic current is also automatically satisfied
once it is identified with the fermionic current due to the
Grassmannian nature of the path integral. Therefore after
integrating out both $\theta$ and $b$, we obtain a path integral
just in terms of $\bar{\chi},\chi$. After some simple algebra and
a redefinition $\psi=\sqrt{I_1(1/T)/I_0(1/T)}\chi$ (and likewise
for $\bar{\psi}$), the total partition function can be exactly
re-written as (up to some unimportant overall factor) \be
\label{exactrelation} Z[A]\sim \int D\bar{\psi}D\psi
e^{-H_{M'}-H_{U'}}, \hspace{10pt}
\frac{M'}{M}=\sqrt{\frac{1+U'}{1+U}}=\frac{I_0(1/T)}{I_1(1/T)},
\ee which is exactly a Wilson fermion with renormalized mass $M'$
and on-site interaction $U'$. If the values of $M'$ and $U'$ leads
to a free Dirac fermion for $\psi$ in the IR, then we conclude
that the original (strongly coupled) lattice $U(1)$ gauge theory
also describes a free Dirac fermion in the IR. We are interested
in $M'=3+\delta M'$ and $U'=0+\delta U'$, so that the $\psi$
fermions are close to form a single gapless Dirac fermion. The
fermion is precisely gapless when $\delta M'+\Sigma=0$ where
$\Sigma$ is the self-energy. To first order it was
evaluated\cite{Chenlattice} to be $\delta U'\approx -8.8\delta
M'$. These statements are reliable for small enough $\delta M',
\delta U'$ since $U'$ is an irrelevant perturbation. There are two
ways to tune the microscopic parameters $M$, $U$ and $T$ to
achieve this. One can simply take $M=M'$, $U=U'$, and $T=0$, and
the physics is very simple from the parton point of view: at $T=0$
the bosons condense and $\psi=e^{-\theta}\chi\approx\langle
e^{-i\theta}\rangle \chi\sim \chi$. Alternatively, we can take
$U=0$ and $M=3-|\delta M|$. For a fixed value of $|\delta M|$,
both $M'$ and $U'$ are tuned by $T$ through
Eq.~\eqref{exactrelation}. What is important here is that Dirac
fermions with effective mass ($\delta M'+\Sigma$) of either sign
can be realized by tuning $T$ -- this is only possible when $M<3$,
since $I_0/I_1$ takes values in $[1,+\infty)$ and $\delta
U'/\delta M'<0$ at the transition point. The theory constructed
this way also has a very simple interpretation: since $U=0$ and
$M<3$, the lattice free fermion $\chi$ is gapped and can be
integrated out. This produces a local effective action for
$b_{\mu}$, for which the leading order term is simply a
Chern-Simons term at level $1$ (see Eq.~\eqref{Chernnumber}). What
remains in the theory is simply an $XY$ boson coupled with a
$U(1)$ gauge theory at Chern-Simons level $1$, and the gapless
Dirac point is accessed by tuning $T$ through a critical point --
exactly the physics described in the bosonization duality
Eq.~\eqref{basicduality} (modulo the strong coupling issue we
mentioned before).

There is yet another way to define the dualities on lattice using
loop models with long-range interactions and statistical angles
mimicking Chern-Simons gauge fields. We refer to
Refs.~\cite{FradkinKivelson,GoldmanFradkin} for readers interested
in this approach.

\subsubsection{Dualities in large-$N$ Chern-Simons-matter theory}
\label{largeN}

One path to field-theoretic dualities is through the the large-$N$
non-Abelian Chern-Simons-matter theories.  These theories involve
matter fields (bosons or fermions) coupled to a $U(N)$ or $SU(N)$
nonabelian Chern-Simons gauge field theory.  The idea of
boson-fermion duality  arose there in a rather curious way.
In 2002, Klebanov and Polyakov proposed~\cite{Klebanov:2002ja} a
holographic duality between the 3D $O(N)$ vector model and
Vasiliev's higher spin theory in AdS$_4$.  Soon it was proposed
that the 3D Gross-Neveu model also has a holographic dual in the
form of a higher spin theory~\cite{Sezgin:2003pt}.  The two higher
spin theories dual to the bosonic $O(N)$ vector model and the
fermionic Gross-Neveu models are called the type-A and type-B
high-spin theories, respectively.  Both theories respect parity;
if this requirement is relaxed, then there exists a one-parameter
family of higher-spin theories interpolating between type-A and
type-B theories.  It was suggested in
Refs.~\cite{Giombi:2011kc,Aharony:2011jz} that
parity-violating higher-spin theories are holographically dual to
the Chern-Simon theories, with bosonic or fermionic matter.  That
leads to the conclusion that there must exist a duality between
the bosonic and fermionic large-$N$ Chern-Simons
theories~\cite{aharony2}.

The large-$N$ duality proposed in Ref.~\cite{aharony2} was
soon confirmed by a substantial number of explicit checks.  In
particular, the correlation
functions~\cite{aharony2,GurAri:2012is}, $2\to2$ scattering
matrices~\cite{Jain:2014nza,Inbasekar:2015tsa}, and
thermodynamics~\cite{Aharony:2012ns,Jain:2013py,Takimi:2013zca,Geracie:2015drf}
match between the two sides of the duality.  The perturbative
calculations of diagrams simplify dramatically in the light-cone
gauge, first used in Ref.~\cite{Giombi:2011kc}.

Already in Ref.~\cite{aharony2} it was suggested that
perhaps the boson-fermion duality is also valid at finite $N$ and
$k$.  In holographic dualities, $1/N$ corrections correspond to
quantum corrections in the bulk theory, and if the bosonic and the
fermion CS theories are different at the $1/N$ order, that would
mean that there are two different quantum versions of the same
classical theory. Another piece of evidence comes from the fact
that the supersymmetric version of the duality, the Giveon-Kutasov
duality~\cite{Giveon:2008zn}, has been tested  at finite $N$.

The concrete form of duality at finite $N$ was proposed only in
late 2015, when Aharony put forward 3 separate
proposals~\cite{aharony3}
\begin{align}
  N_f \textrm{ scalars coupled to } SU(N)_k & \longleftrightarrow N_f \textrm{ fermions coupled to } U(k)_{-N+\frac{N_f}2,-N+\frac{N_f}2} \\
  N_f \textrm{ scalars coupled to } U(N)_{k,k} & \longleftrightarrow N_f \textrm{ fermions coupled to } SU(k)_{N+\frac{N_f}2}\\
  N_f \textrm{ scalars coupled to } U(N)_{k,k+N} &  \longleftrightarrow N_f \textrm{ fermions coupled to } U(k)_{-N+\frac{N_f}2,-N-k+\frac{N_f}2}
\end{align}
At finite $N$ one needs to distinguish $SU(N)$ and $U(N)$ gauge
groups.  In the case of the $U(N)$ gauge group, the Chern-Simons
action has two levels, as reflected in the notation
$U(N)_{k_1,k_2}$, where $k_1$ is the $SU(N)$ level and $k_2$ is
the $U(1)$ level.  The normalization is such that $U(N)_{k,k}$ is
the theory with a trace over the fundamental representation of
$U(N)$


Taking $N_f=N=k=1$, interpreting $SU(1)$ as trivial
\begin{align}
  \textrm{a WF scalar} & \longleftrightarrow \textrm{a fermion coupled to } U(1)_{-\frac12} \\
  \textrm{a scalar coupled to } U(1)_1 &\longleftrightarrow \textrm{a free fermion} \\
  \textrm{a scalar coupled to } U(1)_2 & \longleftrightarrow \textrm{a fermion coupled to } U(1)_{-\frac32}
\end{align}

The first duality is the old
conjecture\cite{wufisher,BarkeshliMcGreevy,seiberg1} of Eqn.
\ref{wffermion1} on the fermionic description of the $3D$ $XY$
fixed point, and the second duality coincides with
(\ref{basicduality}).

\section{Relevance to condensed matter, half-filled Landau level, boundary of 3d TI}

\label{CMApp}

Just like the bosonic charge-vortex duality the fermionic versions
are expected to have powerful applications in condensed matter
physics.  In this section we briefly outline two applications -
one to the theory of the half-filled Landau level and the other to
the theory of strongly correlated surface states of topological
insulators.  Notably, the two applications described
in this section only require the ``weak" form of duality to hold,
namely that the dual Dirac theory in Eqn.~\ref{dualD1} lives in
the ``same Hilbert space" as the free Dirac fermion -- whether or not
it actually flows to the free Dirac fermion in the IR is
irrelevant for the application here. In following sections we
describe generalizations of these dualities that have direct
application to the theory of Landau-forbidden deconfined quantum
critical points in two space dimensions. In those
applications we   actually need the ``strong" dualities, in
the sense that the theories on the two sides of the dualities
actually flow to the same fixed point.

\subsection{The half-filled Landau level}
\label{hfll}

The classic setting for the quantum Hall effect is in a system of
electrons in $2d$ in a strong magnetic field such that a small
number of Landau levels are occupied.  An  important energy scale
is  the Landau level spacing $\hbar \omega_c = {\hbar e B \over
m}$  ($B$ is the magnetic field strength, $e$ the charge of the
electron, and $m$ its mass).  The Landau level filling factor $\nu
= \frac{2\pi \rho}{B}$  (in units where $e = \hbar = 1$) where
$\rho$ is the electron density.  The integer quantum Hall effect
occurs when  $\nu$ is an  integer. Fractional quantum Hall states
occur at certain rational fractional values of $\nu$.  A second
important energy scale is that of the typical strength of the
Coulomb interaction between electrons. This is given by $E_c =
\frac{1}{l_B}$ where the magnetic length $l_B \sim
\frac{1}{\sqrt{B}}$.  In discussing phenomena where only the
Lowest Landau Level (LLL) is partially filled it is interesting to
consider the limit where $E_c \ll \omega_c$.  Then we can ignore
the higher Landau levels and define the problem purely by
projecting the Coulomb interaction to the lowest energy level.
Note that in this limit the kinetic energy  of the electrons has
been quenched.  The only energy scale left in the problem is
$E_c$.

Our interest is in the fate of the system when $\nu =
\frac{1}{2}$.    Empirically this is seen to be a metal albeit a
rather unusual one.  It has non-zero finite values of both
longitudinal $\rho_{xx}$ and Hall $\rho_{xy}$ resistivities.
Theoretically it is interesting to  note that  this metallic
behavior must ultimately derive from the Coulomb energy of
electrons (in the lowest Landau level limit). A classic theory of
this metal - due to Halperin, Lee, and Read (HLR)\cite{hlr} -
describes this as a compressible state   obtained by forming a
fermi surface of ``composite fermions"\cite{jaincf} rather than
the original electrons. In the original HLR theory , the composite
fermions are formed by binding two flux quanta to the physical
electrons. At $\nu = \frac{1}{2}$  this attached flux on average
precisely cancels the external magnetic flux so that the composite
fermions move in effective zero field. This facilitates the
formation of a Fermi surface and leads to an effective field
theory of the metal as a Fermi surface coupled to a fluctuating
gauge field which is then used to describe the physical properties
of this metal.

The HLR theory - and some subsequent refinements -  successfully
predicted many experimental properties.  For instance when the
filling is tuned slightly away from $\frac{1}{2}$, the composite
fermions see a weak effective magnetic field and their
trajectories are expected to follow cyclotron orbits with radii
much larger than the underlying electrons.  These have been
directly demonstrated in
experiment\cite{willett93,kang93,goldman94,smet96} - for reviews
see, e.g., the contribution by Tsui and Stormer in Ref.
\cite{dassarmabook}, and Ref. \cite{willett97}.
Further the composite Fermi liquid acts as a parent for the
construction of the Jain sequence of states\cite{jainbook} away
from $\nu = \frac{1}{2}$: these states are simply obtained by
filling an integer number of Landau levels of the composite
fermions. Finally the composite Fermi liquid yields the
non-abelian Moore-Read quantum Hall state  through pair
``condensation" of the composite fermions\cite{readgrn}.

Despite these successes there were two problems with the HLR
theory. The first is that it is not formulated in the LLL limit.
The standard flux attachment procedure works with the bare kinetic
energy of the electrons rather than with a theory formulated just
in the LLL.  Thus it does not correctly capture the feature of the
LLL that the only energy scale is $E_c$.

A second problem is also apparent once we think about the theory
formulated in the LLL. At half-filling and with a two-body (or
more generally any even body) interaction, there is an extra exact
particle-hole symmetry. This corresponds to viewing the
half-filled Landau level either by starting with an empty Landau
level and adding electrons upto half-filling or by starting with a
filled Landau level and  removing electrons to reach half-filling.
As the Landau orbitals are complex the particle-hole symmetry is
implemented as an {\em antiunitary} symmetry.  We will denote it
as ${\CT}$.

The projection to the LLL plays an important role in numerical
calculations\cite{jainbook} of quantum hall phenomena. At
half-filling both exact diagonalization\cite{HaldRez2000}  and
Density Matrix Renormalization Group studies\cite{geraedtsnum}
strongly support the formation of a compressible metallic state
with the 2-body Coulomb interaction. They also show that this
metallic state is particle-hole symmetric.    Being not formulated
in the LLL, the HLR theory does not keep any such $\CT$ symmetry
manifest.

In 2015, Son proposed\cite{son2015} a modification of the HLR
theory which was manifestly particle-hole symmetric.  He
postulated that the composite fermion is a 2-component Dirac
particle at a non-zero density.   Under the original particle-hole
symmetry operation $\CT$, the composite fermion field $\chi$ is
hypothesized to transform as
 \begin{equation}
 \label{CTchi}
 {\CT}\chi (\CT)^{-1} = i\sigma_y \chi
 \end{equation}
Thus $\chi$ goes to itself rather than to its antiparticle under
$\CT$. Further this transformation implies that the two components
of $\chi$ form a Kramers doublet under $\CT$ (recall that $\CT$ is
anti unitary).

These composite fermions are at a non-zero density
$\frac{B}{4\pi}$ and fill states upto a Fermi momentum $K_f$. This
should be compared with the HLR theory where the prescription for
the composite fermion density is just the electron density $\rho$.
At half-filling we have $\rho = B/4\pi$ and the two prescriptions
agree. However these two prescriptions are different away from
half-filling.

In the particle-hole symmetric theory, the ``Diracness" of the
composite fermion is manifested  as follows: when a composite
fermion at the Fermi surface completes a full circle in momentum
space its wave function acquires a Berry phase of $\pi$.  This is
a ``low-energy" manifestation of the Dirac structure that does not
rely on the specifics of the dispersion far away from the Fermi
surface.

Son's proposal has by now found significant
support\cite{wangsenthil1,maxashvin,wangsenthil2,geraedtsnum,msgmp15,HaldaneBerry1,HaldaneBerry2,SimonBerry,JainBerry}
through a variety of different
ways of thinking about the half-filled Landau level.  In the
context of this review we now show how to justify this proposal
using the fermion-fermion duality  discussed in previous sections.

Consider again the free massless Dirac fermion (Eqn. \ref{freeD}).
As we emphasized  $\T$ and $\CT$ are anomalous symmetries. For the
following discussion we will focus on $\CT$. The fermion density
$\psi^\dagger \psi$ is odd under $\CT$ (and correspondingly so is
$A_0$).  This implies that the Dirac fermions are necessarily at
neutrality.  The $\CT$ anomaly of the theory can be cured if we
regard this theory as living at the boundary of a certain $(3+1)d$
topological insulator with $U(1) \times \CT$
symmetry\footnote{Symmetry Protected Topological (SPT) insulators
with this symmetry are denoted class AIII in the condensed mater
literature. Within free fermion theory class $A III$ insulators
have a $Z$ classification corresponding to $n$ massless Dirac
cones at the surface.  With interactions\cite{3dfSPT2,maxvortex}
this $Z$ classification is reduced to $Z_8$ (so that only $n = 0,
1, ...., 7$ are distinct phases. There is an additional Symmetry
Protected Topological phase which cannot be described within free
fermion theory so that the full classification\cite{3dfSPT2} is
$Z_8 \times Z_2$.   We will henceforth focus on the $n = 1$ free
fermion state which is stable to interactions. }.

Note that background  electric fields are $\CT$-odd while
background magnetic fields are $\CT$-even. We can then consider
the  theory by introducing an external magnetic field while
preserving the $U(1) \times \CT$ symmetry (but not $\T$ symmetry).
The resulting Lagrangian takes the form
\begin{equation}
{\cal L} = \bar{\psi} \left(-i\slashed{\partial } +
\slashed{A}\right) \psi
\end{equation}
with $\vec \nabla \times \vec A = B \hat{z}$ (taking the surface
to lie in the $xy$ plane). The spectrum has the famous Dirac
Landau levels with energy $E_k = \pm \sqrt{2kB}$ with $k \in Z$.
For non-zero $k$ each level comes with a partner of opposite
energy.  Most importantly there is a zero energy Landau level that
has no partner. The $\CT$ symmetry implies that this zeroth Landau
level must be half-filled.

At low energies it is appropriate to project to the zeroth Landau
level. We thus end up with a half-filled Landau level. As usual in
the non-interacting limit this is highly degenerate.  Now we must
include interactions (that preserve the $U(1) \times \CT$
symmetry) between the electrons to resolve this degeneracy.

Thus the surface of this $(3+1)d$ topological insulator maps
exactly to the classic problem of the half-filled Landau level.
Note however that the $U(1) \times \CT$ symmetry of the full
$(3+1)d$ TI maps precisely to the expected $U(1) \times \CT$
symmetry of the half-filled Landau level.  Thus the particle-hole
symmetric half-filled Landau level can be UV completed while
preserving charge conservation and $\CT$ symmetries by placing it
at the surface of a $(3+1)d$ topological insulator.

At any rate the above shows how to realize the half-filled Landau
level by starting with a theory of massless Dirac fermions,
turning on a magnetic field, and then including interactions. Now
let us describe this system using the dual of the free Dirac
theory theory obtained through the fermion-fermion duality.  We
study the dual theory in the presence of a uniform background
magnetic field associated with the $A$ gauge field.  Before doing
so we note that in  relating the results to the standard
half-filled Landau level obtained starting with non-relativistic
fermions, we need to add a background Chern-Simons  term for $A$.
Specifically, at the TI surface, the empty $0$th Landau Level is
assigned a Hall conductivity of $-{1 \over 2}$ while the filled
one is assigned ${1 \over 2}$. In the standard Landau level
problem of non-relativistic fermions, the Hall conductivity
assignments are shifted by ${1 \over 2}$ (so that the empty Landau
level has zero Hall conductivity).  Similar statements apply to
the thermal Hall conductivity. This amounts to adding a term ${1
\over 8\pi}A dA + CS_g$ to both sides of  the fermion-fermion
duality (we use Eqn. \ref{dualD2}).  Thus our proposed theory of
the half-filled Landau level is \beq \label{pfcfl} {\cal L} =i\bar
\chi \slashed{D} _{a}\chi - {2\over 4\pi} bdb + {1\over 2\pi}a db
- {1\over 2\pi}A db ~. \eeq

The physical electric $U_A(1)$ current is \beq \label{physJ} J = -
{1 \over 2\pi} db \eeq We denote the average value of the time
component as $\rho$ (the physical electron density). The equation
of motion of $b$ gives the average effective magnetic field
(denoted $B^*$) seen by the composite fermions \beq \label{Bstar}
 B^* = B - 4\pi \rho
 \eeq
This equation  is identical  to the HLR theory and ensures  that
$B^*$ vanishes at $\nu = {1 \over 2}$.  This is what makes the
fermion-fermion duality (as opposed to, say, the bosonization
dualities) useful in this context.

Varying with respect to $a_0$, we get the condition \beq
\label{agauss} \rho_\chi - {1 \over 4\pi} \epsilon_{ij} \partial_i
(a_j  -  2b_j) = 0 \eeq Here $\rho_\chi$ is the average density of
composite fermions.  The second term is the contribution from the
variation of  the response of the heavy fermion field $\chi_H$
that is implicitly included in  our definition of the Dirac
theory\footnote{Equivalently if we define the fermion partition
function in terms of the $\eta$-invariant, then  though $\eta[a]$
is not identical to the level-$1/2$ CS term, its variation is
identical to the variation of the level-$1/2$ CS term.  Thus for
the purposes of obtaining the equation of motion we can replace
$\eta$ by the level-$1/2$ CS term.}. We thus find that \beq
\label{cfrho} \rho_\chi = {1 \over 4\pi} B \eeq The finite average
density of composite fermions means that (if we ignore the
dynamics of the gauge fields $a$ and $b$) they will form a Fermi
surface. We note that the duality interchanges the role of $\T$
and $\CT$. In particular $\chi$ is now a Kramers doublet under
$\CT$. Further under $\CT$ the composite fermions are Kramers
doublets.  There will thus be a Berry phase of $\pi$ when the
composite fermion goes around the Fermi surface.

These are precisely the key elements of the Dirac composite fermi
liquid theory proposed by Son. We thus see that the
fermion-fermion duality provides a derivation of the Dirac
composite fermi liquid theory.

We close the discussion of the half-filled Landau level with a few
comments. We started the discussion of the composite Fermi liquid
theory by pointing out two problems with the HLR theory - dealing
with the LLL projection, and dealing with the particle-hole
symmetry.  In standard non-relativistic systems these two problems
are coupled together. The particle-hole transformation is a
symmetry of the theory only if the LLL projection is implemented.
In contrast when we start with $\CT$-invariant Dirac fermions in a
field to reach the  half-filled Landau level the issue of
projection to the zeroth Landau level is separated from the issue
of particle-hole symmetry. By construction $\CT$ is an exact
microscopic symmetry even if we do not project to the zeroth
Landau level. When we use the fermion-fermion duality to obtain
the Dirac composite fermi liquid theory we did not implement any
projection to the zeroth Landau level. Thus as written the Dirac
composite fermi liquid action incorporates the $\CT$ symmetry but
does not incorporate additional constraints that may exist  if the
theory lives microscopically in a single half-filled Landau level.

Let us discuss this last issue a bit more. It is convenient now to
reinstate a velocity $v$ and charge $e$ for the original Dirac
fermions (which thus far we set to $1$).  For the massless  Dirac
theory in a magnetic field, the  spacing between the zeroth and
first Landau levels is $v\sqrt{2B} \sim {v \over l_B}$. The
restriction to the zeroth level is legitimate so long as the
Coulomb energy ${e^2 \over l_B} \ll {v \over l_B}$, {\em i.e}, the
``fine structure constant" ${e^2 \over v} \ll 1$. At fixed
electric charge this requires taking the formal limit $v \ra
\infty$. In our derivation of the Dirac composite fermi liquid we
did not explicitly take this limit.  However we expect that this
limit is smooth for the low energy theory near the Fermi surface
of the Dirac composite fermi liquid. In contrast in the HLR theory
the analogous limit corresponds to taking the mass $m$ of the
non-relativistic fermion to zero. But since $m$ appears in the
denominator of the HLR kinetic energy it is not clear what happens
in this limit.

\subsection{Correlated surface states of 3d TIs}
\label{sto}

It is well known that, within free fermion theory, the single
Dirac cone on the surface of a three-dimensional topological
insulator \be \LL=\bar{\psi}i \slashed{\partial}\psi \ee cannot be
gapped without breaking either time-reversal symmetry or charge
conservation. This conclusion is not affected by introducing weak
(short-ranged) interactions since they are RG irrelevant by simple
power counting. It is then natural to ask whether the Dirac
fermion can be gapped by turning on strong electron-electron
interactions, without spontaneously breaking either time-reversal
or charge conservation. Such gapped surface states were found
theoretically in 2013\cite{fSTO2,fSTO4,fSTO1,fSTO3}. These gapped
surface states host intrinsic non-Abelian topological order that
reproduce the parity anomaly of the single Dirac fermion.
Specifically, two distinct topological orders were found. One of
these is equivalent to the Moore-Read Pfaffian
state\cite{MooreMR91} plus a neutral anti-semion topological
order, obtained from the method of ``vortex
condensation"\cite{fSTO1,fSTO3} (there is also a related ``parton"
construction\cite{SeibergWittenSTO}). Another topological order,
called T-Pfaffian\cite{fSTO2,fSTO4} were obtained from an exactly
solvable lattice model\cite{fSTO2}, which apparently bears no
resemblance to a surface Dirac cone in any limit. Moreover, there
are two different varieties of T-Pfaffian topological orders
(called T-Pfaffian$_{\pm}$), distinguished by the action of
time-reversal on the anyons, and it was not clear which one
corresponds to the conventional TI, though it was known that the
other one would corresponds to a conventional TI plus a bosonic
Symmetry Protected Topological state. Similar issues were also
encountered for gapped surface states of topological
superconductors\cite{Fidkowskihe,3dfSPT2,maxvortex}.

It turns out that the T-Pfaffian topological order can be easily
understood using the particle-vortex duality for the Dirac
fermion\cite{mrosscdl15,wangsenthil1,maxashvin} in
Eqn.~\ref{dualD1}. To access a symmetric, gapped phase from the
dual Dirac theory, one can simply pair-condense the dual Dirac
fermions in Eqn.~\ref{dualD1}
$\langle\chi^T\sigma_2\chi\rangle\neq0$ (in the s-wave channel to
preserve time-reversal symmetry). Unlike the original Dirac
fermion $\psi$, this pair condensate does not break the physical
charge $U(1)$ symmetry since $\chi$ does not carry physical
charge. Instead the condensate will gap out both the Dirac fermion
and the gauge field $a_{\mu}$ from Higgs mechanism, leaving behind
a fully gapped vacuum with nontrivial topological order. The
anyonic excitations include the $\chi$ fermion (Bogoliubov
fermion) and various kinds of vortices of the pair condensate --
the single vortex (trapping gauge flux $\pi$) famously carries a
Majorana zero mode and is a non-Abelian Ising-like
anyon\cite{fukane}. Moreover, the dual Dirac description also
fixes the quantum number of time-reversal symmetry on the anyons.
Specifically, the single vortex -- an Ising-like anyon with
physical charge $q=1/4$ -- is a Kramers singlet, i. e.
$\mathcal{T}^2=+1$, leading to the identification of
T-Pfaffian$_+$ as a surface state of the standard topological
insulator\cite{maxashvin, Metlitski2015}.

One can also ask what happens when the dual vortex fermion $\chi$
forms a T-Pfaffian topological order. The resulting state is now a
superconductor (or a paired superfluid), where the low energy
Goldstone mode is described in the dual picture by the free photon
field $a_{\mu}$. The Ising-like anyon in the T-Pfaffian state now
couples to $a_{\mu}$ and should be interpreted as a vortex in the
superconductor. This is nothing but the familiar Fu-Kane
superconductor on the TI surface\cite{fukane}, where a
$\pi$-vortex carries a Majorana zero mode and has non-Abelian
statistics.

Given the close relationship between a Dirac cone and the
half-filled Landau level, as reviewed above in Sec.~\ref{hfll}, it
is obvious that the T-Pfaffian topological order can also be
realized in a half-filled Landau level system. In this case the
relevant symmetry is no longer $\mathcal{T}$ but $\mathcal{CT}$
(  the particle-hole symmetry), and the same
topological order in this context is also known as
PH-Pfaffian\cite{son2015, Fidkowskihe}. In fact the PH-Pfaffian
topological order can be obtained through the traditional HLR
composite fermion formulation by pairing the (non-relativistic)
composite fermions in the $p_x+ip_y$ channel, while the classic
Moore-Read Pfaffian state corresponds to composite fermion pairing
in the $p_x-ip_y$ channel\cite{readgrn}. A recent measurement of
thermal Hall conductance in the $\nu=5/2$ 2DEG
system\cite{thermalhall} appears to be in agreement with
PH-Pfaffian rather than the Pfaffian, or its particle-hole
conjugate (anti-Pfaffian\cite{levinapf,ssletalapf}). This seems to
be in tension with numerical studies over the past two decades
which predicted Pfaffian or anti-Pfaffian as the ground
state\cite{MorfED,WangShengHaldane,Feiguinetal,StorniMorfDasSarma,RezayiSimon,Zaleteletal,RezayiPRL}.
Some theoretical studies have been
attempted\cite{Mross52,Wang52,LianWang,Simon52} but a complete
understanding is yet to be achieved.

\section{A ``miniweb" of descendant dualities: Application to deconfined quantum criticality}
\label{miniweb}

In condensed matter physics, conformal field theories often arise
at critical points separating two distinct phases. In that context
dualities of the kind reviewed here play a crucial role in
describing novel quantum critical points that are beyond the
standard Landau paradigm.  In this section we will use the
elementary dualities described in previous sections  above to
derive many descendant dualities. Within these descendant
dualities, there is a ``miniweb of dualities" that are directly
relevant to a class of quantum critical points that are beyond the
Landau paradigm.

Non-Landau quantum critical points  arise at  several phase
transitions and have been intensely studied in recent years.  One
class of examples  arises when  one or both phases on either side
have ``non-Landau order"\footnote{We will use the term Landau
order to refer to symmetry broken phases where all the non-trivial
physics can be described by the corresponding Landau order
parameter. Symmetry preserving phases within this Landau paradigm
are gapped phases with short range entanglement. Further it is
assumed that their ground state wavefunctions  can be smoothly
deformed into trivial product states while preserving all global
symmetries. Examples of non-Landau phases include Symmetry
Protected Topological phases, intrinsically topologically ordered
phases, algebraic liquid phases with gapless excitations unrelated
to Goldstone physics, etc}. Then since the Landau order parameter
description does not capture the phases it is natural that it does
not describe the phase transition either. More striking are
examples\cite{deconfine1,deconfine2} of Landau-forbidden
continuous phase transitions between Landau-allowed phases.  The
classic example is the phase transition between Neel ordered and
valence bond solid phases of quantum magnets on a square lattice.
The theory for this transition is an example of what is known as a
``deconfined quantum critical point''
\cite{deconfine1,deconfine2}(dQCP).  The critical field theory is
conveniently expressed in terms of ``deconfined''  fractionalized
degrees of freedom though the phases on either side only have
conventional ``confined'' excitations. There are, by now, many
other proposed examples of deconfined quantum critical points in
$2+1$ space-time dimensions~\cite{SU(N)dqcp1, SU(N)dqcp2,
spintexture, SO5, Nahumprx,
Nahumprl,classicalmodel1,classicalmodel2, classicalmodel3,
senthilfisher, SachdevSO5, Kagomeprl, Xutrangle,
HeRG,Honeycombdqcp1,Honeycombdqcp2,Nahum3dloop,MotrunichMC,RibhuSU(N),RibhuCPN}.
Deconfined critical  theories also emerge at phase transitions
between trivial and Symmetry Protected Topological (SPT) phases of
bosons in
$(2+1)d$~\cite{senthilashvin,Tarun_PRB2013,lulee,kevinQSH,Sp(n)prb,
ZYMprx, LuBTPT,SO5}. (For a general introduction to SPT phases,
see for example Ref.~\cite{Wengroupcohomology, Wenscience,
KitaevClassification, Tenfoldway, senthilashvin,NLSM,
Senthilscience, Senthilreview}.)  Interestingly many of the same
CFTs arise both as critical points between trivial and SPT phases
and as critical points between two broken symmetry phases.   Here
we will describe one example in some detail that illustrates the
power of dualities in thinking about such quantum critical
phenomena. Further detail can be found in Ref.~\cite{SO5}.

The first continuum field theory we will consider is the easy
plane non-compact\footnote{We remind the reader that we are using
the condensed matter terminology. Non-compact simply means that
monopole terms are not added to the action though the gauge group
is $U(1)$. The monopole operators exist as local operators in the
theory however. } CP$^1$ (NCCP$^1$) model: \beqn
\mathcal{L}_{\mathrm{CP}^1} =
|D_bz_1|^2+|D_{b-B_1}z_2|^2-|z_1|^4-|z_2|^4+\frac{1}{2\pi}bdB_2 ,
 \label{CP}
\eeqn Here $z_j$ ($j = 1,2$) is a complex scalar field and $b_\mu$
is a dynamical $U(1)$ gauge field. $B_{1,2}$ are background $U(1)$
gauge fields that couple to conserved currents (see below). This
theory arises\cite{deconfine1,deconfine2} as a low energy
description of the phase transition between Neel ordered and
Valence Bond Solid (VBS)  phases of spin-$1/2$ quantum magnets
with $O(2)$ global spin rotation symmetry  (easy plane magnets) in
addition to time reversal and lattice symmetries. We will not
review this realization  in any detail here (see Refs.
\cite{deconfine1,deconfine2}) and will restrict ourselves to
describing the identification of important physical operators in
the microscopic spin system to  local operators of the continuum
field theory.  The $XY$ component of the Neel order parameter
$\vec N$ of the quantum magnet is identified with the two real
components of the operator $z^*_1z_2$. The VBS order parameter
$\vec V$ , on the other hand, is identified with the monopole
operator. As written the Lagrangian in Eqn. \ref{CP} has a
continuous $O(2) \times O(2)$ global symmetry (and  time
reversal). We denote the continuous part of this symmetry
$U_{B_1}(1) \times U_{B_2}(1)$.  Under  $U_{B_1}(1)$  the operator
$z^*_1z_2$ carries charge-$1$. The other $U_{B_2}(1)$ symmetry is
associated with the conservation of gauge flux. The monopole
operator carries charge-$1$ under $U_{B_2}(1)$\footnote{For the
microscopic realization in quantum magnetism, we should supplement
the field theory by adding monopole operators to the Lagrangian to
explicitly break the $U_{B_2}(1)$ flux conservation symmetry. It
is known that the minimum allowed monopole operator (with
continuum angular momentum $l = 0$) has strength-$4$. Further
there is good evidence that at the critical fixed point this
strength-$4$ monopole operator is irrelevant. Thus we will
henceforth study the field theory with the full $U_{B_1}(1) \times
U_{B_2}(1)$ symmetry.}.  In addition to these continuous
symmetries the theory has a $Z_2$ spin flip symmetry ${\cal S}$
under which \be \label{calSdef} {\mathcal S}: \quad z \to \sigma_x
z, \quad b \to b. \ee This combines with $U_{B_1}(1)$ to form one
of the  $O(2)$ symmetries. There is another discrete $Z_2$
symmetry ${\mathcal C}$ which corresponds to charge-conjugation.
Its action is: \be \label{C} {\mathcal C}: \quad z \to \sigma_x
z^*. \quad b \to - b \ee This combines with $U_{B_2}(1)$ to form
the second $O(2)$ symmetry.

By perturbing the theory with a mass term $r \left( |z_1|^2 +
|z_2|^2 \right)$  we can drive a phase transition by tuning $r$.
In the field theory the Neel state is obtained as a `Higgs' phase
when $\langle z_1 \rangle = \langle z_2 \rangle \neq 0$.  This
spontaneously breaks $U_{B_1}(1)$ but preserves $U_{B_2}(1)$. When
$z$ enters a massive phase, the $U_{B_1}(1)$ is preserved but
$U_{B_2}(1)$ is spontaneously broken.  This is because at scales
lower than the $z$-mass, the effective theory is a free Maxwell
theory for $b_\mu$. This corresponds to the monopole condensate
(the resulting photon is the Goldstone boson of the broken
$U_{B_2}(1)$ symmetry). Thus this field theory captures the
possibility of a direct Landau forbidden second order
transition\footnote{ Establishing that the Higgs phase transition
in the field theory is described by a CFT in the IR cannot be done
analytically. Later we will describe supporting numerical evidence
for a second order transition.} between two phases with distinct
broken symmetries.

Next we show that the same theory Eqn. \ref{CP} also describes a
very different phase transition, namely that between trivial and
SPT phases of bosons with $\left(U(1) \times U(1)\right) \rtimes
Z_2$ symmetry.  To access these phases we instead perturb the
theory with a  different ` mass term' \be \label{rprime} r' \left(
|z_1|^2 - |z_2|^2 \right) \ee

Note that this term breaks both ${\mathcal S}$ and ${\mathcal C}$
but preserves their product. Thus the symmetry of the theory is
reduced to $\left(U(1) \times U(1)\right) \rtimes Z_2$. Depending
on the sign of $r'$ either $z_1$ or $z_2$ will condense. Either
condensate will Higgs the dynamical gauge field $b$, and we end up
with seemingly trivial gapped phases. However they are
distinguished as SPT phases.  When $z_1$ condenses (without $z_2$
condensing), at long wavelengths we have $b = 0$. The theory then
has a trivial response to the background gauge fields. On the
other hand when $z_2$ condenses (without $z_1$ condensing), we
have $b = B_1$ and the response becomes \be
 \frac{1}{2\pi} B_1d B_2
\ee This describes an SPT phase of bosons (for instance the
bosonic integer quantum Hall state~\cite{levinsenthil})  protected
by  $U_{B_1}(1) \times U_{B_2}(1)$ symmetry.

Now we consider dual descriptions of this field theory.  A first
observation is that this theory is self-dual, as was discovered
early in Ref.~\cite{ashvinlesik}. This self-duality played
an important role in the proposal of the deconfined quantum
critical point (dQCP)~\cite{deconfine1,deconfine2}. To derive the
self-duality of Eqn. \ref{CP},  we  use the bosonic charge-vortex
duality discussed in section~\ref{intro}. We first perform
particle-vortex duality for each flavor of $z_j$. In the resulting
dual theory the gauge field $b$ is readily integrated out. The
resulting theory is \beq |D_{\hat b} w_1|^2+|D_{\hat
b-B_2}w_2|^2-|w_1|^4-|w_2|^4+\frac{1}{2\pi}\hat bdB_1
\label{CPdual} \eeq
 Eqn.~\ref{CPdual} takes the
exactly the same form as Eq.~\ref{CP}. $w_1$ and $w_2$ are the
vortices of $z_1$ and $z_2$ respectively, and vice versa.

Under the duality, the mass perturbation $r$ is mapped to $ - \hat
r$. The self-duality implies that, at the dQCP where $ r = - \hat
r = 0$, the gauge invariant order parameters $\vec{N}$ and
$\vec{V}$ (the Neel and VBS order parameters) must have the same
scaling dimension, if the critical point of Eq.~\ref{CP} is indeed
a $(2+1)d$ CFT.

Interestingly these same theories also have (at least
weakly\footnote{See Ref. \cite{SO5} for a careful discussion
of the extent to which this can be derived.}) dual descriptions as
theories of fermions coupled to a dynamical $U(1)$ gauge
field\footnote{More precisely a spin$_c$ connection.}.  Consider
standard massless QED with $N_f = 2$  flavors of 2-component
fermions. The Lagrangian is
 \beqn \mathcal{L}_{\mathrm{QED}} = \sum_{j = 1}^2
\bar{\psi}_j \slashed{D}_a \psi_j + \frac{1}{4\pi}ada + 2CS_g,
\label{QED} \eeqn In keeping with our convention, we have
implicitly  assumed that we have defined\footnote{Equivalently we
are defining the phase of the path integral over the light
fermions $\psi$  in terms of the $\eta$ invariant} the massless
Dirac fermion by including two massive Dirac fermions with a large
mass $M > 0$.   The last two terms cancel the contribution of
these massive fermions to the partition function\footnote{The
reader might wonder why we did not choose a different
regularization where we  choose opposite signs for the  $\eta$
invariant for the two fermion species  in defining the massless
Dirac theory.  The reason is that such a  choice breaks the flavor
$SU(2)$ rotation of the fermions. With the present choice, flavor
$SU(2)$ is retained as an exact symmetry. Further note that since
the $\eta$ invariant is exactly cancelled by the added
Chern-Simons terms for $(a,g)$, the partition function is real and
hence the theory is time reversal invariant.}. First let us
identify local operators in this theory. Clearly there are gauge
invariant operators that can be constructed as polynomials of the
fermion fields. In addition there are also monopole operators
which insert multiples of $2\pi$ flux of $a$. An important feature
of this theory is that all these local operators are bosons. Thus,
despite the presence of fermion fields in the Lagrangian, the
theory should be viewed as describing a microscopic system of
bosons. Second let us understand the symmetries of this theory. As
usual the flux conservation leads to an extra global $U(1)$
symmetry under which the monopoles of $a$ are charged. Very
naively there is a flavor $SU(2)$ symmetry that rotates the two
fermions. However rotations in the center of $SU(2)$ can be
absorbed by a gauge transformation so we may be tempted to say
that the physical global flavor symmetry ({\em i.e}, the symmetry
that acts on local operators) is $SU(2)/Z_2 = SO(3)$. This is
incorrect. The theory has monopole operators which transform  in
the spin-$1/2$ representation of the flavor $SU(2)$. To understand
this  we follow the usual strategy of quantizing the theory on
$S^2 \times R$ with a monopole placed at the center of $S^2$.  In
the presence of the acompanying $2\pi$ magnetic flux of $a$, there
will be two zero modes coming from each of the two Dirac flavors.
Gauge invariant states are obtained by occupying one of the two
zero modes. The resulting state is an $SU(2)$ doublet. Thus the
monopole operator transforms as a doublet under the flavor $SU(2)$
of Eqn. \ref{QED}. Further it is a boson.   Combined with the flux
conservation $U(1)$ symmetry the true global flavor
symmetry\footnote{A further detail is that the model also has a
charge conjugation symmetry ${\cal C}_{f}$. Including this, and
time reversal,  the full global symmetry actually becomes  the
group $\frac{SU(2) \times Pin(2)^-}{Z_2} \times Z_2^T$ where the
$-$ sign means that the charge conjugation element squares to
$-1$. See Ref. \cite{SO5}} is thus $U(2) = \frac{SU(2) \times U(1)}{Z_2}$.

A self-duality of this theory can be obtained by  performing the
fermion-fermion particle-vortex duality for each flavor of
$\psi_j$, and eventually integrating out $a_\mu$. Then we again
end up with the same Lagrangian as
Eq.~\ref{QED}~\cite{xudual,mrossduality,seiberg2,SO5}.  This
procedure does not keep the flavor $SU(2)$ symmetry of the theory
manifest. Nevertheless it is an interesting conjecture that the
duality respects this flavor $SU(2)$ symmetry.  Thus $N_f = 2$
massless QED$_3$ with the full global symmetry described above is
conjectured\cite{xudual,mrossduality,seiberg2,SO5} to be self-dual
in the infra-red. But notice that the dual theory also has its own
$SU(2)$ flavor symmetry which is distinct from the $SU(2)$ flavor
symmetry of the original theory. Indeed the dual flavor $SU(2)$
has as a  subgroup the magnetic $U(1)$ symmetry of the original
theory. The conjectured self-duality of  $N_f = 2$ massless
QED$_3$ then implies that the IR physics of the theory has an
emergent $O(4) =  \frac{SU(2) \times SU(2)}{Z_2} \rtimes Z_2$
symmetry\footnote{That it is $O(4)$ and not $SO(4)$ is because
there is an extra $Z_2$ operation (self-duality ``symmetry") that
exchanges the two $SU(2)$ groups. The symmetry is also not Spin(4)
because there are no operators that transform in representations
like $(1/2, 0)$ under the two $SU(2)$ groups.}, in addition to
time reversal.

The infra-red fate of Eqn. \ref{QED} is not entirely clear. There
is some recent numerical evidence\cite{Karthik2016}  however that
it flows to a CFT. As we will see below the duality of this
theory, and the corresponding enlarged symmetries,  makes some
strong predictions that can be checked numerically. Conversely
confirmation of the enlarged symmetry will be a strong test of the
general correctness of the dualities reviewed in this paper. For
the time being we proceed by simply assuming that the theory is
conformal in the IR.

To see the duality  relating Eqn. \ref{QED} to Eqn. \ref{CP}, we
assume that the formal $S$ operation gives a well defined action
on CFTs, and start with the `basic' boson-fermion duality: \be
\mathrm{D} = ST[\mathrm{WF}]. \ee The other boson-fermion duality
then is \be \mathrm{D} = T^{-1} S^{-1} T^{-1}[\mathrm{WF}]. \ee
Multiplying the partition functions on both sides, shifting an
$AdA/4\pi$  to the left, and finally making $A$ dynamical we get a
duality of QED$_3$ to the easy plane CFT defined in Eqn. \ref{CP}.
More explicitly the duality takes the form \bea
&&|D_bz_1|^2+|D_{b-B_1}z_2|^2-|z_1|^4-|z_2|^4+\frac{1}{2\pi}bdB_2
\nn \Longleftrightarrow \hspace{20pt}
&&\bar{\psi}_1i\slashed{D}_a\psi_1+\bar{\psi}_2i\slashed{D}_{a+B_2-B_1}\psi_2+\frac{1}{4\pi}ada+\frac{1}{2\pi}adB_2+\frac{1}{4\pi}B_2dB_2+2CS_g,
\eea

\begin{figure*}[t]
\includegraphics[width=0.55\textwidth]{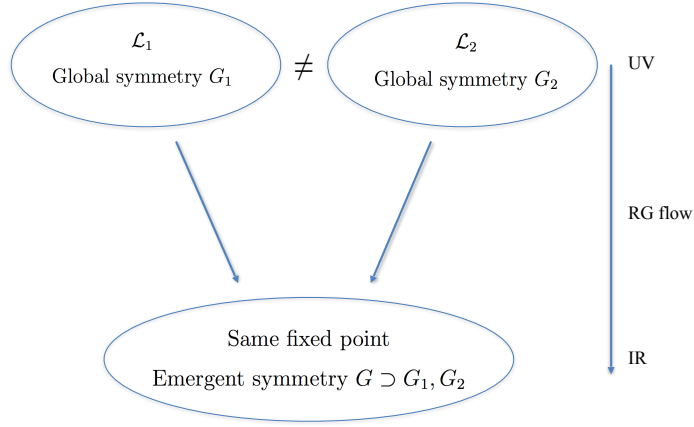}
\caption{The basic logic of emergent symmetries from dualities. }
\label{fig:emergentsymm}
\end{figure*}

Despite the formal appeal of these derivations  they should be
treated with some caution. The manipulations involved treat the
two flavors of Dirac fermions in a non-$SU(2)$ invariant way. Thus
we may worry that the duality is not really faithful to the flavor
$SU(2)$ symmetry.  But let us proceed by making the {\em strong
assumption} suggested by these formal manipulations that Eqn.
\ref{CP} and Eqn. \ref{QED} are both self-dual, and dual to each
other, and that they all flow to the same IR CFT.  This then
immediately has the implication that this IR CFT has an enlarged
$O(4) \times Z_2^T$ symmetry. In particular the easy plane
Neel/VBS transition, when continuous, will have this  enlarged
symmetry.  The basic logic is summarized in
Fig.~\ref{fig:emergentsymm}.

Let us understand the relationship of local operators in the
various descriptions.  The  $\vec N$ and $\vec V$,  correspond  in
QED, to the monopole operators ${\cal M}_a$. We argued before that
they form a doublet under the flavor $SU(2)$ of QED$_3$.  It is
easy to see that they also form a doublet under the flavor $SU(2)$
of the dual QED$_3$ theory. Thus under $O(4)$ the monopole
operators transform in the  $(\frac{1}{2},  \frac{1}{2})$
representation, {\em i.e} as $O(4)$ vectors.  The identification
with the Neel and VBS order parameters means that the $O(4)$
symmetry operation simply rotates the four real components of
$\vec N, \vec V$ into each other.

It is also easy to see that the operator $\bar{\psi }\psi$ maps to
$|z_1|^2 - |z_2|^2$ while $\bar{\psi} \sigma^3 \psi$ maps to
$|z_1|^2 + |z_2|^2$. We invite the reader to check that the
QED$_3$ theory with either of these perturbations added reaches
the same phases as those described by the bosonic theory of Eqn.
\ref{CP}. Note that  in the QED$_3$ description  the mass term
$m\bar{\psi}\psi$ preserves the $SO(4) = \frac{SU(2) \times
SU(2)}{Z_2} $ symmetry on both sides of the transition, but breaks
the $Z_2$ self-duality symmetry. It thus breaks $O(4)$ down to
$SO(4)$. The theory Eq.~\ref{QED} and its self-duality was also
discussed in the context of the boundary of a $3d$ bosonic SPT
state~\cite{xudual}.

To summarize, this miniweb of dualities, in its strongest form,
makes some amazing predictions: the easy-plane dQCP is equivalent
to the bosonic topological transition, and it could have an
emergent O(4) symmetry! This emergent O(4) symmetry of easy-plane
dQCP was first conjectured in Ref.~\cite{senthilfisher}, and
this miniweb of dualities gave another more direct perspective of
this emergent $O(4)$ symmetry.

Note that the $O(4)$ symmetry is not present in the UV Lagrangian
of any of the theories in the duality web that we have discussed.
Thus for $O(4)$ to emerge in the IR it is necessary that, at the
resulting IR fixed point, perturbations that break $O(4)$ to the
appropriate UV symmetries are irrelevant. Now the QED and easy
plane NCCP$^1$ theories have different UV symmetries. Thus for
$O(4)$ to emerge, as per the strong version of the duality,  there
are different perturbations that need to be irrelevant at the
putative $O(4)$ symmetric fixed point.  Alternately it is possible
that QED does flow to an $O(4)$ symmetric fixed point but the easy
plane NCCP$^1$ model does not.  A careful discussion of subtleties
in the arguments for these mini-dualities and the implications for
the enlarged IR symmetry is in Ref. \cite{SO5} to which we
refer the reader.

The Neel/VBS transition in fully $SU(2)$ spin symmetric models  is
described by the NCCP$^1$ model with $SU(2)$ flavor symmetry:
\beqn \mathcal{L} = |D_bz_1|^2+|D_{b}z_2|^2-(|z_1|^2 + |z_2|^2)^2
, \label{SU2CP} \eeqn

Now the Neel order parameter is a 3-component vector $z^\dagger
\vec \sigma z$ while the VBS order parameter continues to be
identified with the monopole operator ${\cal M}_b$.  The
transition associated with the condensation of $z$, if second
order, will describe a continuous Landau forbidden Neel-VBS
transition of $SU(2)$ invariant spin-$1/2$ magnets on the square
lattice.  Numerical work sees good evidence for a second order
transition (albeit with poorly understood drifts in exponents);
remarkably numerical work also unearthed\cite{Nahumprl} the
emergence of an $SO(5)$ symmetry that rotates the Neel and VBS
order parameters into one another (a possibility raised earlier by
a sigma model formulation in Refs.
\cite{berryphaseprl,senthilfisher}). Inspired by this, Ref.
\cite{SO5} proposed that the $SU(2)$ invariant NCCP$^1$
model at its critical point is itself self-dual. This generalizes
the self-duality of its easy plane counterpart. Indeed the easy
plane self duality follows as a consequence if Eqn. \ref{SU2CP} is
self-dual. Importantly the proposed self-duality of Eqn.
\ref{SU2CP} implies that the critical fixed point has an emergent
$SO(5)$ symmetry. The argument is somewhat similar to the one used
for the emergent $O(4)$ symmetry of the $N_f = 2$ QED$_3$ theory.
The manifest continuous symmetry in eqn. \ref{SU2CP} is only
$SO(3) \times U(1)$. This is also the manifest continuous symmetry
group of the dual theory. However the $U(1)$ of either theory is a
subgroup of the $SO(3)$ of the corresponding dual theory.  Thus if
the duality is right, the critical fixed point must have enlarged
$SO(5)$ symmetry that is not manifest in either description. The
numerical observation of such enlarged symmetry is support for
such a self-duality. Ref. \cite{SO5} also proposed a
fermionic dual for Eqn. \ref{SU2CP}. This theory - dubbed the
QED-Gross Neveau model - is the $N_f = 2$ QED$_3$ theory augmented
with a critical scalar that couples to a fermion bilinear through
a Yukawa coupling. This fermionic theory is itself also self-dual
and there is then an interesting (though conjectural) web of
dualities involving Eqn. \ref{SU2CP}.

\section{Numerical evidence for the duality}
\label{Numerics}

The easy-plane NCCP$^1$ model arises as  the field theory
describing  the dQCP between an in-plane (XY) antiferromagnet
(AFM) and a VBS phase in spin-$1/2$ quantum magnets on a square
lattice.  In the presence of full $SO(3)$ spin rotation symmetry
this transition  has been explicitly realized in lattice models
which have been extensively simulated numerically using unbiased
QMC
techniques\cite{Sandvik2002,Sandvik2007,Lou2009,Sen2010,Sandvik2010,Nahum3dloop,Harada2013,Pujar2013,Pujari2015,Nahumprx,Nahumprl,Shao2016}.
An important such lattice model is the sign-problem-free $J-Q$
model which consists of both a nearest neighbor antiferromagnetic
Heisenberg interaction (strength $J$ and a four-spin plaquette
term (strength $Q$).  Naturally the easy plane version can also be
studied by adding suitable anistropies to the exchange constants.
Although there are studies that indicate that some version of the
$J$-$Q$ model with an in-plane spin symmetry and other U(1)
symmetric models should lead to a first order transition
\cite{Kuklov08,Geraedts2012,Jonathan2016,Jonathan2017},
Ref.~\cite{epjq} demonstrates that a different model -  the
Easy Plane $J-Q$ model (EPJQ model), instead leads to a continuous
transition in some region of its parameter space. We also note
there is a recent QMC work on the extended Hubbard model of
hardcore bosons on the kagome lattice, suggesting a similiar
continuous easy-plane phase transition~\cite{XFZhang2017}.

The $N_f=2$ QED action has recently been simulated directly using
a lattice QED model~\cite{Karthik2016}. The numerical results are
consistent with an IR CFT.  The scaling dimension of $\bar{\psi}
\vec \sigma \psi$ was computed.  Further, as discussed above, $N_f
= 2$ QED$_3$ gauge field arises also as the effective theory that
describes the transition between a bosonic symmetry protected
topological (BSPT) state and a trivial Mott state in
$2d$~\cite{Tarun_PRB2013,lulee}. This transition was also realized
in an interacting fermion model on a bilayer honeycomb (BH)
lattice introduced in Refs.~\cite{kevinQSH,mengQSH2} and
simulated~\cite{kevinQSH,mengQSH2} with a determinantal QMC method
(DQMC).  We describe  the Hamiltonian  in detail below. The
lattice model of Ref.~\cite{mengQSH2} has an exact SO(4)
symmetry that precisely corresponds to the proposed emergent
symmetry of the $N_f=2$ QED. Note that the fermions in the BH
model do not directly correspond to the Dirac fermions of the
$N_f=2$ QED action, because the former are not coupled to any
dynamical gauge field. The relation between the two systems
instead arises from the correspondence of the gauge invariant
fields of $N_f = 2$ QED to the low-energy bosonic excitations of
the BH model.

Using the BH model and the EPJQ model, the IR duality between the
$N_f = 2$ QED and the easy plane NCCP$^1$ field theories can be
explored on the lattice. A four component order parameter
$\boldsymbol{n}$ which transforms as a vector under the O(4)
symmetry (as a consequence of the duality) can also be
conveniently defined in both lattice models, with explicit forms
that we explain below. Thus, the two systems can be investigated
and compared via unbiased large-scale QMC simulations~\cite{epjq}.

\subsection{Bilayer honeycomb model for BTT} \label{sec:BHmodel}

The BH model is a fermionic model defined on a honeycomb lattice
\cite{kevinQSH,mengQSH2}. On each site, we define four flavors of
fermions (two layers $\times$ two spins);
\begin{equation}
c_{i}=(c_{i1\uparrow},c_{i1\downarrow},c_{i2\uparrow},c_{i2\downarrow})^\intercal.
\end{equation}
The Hamiltonian is
\begin{equation}
\label{eq:BH} H_\text{BH}=H_\text{band}+H_\text{int},
\end{equation}
where the band and interaction terms are given by
\begin{subequations}
\begin{align}
  H_\text{band}&=-t\sum_{\langle ij\rangle}c_{i}^\dagger c_{j}+\lambda\sum_{\langle\!\langle i j\rangle\!\rangle}
  \ii \nu_{ij} (c_{i}^\dagger \sigma^3 c_{j}+\text{h.c.}),\\
  H_\text{int}&=V\sum_{i}(c_{i1\uparrow}^\dagger c_{i2\uparrow}
   c_{i1\downarrow}^\dagger c_{i2\downarrow}+\text{h.c.}),
\end{align}
\end{subequations}
where $\langle i j\rangle$ and $\langle\!\langle i
j\rangle\!\rangle$ denote nearest-neighbor intra- and inter-layer
site pairs, respectively. The band Hamiltonian $H_\text{band}$ is
just two copies of the Kane-Mele model~\cite{kane2005a}, which
drives the fermion into a quantum spin Hall state with spin Hall
conductance $\sigma_\text{sH}=\pm2$ (depending on the sign of the
spin-orbit coupling $\lambda$). Including a weak interaction $V$,
the bilayer quantum spin Hall state automatically becomes a BSPT
state~\cite{mengQSH2,Wu2016,youspn}, where only the bosonic O(4)
vector $\boldsymbol{n}$ remains gapless (and protected) at the
edge, while the fermionic excitations are gapped out. However, a
strong interlayer pair-hopping interaction $V$ eventually favors a
direct product state of anti-bonding Cooper pairs. In the strong
interaction limit ($V\to\infty$), the ground state of the BH model
reads
\begin{equation}\label{eq:gs}
|\text{GS}\rangle=\prod_{i}(c_{i1\uparrow}^\dagger
c_{i1\downarrow}^\dagger-c_{i2\uparrow}^\dagger
c_{i2\downarrow}^\dagger)|0_c\rangle,
\end{equation}
with $|0_c\rangle$ being the fermion vacuum state. This state has
no quantum spin Hall conductance, i.e., $\sigma_\text{sH}=0$, and,
more importantly, it is a direct product of local wave functions.
Hence we call it a trivial Mott insulator state. It was found
numerically that there is a direct continuous transition between
the BSPT and the trivial Mott phases at
$V_c/t=2.82(1)$~\cite{mengQSH2}, where the single-particle
excitation gap does not close but the excitation gap associated
with the bosonic O(4) vector closes and the quantized spin Hall
conductance changes from $\pm2$ to $0$.

The low-energy bosonic fluctuations around the critical point form
an O(4) vector
$\boldsymbol{n}=(\Re\Sigma,\Im\Sigma,\Re\Delta,\Im\Delta)$ with
$\Sigma, \Delta$ defined as

\begin{subequations}
\begin{align}
\Sigma_i&=(-1)^i(c_{i1\uparrow}^\dagger c_{i2\downarrow}+c_{i2\uparrow}^\dagger c_{i1\downarrow}), \label{eq:O4vector1}\\
\Delta_i&=(c_{i1\downarrow}c_{i1\uparrow}-c_{i2\downarrow}c_{i2\uparrow}),
\label{eq:O4vector2}
\end{align}
\label{eq:O4vector}
\end{subequations}
\noindent Thus $\Sigma$ carries spin and $\Delta$ carries charge.
The BH model Eq.~\ref{eq:BH} respects the global $\text{SO}(4)$
symmetry of the vector $\boldsymbol{n}$. If the symmetry is
lowered to
$\text{U}(1)_\text{spin}\times\text{U}(1)_\text{charge}$, then,
based on the analysis of $N_f = 2$ QED, in principle the mass term
$M \bar{\psi} \sigma^3 \psi$ is allowed; hence the BSPT-Mott
transition is unstable towards spontaneous symmetry-breaking of
the remaining symmetries. The symmetry of the mass term $M
\bar{\psi} \sigma^3 \psi$ is identical to the following
Hubbard-like interaction (both forming a $(1,1)$ representation of
the SO(4)):
\begin{equation}
\frac{U}{2}\sum_i(\Delta_i^\dagger\Delta_i+\Delta_i\Delta_i^\dagger-\Sigma_i^\dagger\Sigma_i-\Sigma_i\Sigma_i^\dagger)=U\sum_{i}\rho_{i\uparrow}\rho_{i\downarrow}.
\end{equation}
Here $\rho_{i\sigma}$ is the density operator (for
$\sigma=\uparrow,\downarrow$ spins),
\begin{equation}
\rho_{i\sigma}=(c_{i1\sigma}^\dagger
c_{i1\sigma}+c_{i2\sigma}^\dagger c_{i2\sigma}-1),
\end{equation}
which counts the number of $\sigma$-spin fermions in both layers
on site $i$ with respect to half-filling. The repulsive $U>0$ (or
attractive $U<0$) interaction drives spin
$\langle\Sigma\rangle\neq0$ (or charge
$\langle\Delta\rangle\neq0$) condensation, leading to a
spin-density wave (SDW)~\cite{Wu2016} (or superconducting) phase
that breaks the $\text{U}(1)_\text{spin}$ (or
$\text{U}(1)_\text{charge}$) symmetry spontaneously.

\subsection{EPJQ model for the easy-plane dQCP} \label{sec:JQmodel}

The EPJQ model is a spin-$1/2$ system with anisotropic
antiferromagnetic couplings which we here define on the simple
square lattice of $L^2$ sites and periodic boundary conditions. It
is a ``cousin''  of the previously studied
SU(2)$_\text{spin}$ $J$-$Q_3$ model
\cite{Lou2009,Sen2010,Sandvik2010}which in turn is an extension of
the original $J$-$Q$, or $J$-$Q_2$, model \cite{Sandvik2007}.
Starting from the spin-$1/2$ operator $\boldsymbol{S}_{i}$ on each
site $i$, we define the singlet-projection operator on lattice
link $ij$;
\begin{equation}
P_{ij}=\tfrac{1}{4}-\boldsymbol{S}_i\cdot\boldsymbol{S}_j,
\end{equation}
then the model Hamiltonian reads
\begin{equation}
H_\text{JQ}=-J\sum_{\langle ij\rangle}(P_{ij}+\Delta
S_i^zS_j^z)-Q\hskip-2mm\sum_{\langle ijklmn \rangle}\hskip-2mm
P_{ij}P_{kl}P_{mn}, \label{eq:JQmodel}
\end{equation}
where the $\Delta S_i^zS_j^z$ term for $\Delta \in (0,1]$
introduces the easy-plane anisotropy that breaks the
$\text{SU}(2)_\text{spin}$ symmetry down to
$\text{U}(1)_\text{spin}$ explicitly. In the $Q$ term the index
pairs $ij$, $kl$, and $mn$ correspond to links forming columns on
$2\times 3$ or $3\times 2$ plaquettes, as illustrated in Fig.~1 of
Ref.~\cite{Lou2009}.

To study the columnar VBS (dimer) order realized in the EPJQ
model, we define
\begin{subequations}
  \label{dops}
  \begin{align}
  & D^x_{i}=(-1)^{x_i}{\bf S}_i \cdot {\bf S}_{i+\hat x},\\
  & D^y_{i}=(-1)^{y_i}{\bf S}_i \cdot {\bf S}_{i+\hat y},
  \end{align}
\end{subequations}
where ${i+\hat x}$ and ${i+\hat y}$ denote neighbors of site $i$
in the positive $x$ and $y$-direction, respectively. At the
critical point, the proposed self-duality (through the putative
duality with $N_f = 2$ QED) implies that the $C_4$ rotation
symmetry and the $\text{U}(1)_\text{spin}$ symmetry are enlarged
into an emergent $\text{O}(4)$ symmetry, such that the components
of the $\text{O}(4)$ vector order parameter (after some proper
normalization)
\begin{equation}
\boldsymbol{n}=(D^x,D^y,S^x,S^y),
\end{equation}
should all have the same scaling dimension~\cite{SO5}.
Ref.~\cite{epjq} demonstrated that when $\Delta = 1/2$,
Eq.~\ref{eq:JQmodel} hosts a direct second order phase transition
from Neel  to VBS  states.

\subsection{Duality relations}

\begin{figure*}[t]
\includegraphics[width=\textwidth]{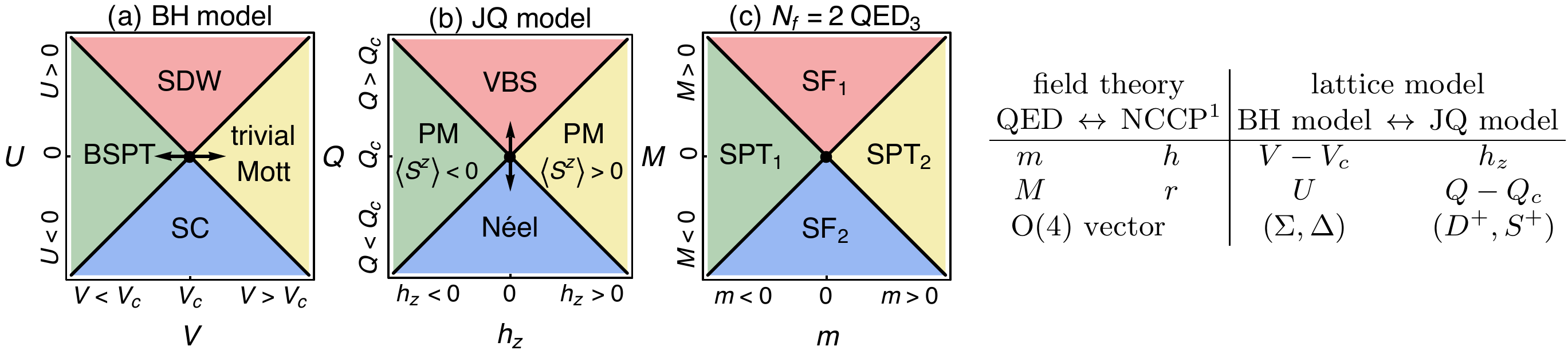}
\caption{Schematic phase diagrams of (a) the bilayer honeycomb
(BH) model, (b) the easy-plane $J$-$Q$ (EPJQ) model, and (c) the
$N_f=2$ QED theory. In (a), the BH model contains two
symmetry-breaking phases: the spin-density-wave (SDW) and
superconducting (SC) phases, and two symmetric phases: the bosonic
symmetry-protected topological (BSPT) and the trivial
Mott-insulating phases. In (b), the EPJQ model also contains two
symmetry-breaking phases: the N\'eel antiferromagnetic (AFM) phase
and the valence-bond solid (VBS) phase, and two spin-polarized
phases induced by an external staggered field. In (c), as was
shown in Ref.~\cite{Tarun_PRB2013,lulee,SO5}, when tuning
the two masses $m \bar{\psi} \psi$ and $M \bar{\psi} \sigma^3
\psi$, the $N_f = 2$ QED theory also has two symmetry-breaking
(SB) phases and two symmetric (SY) phases, one of which is the
BSPT state. The two mass terms $m \bar{\psi} \psi$ and $M
\bar{\psi} \sigma^3 \psi$ are dual to the staggered external
magnetic field $h (|z_1|^2 - |z_2|^2)$ and the tuning parameter $r
(|z_1|^2 + |z_2|^2)$ in the easy plane NCCP$^1$ theory
respectively.} \label{fig:phasediagrams}
\end{figure*}

Fig.~\ref{fig:phasediagrams} summarizes the intuitive duality
relations among the BH, EPJQ, and QED models. To numerically prove
the validity of these duality relations, Ref.~\cite{epjq}
investigated the following critical behavior at the BSPT--Mott
transition in the BH model:
\begin{subequations}
\label{eq:BHexponents}
\begin{align}
\xi&\sim|V-V_c|^{-\nu_\text{BH}}, \label{eq:BHexponents1}\\
\langle\rho_{i\uparrow}\rho_{i\downarrow}\rho_{j\uparrow}\rho_{j\downarrow}\rangle&\sim|\boldsymbol{r}_{ij}|^{-1-\eta^{\rho}_{\text{BH}}}, \label{eq:BHexponents2} \\
\langle\Delta_{i}^\dagger
\Delta_j\rangle&\sim|\boldsymbol{r}_{ij}|^{-1-\eta^{\Delta}_{\text{BH}}};
\label{eq:BHexponents3}
\end{align}
\end{subequations}
where $\boldsymbol{r}_{ij}$ is the lattice vector separating the
sites $i,j$, $\xi$ is the correlation length of the critical O(4)
bosonic modes of the system, and the density $\rho_{i\sigma}$ and
pairing $\Delta_{i}$ operators have been defined in
Sec.~\ref{sec:BHmodel}. Ref.~\cite{epjq} also studied the
following expected critical scaling behavior at the AFM--VBS
transition in the EPJQ model:
\begin{subequations}\label{eq:JQexponents}
\begin{align}
\xi&\sim|Q-Q_c|^{-\nu^{xy}_\text{JQ}},\label{eq:JQexponents1} \\
\langle S_i^zS_j^z\rangle&\sim|\boldsymbol{r}_{ij}|^{-1-\eta^{z}_{\text{JQ}}},\label{eq:JQexponents2}\\
\langle
S_i^+S_j^-\rangle&\sim|\boldsymbol{r}_{ij}|^{-1-\eta^{xy}_{\text{JQ}}}.
\label{eq:JQexponents3}
\end{align}
\end{subequations}
where $\xi$ is the correlation length of the easy-plane spins.

If the strong duality web for the easy plane NCCP$^1$ model  is
correct, and provided that $N_f = 2$ QED is indeed the theory for
the BSPT--Mott transition, then the exponents defined above must
satisfy the following relationships \cite{SO5}:
\begin{subequations}
\begin{align}
&3-\frac{1}{\nu_\text{BH}}=\frac{1+ \eta^z_\text{JQ}}{2},
\label{eq:dualityrelations1}\\
&3-\frac{1}{\nu^{xy}_\text{JQ}}=\frac{1+\eta_\text{QED}}{2} =
\frac{1+\eta^{\rho}_{\text{BH}}}{2},
\label{eq:dualityrelations2}\\
&\eta^{\Delta}_\text{BH}=\eta^{xy}_\text{JQ}.
\label{eq:dualityrelations3}
\end{align}
\label{eq:dualityrelations}
\end{subequations}

Here $\eta_\text{QED}$ is the anomalous dimension of the fermion
mass $\bar{\psi}\sigma^3\psi$ 
which was numerically estimated in the recent lattice QED
calculations in Ref.~\cite{Karthik2016}.
Eqs.~(\ref{eq:dualityrelations}) essentially mean that the gauge
invariant operators that map to each other under the duality
transformation must have the same scaling dimension at the
critical point.

Within the error bar, all three key predictions in
Eq.~\ref{eq:dualityrelations} have been confirmed in
Ref.~\cite{epjq}. Thus the ``miniweb of dualities" discussed
in the previous section has received very promising support from
the numerics. Another recent numerical simulation of $N_f = 2$
noncompact QED also supports the emergence of the O(4) symmetry of
the theory~\cite{karthik2017}.

As the dualities of the ``miniweb" were derived (at least as weak
dualities) based on the assumption of the elementary
dualities~\cite{son2015,wangsenthil1,wangsenthil2,maxashvin,wufisher,seiberg1},
a proof of the ``miniweb" indirectly also proves the latter. In
principle this result can lead to a large number of further
descendant dualities.

We also point out that the recent development of conformal
bootstrap methods has given us rigorous bounds for CFTs with
$O(N)$ symmetry (see for example Ref.~\cite{bootstrap}).
These bounds seem incompatible with the numerically measured
scaling dimension of the antiferromagnet and VBS order parameters
at the deconfined QCP with  full SU(2) spin symmetry. As
discussed earlier, this deconfined QCP was conjectured to possess
an emergent SO(5) symmetry~\cite{senthilfisher}, which is
apparently confirmed by
 numerical simulations with
a fairly large system size~\cite{Nahumprl}. The inconsistency
between numerical simulation and bootstrap methods leads to the
question of whether the deconfined QCP is truly a second order
phase transition, or a ``generic" weak first order transition,
which has a rather large correlation length. But its true nature
evades the capability of the finite size numerical simulations.

Indeed, although no sign of typical first order transition was
observed in these numerical simulations, unusual scaling behaviors
were indeed noted~\cite{Nahumprx,Sandvik2010,Shao2016}, which
raised concerns of conformal invariance in the infrared limit.
Ref.~\cite{SO5} proposed that the deconfined QCP may actually be a
``pseudocritical point", similar to what is known in some
classical statistical mechanics models (like the $2d$ 5-states
Potts models, see, for eg,
Ref.~\cite{nienhuispotts,nauenbergscalapino,cardynauenbergscalapino,RychkovWalking1,RychkovWalking2})
and some quantum field theories\cite{Gies,merge,gukov2017rg}.
Although it is not a true CFT in the infrared limit, it is
generically close to the case where two true CFTs merge together,
as was studied in, eg, Ref.~\cite{merge}. There is a dimensionless
parameter $\alpha$: when $\alpha > \alpha_c$, there are two true
CFTs, and the two CFTs merge and annihilate each other at $\alpha
= \alpha_c$. Ref.~\cite{merge} demonstrated that when $\alpha$ is
slightly smaller than $\alpha_c$, though a CFT no longer exists,
the correlation length of the system generally scales with
$|\alpha - \alpha_c|$ in the same way as the Kosterlitz-Thouless
transition: $\xi \sim \exp(\pi/\sqrt{|\alpha - \alpha_c|})$, which
can be very large when $\alpha$ is close to $\alpha_c$. In fact,
the well-known KT transition fixed point itself can be interpreted
this way~\cite{merge}.

One can naturally ask if the single-flavor dualities discussed
earlier in this review can be checked numerically. Unfortunately
most of the field theories involved come with sign-problems (due
to Chern-Simons gauge couplings) that forbids large-scale
Monte-Carlo simulations (except for the original boson-vortex
duality which has been simulated\cite{sudbo99,Kajantie04}).
Recently a particular sign-free lattice theory that is believed to
be dynamically (but not topologically) equivalent to
Eqn.~\ref{dualD1} has been simulated in Ref.~\cite{KNDDual}. No
critical point was seen in that simulation. Rather it was found
that a Dirac mass seemed to be generated spontaneously, breaking
time-reversal symmetry, in contrast with the expectation from the
strong duality. If this result indeed holds for Eqn.~\ref{dualD1},
then the Dirac-Dirac duality would only be a weak duality but not
a strong one. More numerical explorations of these field theories
are clearly called for.

\section{Duality and Bosonization of Majorana fermions}
\label{MajoranaBosonization}

\subsection{Duality of a single Majorana fermion}

In this section we will review a proposed dual description of
$(2+1)d$ Majorana fermions~\cite{xumajorana,seiberg3}.
Ref.~\cite{seiberg3} studied more general theories whose
both sides of the duality can couple to $SO(N)$ and $USp(N)$ gauge
fields; here we will mainly follow Ref.~\cite{xumajorana}
which discussed duality of free Majorana fermions which is more
relevant to condensed matter systems.

Consider a single two-component
Majorana fermion in $(2+1)d$, 
\begin{equation}
{\mathcal L}_\xi =\bar{\xi} (i \slashed{\partial}  + m)\xi.
\label{Lchi}
\end{equation}
Let us consider the two phases of the Majorana fermion by tuning
the mass term $m$ from positive to negative value. We regularize
Eqn.~\ref{Lchi} so that for $m > 0$, $\xi$ realizes a trivial
phase with no edge state, or in other words its edge state has
chiral central charge $c_-=0$; while for $m < 0$ it realizes a
$p_x+ip_y$ superconductor with $c_- = 1/2$. Such a regularization
can be provided by starting with a lattice model with two gapless
Majorana cones and initially gapping one of them out to produce
Eqn.~\ref{Lchi} with $m =0$. Thus, the massless Majorana cone
corresponds to the critical point between a trivial phase and a
$p_x+ip_y$ superconductor. Now, the idea is to produce a dual
theory that realizes the same two phases by tuning a parameter.
Then at the critical point we may conjecture the dual theory has
the same infrared behavior as Eqn.~\ref{Lchi}. Observe that we can
also capture the same pair of gapped phases with the following
theory of a boson coupled to an $SO(N)_1$ CS gauge field:
\begin{equation}
{\mathcal L}_b = |D_a \phi|^2 - r |\phi|^2 - g |\phi|^4 +
\mathrm{CS}_{SO(N)}[a]_1+N \cdot \mathrm{CS}_g \label{SON}
\end{equation}
Here, $\phi$ is an $N$-component real vector, $a$ is an $SO(N)$
gauge field and \beq {\rm CS}_{SO(N)}[a]_1 = \frac{1}{2\cdot 4\pi}
\mathrm{tr}_{SO(N)}\left(a\wedge da - \frac{2 i}{3} a \wedge a
\wedge a\right). \label{CSO}\eeq The trace in Eq.~(\ref{CSO}) is
in the vector representation of $SO(N)$.

Let's analyze the mean field phase diagram of Eqn.~\ref{SON}. When
$r > 0$, $\phi$ is gapped. The theory $SO(N)_1$ coupled to gapped
bosonic matter gives a state with no intrinsic topological order;
the vector boson $\phi$ is transmuted to a fermion by the CS
field.\footnote{The $SO(3)_1$ example might be the most familiar:
this theory is the same as $SU(2)_2 = \{1, \sigma, f\}$ restricted
to integer spin, $i.e.$ to $\{1, f\}$. The chiral central charge
is $c_-= -3/2$. Notice that for $SO(N)_1$ CS theory, we forbid
excitations with projective representation (such as spinor) of
$SO(N)$, otherwise the bulk would have topological order, like the
case studied in Ref.~\cite{wenso5}.} The $SO(N)_1$ CS gauge
theory by itself has a chiral central charge $c_- = -N/2$, which
exactly cancels the background gravitational CS term in
Eqn.~\ref{SON}. So the $r
> 0$ phase is a trivial state with total $c_- = 0$. On the other hand,
for $r < 0$, $\phi$ condenses, which breaks the gauge group
$SO(N)$ to $SO(N-1)$. At low energies, we can then take $a$ to be
an $SO(N-1)$ gauge field, obtaining an $SO(N-1)_1$ CS theory.
Again, this is a state with no intrinsic topological order, but
the background gravitational term in Eqn.~\ref{SON} is no longer
fully cancelled, rather: $c_- = N/2 - (N-1)/2 = 1/2$. So the $r <
0$ phase is a $p_x+ip_y$ topological superconductor. The proposal
is that the strongly interacting field theory Eqn.~\ref{SON} in
the IR limit at $r = 0$ is dual to a single noninteracting
$(2+1)d$ Majorana fermion with $m = 0$.

\subsection{Parton approach to Majorana duality}

Here we consider a lattice model for Majorana fermion $c_j$
($c_j^\dagger = c_j$). We introduce on each site $j$, $N$ colors
of slave Majorana fermions $\chi_{j,\alpha}$ ($\alpha = 1 \ldots
N$) for odd $N$ such that \beqn c_j = (i)^{\frac{N-1}{2}}
\prod_{\alpha = 1}^N \chi_{j,\alpha}. \label{parton1}\eeqn By
construction $\chi_{\alpha}$ is coupled to a dynamical $SO(N)$
gauge field. Now we design an identical mean field $p_x+ip_y$
superconductor band structure for each color of $\chi_\alpha$. At
the mean field level, there are $N$ chiral Majorana fermions at
the boundary, which in total leads to chiral central charge $c_- =
N/2$. However, if the $SO(N)$ gauge symmetry is unbroken, after
coupling to the $SO(N)$ gauge field there will be no gauge
invariant degrees of freedom left at the boundary, so we are left
with $c_- =0$. Also, integrating out $\chi_{\alpha}$ would
generate a CS term at level $k = 1$ for the $SO(N)$ gauge field,
as well as a gravitational CS term at level $N$, as in
Eqn.~\ref{SON}. As already discussed, there is no topological
order in the bulk, thus this state is again a trivial state of the
physical Majorana fermion $c_j$.

Using the slave particles $\chi_{j,\alpha}$ we can also define a
$SO(N)$ vector boson $ \hat{\phi}_{j,\alpha}$: \beqn
\hat{\phi}_{j, \alpha} \sim (i)^{\frac{N-1}{2}} \epsilon_{\alpha
\alpha_1, \cdots \alpha_{N-1}} \chi_{j, \alpha_1} \cdots
\chi_{j,\alpha_{N-1}}. \eeqn
When $\hat{\phi}_{\alpha}$ condenses, it breaks the $SO(N)$ gauge
group down to $SO(N-1)$, and one of the slave fermions (say
$\chi_N$) is no longer coupled to any gauge field, thus its
topological band structure implies that the entire system is
equivalent to one copy of $p_x + ip_y$ topological superconductor.
Likewise, the edge mode associated to $\chi_N$ sees no gauge field
and survives as a true $c_- = 1/2$ edge mode of a $p_x + i p_y$
superconductor. All other edge modes are, as before, eliminated by
$SO(N-1)$ gauge field fluctuations. Thus, by coarse-graining
$\hat{\phi}$ into a continuum field $\phi$, we can describe a
transition between a trivial state and a $p_x + ip_y$
superconductor with Eqn.~\ref{SON}.

Notice that the integer $N$ in the dual theory Eqn.~\ref{SON}
needs not be odd. A slightly more involved parton construction for
even $N$ was given in Ref.~\cite{xumajorana}.


\subsection{The dictionary}

How do we represent the physical Majorana fermion in the dual
theory Eqn.~\ref{SON}? Recall that the magnetic flux of the
$SO(N)$ gauge field is classified by $\pi_1(SO(N)) = Z_2$. Indeed,
imagine the system on a spatial sphere $S^2$. As usual, we place a
magnetic flux through the sphere by dividing it into two
hemispheres, and gluing the fields in the two hemispheres along
the equator with a gauge transformation $g(\theta)$, $\theta \in
[0, 2 \pi]$. Such gauge transformations are classified by $\pi_1$
of the gauge group. In the case of the $SO(N)$ group, a simple
representative for the single non-trivial magnetic flux sector on
$S^2$ is obtained by considering an ordinary flux $2 \pi$ Dirac
monopole in the $SO(2)$ subgroup of $SO(N)$. Note that by an
$SO(N)$ rotation we can invert the magnetic flux in the $SO(2)$
subgroup, so the $SO(2)$ magnetic flux is, only defined modulo
$4\pi$. The magnetic flux breaks the $SO(N)$ group down to $SO(2)
\times SO(N-2)$ and the state on $S^2$ must be neutral under this
reduced gauge group. As in the Abelian case, the CS term in
Eqn.~\ref{SON} leads to the monopole carrying an $SO(2)$ charge
$1$, so to make the monopole neutral we must attach to it the
boson $\phi_1 + i \phi_2$, which makes the neutral monopole a
fermion. Also, the angular momentum of the neutral monopole is
half-odd-integer because of the $SO(2)$ flux. We conclude that the
$SO(N)$ monopole on $S^2$ carries charge under fermion parity, and
identify the $SO(N)$ space-time monopole $V_M$ with the Majorana
fermion operator $\xi$. In particular, this discussion means that
dynamical $Z_2$ $SO(N)$ monopoles are prohibited in the partition
function of dual theory Eqn.~\ref{SON}, as they violate fermion
parity conservation.

One can also see that the $SO(N)$ monopole on $S^2$ will have a
non-trivial fermion parity from the parton construction. Indeed,
when there is a $2\pi$ flux of $SO(2)$ through the sphere, it will
be seen by  partons $\chi_1, \chi_2$, while $\chi_\alpha$, $\alpha
= 3\ldots N$ will see no flux. The ground state will then have
$SO(2)$ charge $1$. Since $\chi_\alpha$ carry fermion parity, the
ground state also carries $(-1)^F = -1$. The $SO(2)$ charge gets
neutralized by adding a boson $\phi_1 + i \phi_2$. However, since
$\phi$'s carry no fermion parity, $(-1)^F = -1$ is not affected.

Having established the equivalence of phases and operators in the
free Majorana theory Eqn.~\ref{Lchi} and the dual theory
Eqn.~\ref{SON}, we conjecture that they are actually dynamically
equivalent at their respective IR fixed points.
\begin{table}[htp]
\caption{Duality Dictionary}
\begin{center}
\begin{tabular}{|c|c|}
\hline
Fermionic Theory & Bosonic Theory \\
\hline

Majorana: $\xi$ & Monopole: $V_M$ \\
``m": $\bar{\xi}\xi$ & ``r": $\phi \cdot \phi $\\
\hline
\end{tabular}
\end{center}
\label{default}
\end{table}%
We note that the level-rank duality of Chern-Simons-matter gauge
theories with $O(N)_k$ gauge group in the large-$N$, large-$k$
limit has been proven in Ref.~\cite{aharony2}. Our
conjecture Eqn.~\ref{Lchi} $\leftrightarrow$ Eqn.~\ref{SON}
amounts to a statement that the duality continues to hold when $k
=1$ and $N$ is finite.

A dual bosonized description of multiple flavors of Majorana
fermions was also given in Ref.~\cite{xumajorana}.

\subsection{Dual description of a supersymmetric fixed point}

The critical point of the $SO(N)$-Higgs-CS theory Eqn.~\ref{SON}
is an infrared (IR) fixed point of an ultraviolet (UV) fixed
point, which we call the $SO(N)$-Tricritical-CS theory. This
theory corresponds to tuning $g$ in Eqn.~\ref{SON} to a critical
value $g_c$: at mean field $g_c = 0$ (we assume that the action is
still bounded from below due to the existence of higher order
terms in the polynomial of $|\phi|$), which is analogous to the
tricritical Ising fixed point.

On the fermion side, a natural UV fixed point that flows to the
free Majorana fermion in the IR is the Gross-Neveu-Yukawa fixed
point: \beqn \mathcal{L} = \bar{\xi} i \slashed{\partial} \xi +
(\partial_\mu \sigma)^2 + \lambda \sigma \bar{\xi}\xi -
\frac{\tilde{\lambda}^2}{4} \sigma^4. \label{gny}\eeqn where
$\sigma$ is a real scalar. The relevant perturbation $ - s
\sigma^2 $ in Eqn.~\ref{gny} is dual to $- (g-g_c) |\phi|^4$ in
Eqn.~\ref{SON}: when $s, \, g-g_c > 0$, Eqn.~\ref{gny} and the
$SO(N)$-Tricritical-CS theory respectively flow to
Eqns.~\ref{Lchi} and \ref{SON}. On the other hand, when $s, \,
g-g_c < 0$, both theories have a first order transition between a
trivial phase and a $p_x + i p_y$ superconductor.

An exact renormalization flow of Eq.~(\ref{gny}) is difficult to
compute, but if there is only one fixed point with nonzero
$\lambda$ and $\tilde{\lambda}$, then this fixed point must have
$\lambda^\ast = \tilde{\lambda}^\ast$, and it is a supersymmetric
$\mathcal{N} = 1$ conformal field
theory~\cite{grovervishwanath1,grovervishwanath2,klebanov}. Thus,
our construction also conjectures that this $\mathcal{N} = 1$
supersymmetric conformal field theory is dual to the
$SO(N)$-Tricritical-CS theory. The supersymmetry makes the
following prediction about the scaling dimensions of the
$SO(N)$-Tricritical-CS theory: \beqn \Delta[V_M] -
\Delta[|\phi|^2] = 1/2.\eeqn

An analogous duality between the Dirac fermion Gross-Neveu fixed
point and the $U(1)$-Tricritical-CS fixed point was conjectured in
Ref.~\cite{karchtong}.

The technique of lattice duality we reviewed before can be
naturally generalized to the case with Majorana fermion and
$SO(N)$-CS-matter field theory discussed in this section. This was
shown explicitly in Ref.~\cite{chen2018,jian2018}.

Our dual theory Eqn.~\ref{SON} obviously breaks time-reversal
symmetry, while the single Majorana cone Eqn.~\ref{Lchi} could
preserve the time-reversal symmetry, if the system is defined on
the $2d$ boundary of a $3d$ topological superconductor in class
DIII (the topological phase with index $\nu=1$ is believed to be
realized by the B-phase of superfluid He$^3$). For the Dirac
fermion, the dual U(1)-WFCS theory is believed to have an emergent
time-reversal symmetry in the IR, which transforms the matter
field $\phi$ into its vortex~\cite{seiberg1}. However, this simple
solution does not apply to our $SO(N)$ gauge theory, as there is
no known analogue of the boson-vortex duality for an $SO(N)$
matter field with $N \ge 3$. Thus, we do not yet understand how
time-reversal is hidden in the dual theory Eqn.~\ref{SON}.

\section{Discussion}
\label{discussion}

We reviewed recent advances in understanding dualities of
$2+1$-dimensional quantum field theories and their impact on some
problems in condensed matter physics.  We conclude by briefly
outlining several related developments and future directions.

{\em Other field theory dualities}:

We mostly focused on some basic examples of dualities in $(2+1)d$.
There is a rich structure of inter-connected dualities that have
been proposed, mostly in the high energy literature, of theories
with many different gauge groups including non-abelian ones, and
with different matter contents. As we briefly discussed in Sec.
\ref{largeN} in some cases it has been possible to obtain direct
proof in suitable large-$N$ limits through explicit computations.
Away from these solvable limits the dualities remain conjectural
but seem to be internally consistent with each other (similar to
the basic examples reviewed in the paper). A major open question
is which of  these field theories actually flow to CFTs in the IR
limit.  Indeed some of the consistency checks of the dualities
assume conformal invariance of the IR fixed points of the
considered Lagrangians.

{\em Weak versus strong dualities:}

Weak dualities of two theories are statements that they have the
same local operators, the same global symmetries (and possibly
anomalies), and phase diagram. They can often be unambiguously
derived. Indeed all the examples we have discussed can be proven
as weak dualities. They also open up the possibility of strong
dualities where the two theories in question flow to the same CFT
in the IR. It is the strong dualities that mostly remain
conjectural at present. It is of course conceivable that some
duality holds only  in its weak but not in its strong form.
Several examples of this phenomenon were recently described in
Ref. \cite{bits2018}. As we have discussed, in many condensed
matter contexts, the weak form is already extremely useful and
leads to powerful results whether or not the strong form holds.

{\em  Relevance to experiments}:

Duality ideas have played an important role in shaping the
thinking on  two classic experimental problems in quantum
criticality in condensed matter physics. The first is the
superconductor-insulator transition in two dimensional thin films
driven either by controlling the film thickness or by tuning an
external magnetic field. The second is the integer and fractional
quantum Hall plateau transitions in two dimensional electron gases
in strong magnetic fields. Early theoretical work proposed
connections between these two problems based on flux attachment
ideas.  The theoretical developments we have reviewed clarify
these connections considerably at least in cases where these
transitions are realized in simplified theoretical models.
Specifically effects like disorder, and long range Coulomb
interactions, which are certainly present in the real system and
are potentially important in determining the universality class
are absent in  the  conformally invariant  field theories that we
have discussed. Nevertheless as concrete examples of interacting
field theories that display the same phase transitions as the
experimental systems, progress on these theories gives us
potentially useful insight into the more difficult situations
presented in the experiments.

With this caveat, we highlight some fairly recent experimental
results that hint toward an important role played by dualities.
Ref.\ \cite{breznay2015} studied longitudinal and transverse
resistivities at the magnetic field tuned superconductor-insulator
transition in Indium-Oxide thin films.  Theoretically it has long
been appreciated that the resistivity tensor at this quantum
critical point is universal, and both components are of order the
resistance quantum $\frac{h}{e^2}$.  Further in the vicinity of
the transition the resistivity data is expected to collapse into
universal scaling forms that govern the crossover away from the $T
= 0$ critical point. Interestingly Ref. \cite{breznay2015}
found evidence that at the critical point \be \label{bsitrho}
\rho_{xx} \approx \frac{h}{4e^2}, ~~~\rho_{xy} \approx 0 \ee If
exact this is precisely what is expected if the critical point is
described by a self-dual theory where charge and vortex
descriptions are equivalent. Quite generally charge-vortex duality
implies that the electrical conductivity tensor $\sigma$ is
proportional to the vortex resistivity tensor $\rho_v$: \be \sigma
= \left(\frac{4e^2}{h} \right)^2 \rho_v \ee If the critical system
is assumed to be self-dual, then right at the critical point we
must have $\sigma^T = \left(\rho_v \right)^{-1}$  which implies
that
\begin{equation}\label{sxx2sxy2}
  \sigma_{xx}^2 + \sigma_{xy}^2 = \left(
\frac{4e^2}{h}\right)^2.
\end{equation}
If further $\sigma_{xy}$ is assumed to
be continuous across the transition then as it is zero in the
insulator it will be zero at the critical point itself. The
behavior in Eqn. \ref{bsitrho} then follows.

It is not known theoretically if these assumptions are correct but
if the experimental result in Eqn. \ref{bsitrho}, is taken at face
value, it suggests examining this possibility in theoretical models.
In this context we note that the standard Wilson-Fisher theory
(the $3D$ XY universality class) - which is one simplified model
for this transition -  is {\em not} expected to be self-dual. This
model builds in both $\T$ and $\CT$ symmetries due to which the
Hall resistivity is exactly zero. In the experiments neither of
these are good symmetries. Thus it is interesting to ask if there
are models for the superconductor-insulator transition without
these symmetries included at the microscopic level where the
assumptions on the self-duality and the Hall conductivity can be
evaluated.

Refs. \cite{mmsr2016,mm2017} discuss a model of bosonic Cooper
pairs at a Landau level filling $\nu = 1$ for which a composite
fermion description is described, and suggest it to have a
self-dual response in transport.  The conclusion was drawn,
however, from a simple RPA calculation, whose reliability is not
guaranteed. Ref.~\cite{Hsiao:2018fsc} studied a model consisting
of a Wilson-Fisher boson coupled to a $(3+1)d$ gauge field where
self-duality can be demonstrated explicitly at a given value of
the gauge coupling.  In the model of Ref.~\cite{Hsiao:2018fsc},
exact relationships including Eq.~(\ref{sxx2sxy2}) can also be
derived from the composite fermion representation, but only at the
self-dual gauge coupling: at the generic value of the latter there
are higher-order diagrams that lead to violation of these
relationships. Thus the self duality advocated in
Ref.~\cite{mmsr2016,mm2017} does not seem to be exact.

As mentioned above, there are many non-abelian extensions of the
dualities that are beyond the scope of our review, and some  of
those dualities may have potential applications in condensed
matter physics.  One such example is the fractional quantum Hall
transition, from bosonic $\nu=1/2$ Laughlin state to a trivial
boson insulator, realized on lattice systems without disorder. The
standard composite boson description of such a transition looks
like \be \LL=|D_b\phi|^2-|\phi|^4-\frac{2}{4\pi}bdb, \ee where we
have suppressed external field $A$ for simplicity. Through
level-rank dualities, it was proposed\cite{seiberg2} that this is
dual to an $SU(2)$ gauge theory \be
\LL=\bar{\Psi}i\slashed{D}_a\Psi+CS[a], \ee where $a$ is an
$SU(2)$ gauge field, $\Psi$ is a Dirac fermion in the fundamental
representation of the gauge $SU(2)$, and $CS[a]$ is the standard
Chern-Simons term for $a$ at level $k=1$. The $\Psi$ fermion is
regularized so that with one sign of Dirac mass it induces a
Chern-Simons term at level $k=-1$ and with the other sign no
Chern-Simons term is induced. This duality predicts that at the
transition a global $SO(3)$ symmetry (manifest on the fermion side
as a flavor symmetry) emerges\cite{seiberg2}, which in the
composite boson theory rotates the monopole operator
$\mathcal{M}_b$ (now a Lorentz vector due to the Chern-Simons
coupling) to the flux density $\nabla\times b$ (also a Lorentz
vector)\cite{Leeetal}. This implies that the physical boson creation
operator $\Phi$ will have correlation function of the form $\langle
\Phi^{\dagger}(r)\Phi(0)\rangle\sim 1/r^4$ at the critical point, since
it is related to a conserved charge by an emergent symmetry. This
could possibly be checked in future numerical calculations though
current numerical methods do not give convenient access to scaling
dimensions at such quantum Hall  transitions.

Some of these non-abelian dualities were also used, in a recent
theoretical attempt\cite{MulliganSuperUniversality}, to shed light
on the long-standing puzzle of ``superuniversality" in quantum
Hall plateau transitions -- the apparent phenomenon that critical
exponents like $\nu$ are identical (within error bars) in all the
observed quantum Hall (integer or fractional) plateau transitions.
Specifically, Ref.~\cite{MulliganSuperUniversality} discussed a
series of critical points (without disorder or Coulomb long-range
interaction) that have the same $\nu$. It is interesting to see if
this line of thinking can be pushed to problems with disorder or
long-range interaction, which are clearly important for the
experimental observations.

{\em Novel quantum criticality}:

Despite the work of many decades, our understanding of what kinds
of continuous quantum phase transitions are possible and their
description remains very poor.  Dualities play an important
conceptual role in broadening the range of allowed continuous
quantum phase transitions. We have already discussed their
relevance to Landau-forbidden deconfined quantum critical points.
Here we describe a different example, namely a transition between
a  free massless Dirac fermion and the topologically ordered
T-Pfaffian state (see Sec. \ref{sto}).  Both these are possible
surface states of the standard $3d$ topological insulator. However
when the Dirac fermion is described with the usual fermion
variables it is very hard to see how there can ever be a
continuous transition to the T-Pfaffian state. On the other hand
suppose the fermion-fermion duality holds in its strong form. Then
in the dual fermion description  there is a very simple field
theory that we can write down for such a continuous transition.
Simply deform the dual fermion theory by including a critical
scalar that carries charge-$2$ under the dynamical gauge field,
and that couples to the dual fermions through a Yukawa coupling.
We do not know if this theory really flows to a CFT in the IR but
it provides a formulation which makes the possibility of a direct
second order transition feasible.

{\em Dualities in other dimensions}:

Finally though we have focused on dualities in $(2+1)d$ it is
interesting to ask about generalizations to $(3+1)d$. There are
many well known examples of supersymmetric dualities in $(3+1)d$.
However at the time of writing there are almost no reasonably
established examples without supersymmetry. Recent
work\cite{bits2018} has proposed the possibility that $SU(2)$
gauge theory with a single adjoint Dirac fermion in $(3+1)d$ may
be dual in the IR to a theory of a free massless Dirac fermion
augmented with a topological field theory.  The possibility that
this gauge theory may flow to a free Dirac fermion was proposed in
Ref.\ \cite{poppitz2018} but a careful analysis of the
anomalies of the gauge theory in Ref.\ \cite{cordova2018}
showed that the free Dirac theory is not enough to match all
anomalies. The inclusion of the gapped TQFT in Ref.\
\cite{bits2018} along with the free massless fermion enables
matching all anomalies while preserving the  UV symmetries of the
gauge theory. The  proposal in Ref. \cite{bits2018} can be
shown to hold  at the level of a weak duality but it is not
presently known if it holds in strong form.

One possible approach\cite{Aitken:2018joz} to $(3+1)d$ dualities is
through the procedure of ``deconstruction,'' which was originally
proposed as a way to generate a fifth dimension by stacking up
four-dimensional theories\cite{ArkaniHamed:2001ca}.  While such
procedure can be applied with success in supersymmetric theories,
without supersymmetry it suffers from the lack of analytical control.
Similarly, one can derive supersymmetric $(1+1)d$ dualities by
compactifying $(2+1)d$ pairs of dual theories, but one encounters the
problem of strong coupling if supersymmetry is not
present\cite{Karch:2018mer}.


\section*{Acknowledgements}
Our understanding of the topic reviewed here was influenced by
collaborations and conversations with several people.  We
particularly thank Maissam Barkeshli, Matthew Fisher, Yin-Chen He,
Max Metlitski, Adam Nahum, Andrew Potter, Nathan Seiberg, David
Tong, Ashvin Vishwanath, and Edward Witten for
collaborations/discussions. TS was supported by NSF Grant No.
DMR-1608505, and partially through a Simons Investigator Award
from the Simons Foundation.  D.S. was supported, in part, by the
DOE Grant No.\ DE-SC0009924 and a Simons Investigator Award from
the Simons Foundation. CW was supported by the Harvard Society of
Fellows. CX is supported by the David and Lucile Packard
Foundation.

\appendix
\section{Ising/Majorana dualities in $(1+1)d$}
\label{1dnotes}

In this appendix we discuss the continuum version of the $(1+1)d$
Kramers-Wannier duality and Jordan-Wigner duality, paying special
attention to various $\mathbb{Z}_2$ gauge fields (dynamical or
background). The results here are of very little practical use
since $(1+1)d$ stories are quite well understood. The main goal
here is to display the amusing structures of $(1+1)d$ dualities
that are in parallel with those in $(2+1)d$. In particular, if we
do the following substitution, then we can translate $(2+1)d$
stories to $(1+1)d$ almost word for word:
\begin{enumerate}

\item Dirac fermion $\Longrightarrow$ Majorana fermion

\item $O(2)$ Wilson-Fisher $\Longrightarrow$ Ising scalar

\item $\frac{1}{2\pi}AdB$ (response of $U(1)\times U(1)$ boson
integer quantum hall) $\Longrightarrow$ $\pi A\wedge B$ (response
of $\Z_2\times \Z_2$ Haldane chain)

\item $\frac{1}{4\pi}AdA$ (response of fermion integer quantum
Hall) $\Longrightarrow$ $\pi {\rm Arf}(A\cdot\rho)$ (response of
Kitaev chain, to be explained in more detail below)

\end{enumerate}

Let us begin with the Kramers-Wannier ``self-duality" -- the
quotation mark is to indicate that it is really not a
self-duality. The duality should be properly written as below:
\beq
(D_B\phi)^2-\phi^4\hspace{10pt}\Longleftrightarrow\hspace{10pt}(D_b\tilde{\phi})^2-\tilde{\phi}^4+\pi
b\wedge B, \eeq where $\phi$ is the Ising scalar in the continuum,
$\tilde{\phi}$ corresponds to the ``kinks" of $\phi$, $B$ is a
background (probe) $\Z_2$ gauge field that couples to the Ising
charge, and $b$ is a dynamical $\Z_2$ gauge field. Both gauge
fields take value in $H^1(X,\Z_2)$ where $X$ is the space-time
manifold (assumed to be orientable), normalized in such a way that
a nontrivial flux takes value $1$ (instead of $\pi$). The last
term $\pi b\wedge B$ assigns a nontrivial global $\Z_2$ charge to
each $\Z_2$ instanton of $b$, hence identifying the $\Z_2$
instanton of the $b$ gauge field with $\phi$ on the left side.
This term is the response of a $\Z_2\times \Z_2$ Haldane chain,
and plays a very similar role with the $BF$ term
$\frac{1}{2\pi}AdB$ in $(2+1)d$. The $\Z_2$ gauge field $b$ is
flat, with instantons suppressed because of the global $\Z_2$
symmetry. Therefore $b$ has no nontrivial dynamics (except on
imposing a global constraint), and is often neglected, making the
duality appears to be a ``self-dual". But $b$ is important
topologically, and properly including it makes the above duality
more similar to the boson-vortex duality in $(2+1)D$.

Now consider the Jordan-Wigner duality. Let's start with a duality
involving a free Majorana fermion \beq
\bar{\chi}i\slashed{D}_{A\cdot\rho}\chi\hspace{10pt}\Longleftrightarrow\hspace{10pt}(D_b\phi)^2-\phi^4+\pi
\left[{\rm Arf}(b\cdot\rho)+{\rm Arf}(\rho)\right]+\pi b\wedge A.
\eeq This form of the duality has been discussed in
\cite{KTBosonization}, and we discuss more details here. We
briefly explain some notations and concepts here: $A$ is a
background $Z_2$ gauge field and $\rho$ is a reference spin
structure on $X$, and $A\cdot\rho$ is another spin structure
obtained from $\rho$ by superposing $A$ on it (so despite the
notation, $0\cdot\rho=\rho$). Physically this is nothing but to
gauge the fermion parity -- the only subtlety is that unlike
ordinary $\Z_2$ gauge field, there is no canonical choice on the
spin structure $\rho$ (for ordinary $\Z_2$ gauge field the
canonical choice would be the trivial bundle). On the right hand
side, $b$ is a dynamical $\Z_2$ gauge field, and the term $\pi
{\rm Arf}[b\cdot\rho]$ is a $\Z_2$-valued topological invariant of
the spin structure $b\cdot\rho$ known as the Arf invariant. We
will not repeat the mathematical definition here (see for example
\cite{spincobordism, RyuArf}), but we shall list some of its
mathematical and physical properties that will be useful for later
discussions:
\begin{enumerate}
\item If we integrate out the Kitaev Majorana chain on a
orientable manifold with spin structure $\rho$, the partition
function is given by $(-1)^{{\rm Arf}(\rho)}$. In particular, on a
space-time torus, the partition function is $-1$ if and only if
the fermions have periodic boundary conditions in both the space
and time directions.

\item From the physics of the Kitaev chain, we immediately
conclude that if a $\Z_2$ gauge field $b$ has the partition
function $(-1)^{{\rm Arf}(b\cdot\rho)}$, its $\Z_2$ instanton will
(a) carry a nontrivial $\Z_2$ charge and (b) become a fermion. In
the context of the above duality, this means that the $\Z_2$
instanton of $b$ on the right hand side (bound with $\phi$)
corresponds to the free Majorana fermion on the left side.

\item The following two identities hold: \beqn \label{identities}
\sum_{b\in H^1(X,\Z_2)}(-1)^{{\rm Arf}(b\cdot\rho)+{\rm
Arf}(\rho)+\int b\wedge A} & \sim& (-1)^{{\rm Arf}(A\cdot\rho)},
\nn (-1)^{{\rm Arf}(\rho)+{\rm Arf}(A\cdot\rho)+{\rm
Arf}(B\cdot\rho)+{\rm Arf}[(A+B)\cdot\rho]}&=&(-1)^{\int A\wedge
B}, \eeqn where the first identity, useful when integrating out
$b$, is true up to a normalization constant. This identity enables
the matching of the phase diagram for the Jordan-Wigner duality:
with an appropriate mass term, the left hand side can go into a
Kitaev phase, with partition function $(-1)^{{\rm
Arf}(A\cdot\rho)}$. This, on the right hand side, corresponds to a
phase in which $\phi$ is gapped, and a pure gauge theory remains
in the IR. Integrating out $b$ using the first of
Eq.~\eqref{identities} produces the right partition function.

The second identity of Eq.~\eqref{identities} has a simple
physical origin: consider four copies of Kitaev Chains with a
global $\Z_2\times \Z_2$ symmetry acting in the following way: the
first $\Z_2$ acts as
$(\chi_1,\chi_2,\chi_3,\chi_4)\to(\chi_1,-\chi_2,\chi_3,-\chi_4)$,
while the second $\Z_2$ acts as
$(\chi_1,\chi_2,\chi_3,\chi_4)\to(\chi_1,\chi_2,-\chi_3,-\chi_4)$.
It is then easy to check (for example, by studying the Majorana
end states) that this system is equivalent to a bosonic SPT,
namely a Haldane chain protected by the $\Z_2\times \Z_2$
symmetry. This is exactly what the second equation in
Eq.~\eqref{identities} displays.

\end{enumerate}

Point (2) and (3) above show the strong analogy between the Arf
invariant in $(1+1)d$ and the Chern-Simons term
$\frac{1}{4\pi}ada$ in $(2+1)d$. The Jordan-Wigner duality, in
this sense, is very similar to the $(2+1)d$ bosonization.

Now consider another global unitary $\Z_2$ symmetry on the
Majorana $\mathcal{S}: \chi\to \gamma_0\gamma_1\chi$ (the chiral
symmetry). This symmetry forbids the mass term $\bar{\chi}\chi$
and is anomalous: under this symmetry transform the Lagrangian
gains an additional term \beq \mathcal{S}:
\bar{\chi}i\slashed{D}_{A\cdot\rho}\chi \to
\bar{\chi}i\slashed{D}_{A\cdot\rho}\chi+{\rm Arf}(A\cdot\rho).
\eeq Physically this is because when the fermions are gapped, the
operation $\mathcal{S}$ exchanges a trivial superconductor with a
Kitaev chain.

Now how is $\mathcal{S}$ implemented on the Ising side? Two
requirements must be satisfied: (1) the Ising mass $m\phi^2$
should be odd under $\mathcal{S}$, and (2) an additional term
${\rm Arf}(A\cdot\rho)$ must be added to the Lagrangian after the
transform, in another word, the anomaly should be properly
captured. It turns out that an Ising-kink duality on $\phi$ does
the job: \beqn &&(D_b\phi)^2-\phi^4+\pi \left[{\rm
Arf}(b\cdot\rho)+{\rm Arf}(\rho)\right]+\pi b\wedge A \nn
&\to&(D_{\tilde{b}}\tilde{\phi})^2-\tilde{\phi}^4+\pi
b\wedge\tilde{b}+\pi \left[{\rm Arf}(b\cdot\rho)+{\rm
Arf}(\rho)\right]+\pi b\wedge A \nn
&\to&(D_{\tilde{b}}\tilde{\phi})^2-\tilde{\phi}^4+\pi\left[{\rm
Arf}(\tilde{b}\cdot \rho)+{\rm
Arf}(\rho)\right]+\pi\tilde{b}\wedge A+\pi{\rm Arf}(A\cdot \rho),
\eeqn where the last line comes from integrating out $b$ using
Eq.~\eqref{identities}. This is obviously parallel to how the
parity-anomaly is realized in the $(2+1)d$ bosonization.

Similar to $(2+1)d$, we can define $S$ and $T$ operations on the
theories: \beqn S: \mathcal{L}[A] &\to& \mathcal{L}[a]+\pi a\wedge
A, \nn T: \mathcal{L}[A\cdot\rho] &\to&
\mathcal{L}[A\cdot\rho]+\pi{\rm Arf}(A\cdot \rho), \nn
\mathcal{L}[A] &\to& \mathcal{L}[A]+\pi\left[{\rm Arf}(A\cdot
\rho)+{\rm Arf}(\rho)\right], \eeqn where the first and second
line for the $T$ transform differ slightly, depending on whether
the theory before the transform requires a spin-structure or not.
Due to the $\Z_2$ nature of the gauge fields, the group spanned by
$S$ and $T$ is $SL(2,\Z_2)$, much smaller than the $SL(2,\Z)$ in
$(2+1)d$. In particular, by using Eq.~\eqref{identities} we can
explicitly check that $S^2=T^2=(ST)^3=1$, and there are only six
different group elements: \beq \{1, S, T, ST, TS, STS=TST\}. \eeq
Similar to $(2+1)d$, this can be understood in terms of
electric-magnetic dualities of a gauge theory in one dimension
higher. In this case the relevant theory is the $\Z_2$ gauge
theory in $(2+1)d$, often labeled as $\{1,e,m,\epsilon\}$. The six
elements in $SL(2,\Z_2)$ correspond to the six ways to permute the
$e,m,\epsilon$ particles in the $Z_2$ gauge theory.

Now we can obtain the usual Jordan-Wigner duality (the inverse of
the one mentioned before): \beq
(D_B\phi)^2-\phi^4\hspace{10pt}\Longleftrightarrow\hspace{10pt}\bar{\chi}i\slashed{D}_{a\cdot\rho}\chi+\pi
a\wedge B+\pi\left[{\rm Arf}(B\cdot\rho)+{\rm Arf}(\rho)\right].
\eeq

Unfortunately due to the $\Z_2$ nature of the $T$ transform, there
does not seem to be an analog of fermion-fermion duality in
$(1+1)d$.

\section{Some useful formal structures}
\label{formalstuff}

Here we collect together some definitions and explanations of some
formal structures that are useful in precise definitions of some
of the field theories we work with.  More detailed explanations
can be found in, e.g., Refs.
\cite{nakaharabook,wittenreview,Metlitski2015,SeibergWittenSTO}.

We begin with the definition of the free massless Dirac fermion in
$2+1$-D in Eqn. \ref{freeD}.  As we emphasized it is convenient to
include both a background gauge field $A$ and to place the theory
on an arbitrary oriented manifold with a metric $g$.

Before defining the Dirac theory on such a general manifold it is
necessary to  first define the  concept of spin and spin$_c$
structures. Note that for a general $D$-dimensional manifold it is
not possible to choose a coordinate system globally. Rather we
divide the manifold into overlapping patches such that within each
patch a smooth coordinate system can be chosen ({\em i.e} a smooth
orthonormal  basis for the tangent space).  To go between two
patches $P_1$ and $P_2$ in their overlap region $P_1  \cap P_2$,
we perform a rotation $V_{12}$. For an orientable manifold we can
take $V_{12}$ to be an element of $SO(d)$ ($d$ is the space-time
dimension).  On triple overlaps of patches $P_1 \cap P_2 \cap P_3$
we require \be V_{12}V_{23}V_{31} = 1 \ee This is known as the
cocycle condition.  To define spinors ({\em i.e} fermions) in this
manifold we need to lift the transition matrices $V$ to elements
$U$ of the double cover $Spin(d)$ of $SO(d)$.  Both $U$ and $-U$
of $Spin(d)$ correspond to the same $V$ of $SO(d)$. In the spinor
representation the cocycle condition on triple overlaps implies
\be U_{12}U_{23}U_{31} = f_{123} \ee where $f_{123} = \pm 1$ and
arises from the sign ambiguity in choosing $U$. If we can choose
the signs of $U$ for all transitions such that for all triple
overlaps $f_{123} = 1$, then   we can define spinors consistently
globally.  A specific choice of such signs is known as a spin
structure. In general  a  manifold will admit more than one spin
structure ({\em i.e} more than one choice of signs of transition
matrices $U$). Note that the difference between two spin
structures can be thought of as a $Z_2$ gauge field.  A manifold
with a specific spin structure is known as a spin manifold. All
oriented space-time manifolds with $d < 4$ admit  spin structures.
For $d = 4$ there are however manifolds that do not admit any spin
structure (a well known example is $CP^2$).  The functions
$f_{ijk}$ are symmetric in the 3 indices and are invariant under a
$Z_2$ gauge transformation of the spinors in different patches:
this  takes  $U_{ij} \rightarrow s_i U_{ij} s_j $ with $s_{i,j} =
\pm 1$. Furthermore we clearly have $f_{ijk}f_{jkl}f_{kli} = 1$ as
then the sign ambiguity of any single $U$ cancels out.  Such
functions $f_{ijk}$  define elements $w_2$ of a cohomology class
$H^2(M, Z_2)$ called the second Stiefel-Whitney class.

Even if the manifold does not admit (or we do not specify) a spin
structure, we may still be able to define fermions if they couple
to other gauge fields. The cases  pertinent to us is when the
fermions couple to a $U(1)$ gauge field $A$. In this case we can
compensate any failure of the cocycle condition on $U$ by
combining with fluxes of the $U(1)$ gauge field, {\em i.e}
whenever $w_2$ is non-zero we place a $\pi$ flux of $A$.  Such a
field $A$ is known as a spin$_c$ connection and satisfies the
modified flux quantization condition
\begin{equation}
\int_C \frac{dA}{2\pi} = \int_C \frac{w_2}{2} ~~(mod~ Z)
\end{equation}
for every oriented 2-cycle $C$. (Formally in this case we work
with transition functions in the group $\frac{U(1) \times
Spin(d)}{Z_2}$ rather than just in $Spin(d)$). Note that the
ability to define fermions by coupling to a spin$_c$ connection
assumes that there are no bosons that couple to $A$ with
charge-$1$. Thus demanding that the fermions are coupled to a
spin$_c$ connection is a good book-keeping device to track that
all odd-charge (under $A$) fields are fermions while even-charge
ones are bosons.

Let us now turn to the definition of the free massless Dirac
theory in eqn. \ref{freeD}. We  regularized the theory by
including a massive fermion $\psi_H$ with a heavy mass $M < 0$.
Consider the partition function of this theory obtained by
evaluating the free fermion path integral. Clearly this takes the
form \be Z_\psi = det (D) det(D - i M) \ee where $D$ is the
covariant derivative defining the Dirac operator (that includes
the coupling to $(A,g)$, and $A$ is taken to be a spin$_c$
connection.  With  Euclidean signature $D$ is hermitian and hence
has real eigenvalues: \be D \psi_i = \lambda_i \psi_i \ee Thus
formally we have \bea
Z_\psi & =  & \Pi_i \lambda_i \left( \lambda_i - i M \right) \\
& = & \Pi_i \frac{\lambda_i }{\lambda_i + i M} (\lambda_i^2 + M^2)
\eea We will be interested primarily in the phase of $Z_\psi$ for
which only the first term in the product in the second line
matters. For $\lambda_i > 0$, this term contributes a factor $1/i$
to the phase while for $\lambda_i < 0$ it contributes a factor of
$i$.  It follows that we can write \bea
Z_\psi & =  & |Z_\psi| e^{-i \frac{\pi}{2} \sum_i sgn(\lambda_i)} \\
& = & |Z_\psi| e^{-i \frac{\pi}{2} \eta} \eea Here $\eta$ is
defined to be the regularized sum over $sgn(\lambda_i)$ that
appears in the first line, and is known as the $\eta$ invariant.
$\eta$ is a function of $(A,g)$.

Note that for any unitary quantum field theory on an orientable
spacetime $X$, the partition function $Z \rightarrow Z^*$ when the
orientation is reversed. In a time reversal invariant theory,
orientation reversal is a symmetry and thus $Z$ must be real.  For
the present theory, $Z_\psi$ is complex and hence not time
reversal invariant. This is of course due to the choice of
regulator which included the heavy fermion. The parity anomaly is
the statement that the partition function cannot be made real with
any ``local" regulator.  On the other hand $Z_\psi$ can be
rendered real if we regard the theory as living at the boundary of
a free fermion topological insulator in $3+1$-spacetime
dimensions. Integrating out the fermions in the bulk of the
topological gives the well-known $\theta$ term contribution to the
action  which we write as \be Z_{bulk} = e^{i \pi \left(
\frac{1}{2} \int d^4x {F \over 2\pi} \wedge {F \over 2\pi} +  {1
\over 192 \pi^2} \int d^4x\ \text{tr}( R \wedge R)\right)} \ee
Here $F = dA$ and $R$ is the Riemann curvature tensor  The net
partition function - bulk + boundary - is \be |Z_\psi| e^{-i
\frac{\pi}{2} \eta}e^{i \pi \left( \frac{1}{2} \int d^4x {F \over
2\pi} \wedge {F \over 2\pi} +  {1 \over 192 \pi^2} \int d^4x\
\text{tr}( R \wedge R)\right)} \ee The combination of the two
exponentials in this product is known (by the Atiyah-Patodi-Singer
index theorem) to equal \be (-1)^J \ee where $J$ is an integer and
is a topological invariant.  Thus the combined boundary-bulk
partition function is real: this is the formal characterization of
the time reversal invariance of the massless Dirac fermion when it
lives at the boundary of a $3+1$-D topological insulator.

Note that the time reversed boundary theory $Z_\psi^* $ is related
to  $Z_\psi$  through \be Z_\psi^* = | Z_\psi| e^{-i\pi \eta[A,g]}
\ee Now it follows from the index theorem mentioned above that \be
\frac{\eta[A,g]}{2} = \frac{1}{2} CS[A,g] + J \ee with \be CS[A,g]
= \frac{1}{4\pi} AdA + 2CS_g \ee To obtain this form use the well
known result that the bulk $\theta$ term is a total derivative and
yields the Chern-Simons term at the boundary. The gravitational
Chern-Simons term is written in terms of the Levi-Civita spin
connection $\omega$ on the tangent bundle: \be CS_g = \frac{1}{192
\pi} \int d^3x Tr (\omega d\omega + \frac{2}{3} \omega^3) \ee (The
normalization is such that a $p_x+ip_y$ superconductor in $2+1$-d
will have in its response a term $CS_g$ with coefficient $1$.
Physically this corresponds to a thermal hall effect. Thus $2CS_g$
corresponds to the thermal Hall conductance of the $\nu = 1$
integer quantum hall state. ) Thus we have \be
 e^{-i\pi \eta[A,g]} = e^{iCS[A,g]}
 \ee
This is precisely the result we wrote down on physical grounds in
the main text.

\section{Fermionic dual of the $XY$ Wilson-Fisher fixed point}
\label{fparton}

Consider a theory of a single complex boson $\phi$ (possibly on a
lattice) coupled to a background gauge field $B$. We represent
this system as a theory of two fermions $f_1$ and $f_2$ each
coupled to the same $U(1)$ gauge field $a$ (strictly speaking a
spin$_c$ connection but with opposite charges $\pm 1$. We also
assign $U_B(1)$ charges $1/2$ to each of the $f_I$.  The
(schematic) Lagrangian is
\begin{equation}
\label{basicP} {\cal L}_0 = {\cal L}[f_1, a + \frac{B}{2}] + {\cal
L}[f_2, -a + \frac{B}{2}]
\end{equation}
The physical boson $\phi = f_1 f_2$ is gauge invariant and carries
$U_B(1)$ charge-$1$.  ${\cal L}_0$ is a faithful representation of
the original boson system so long as we also allow monopole
operators in $a$ as part of the Lagrangian\footnote{ It is common
in the CM literature to ``justify" this description by first
writing $\phi = f_1 f_2$ on the lattice. This representation is
redundant: the gauge redundancy is $SU(2)$ corresponding to
rotations $f_I \rightarrow U_{IJ} f_J$ with $U \in SU(2)$.  Next
we imagine a ``mean field" theory in terms of the $f_I$ which
breaks the $SU(2)$ gauge symmetry down to $U(1)$.  (Other mean
fields are possible and will lead to other effective theories of
the same physical system which may be convenient to access other
phenomena.). The theory of fluctuations about this `mean field'
will then include a $U(1)$ gauge field and will lead to Eqn.
\ref{basicP}.}.

Now denote by $m_a$ the flux quantum for $a$  and $m_B$ the flux
quantum for $B$. These must satisfy the conditions
\begin{eqnarray}
m_a + \frac{m_B}{2} & = & n_1 \\
-m_a + \frac{m_B}{2} & = & n_2
\end{eqnarray}
with $n_{1,2} \in Z$. It follows that $m_B \in Z$ (as required),
$2m_a \in Z$ and $2m_a - m_B = 0 (mod 2)$. These conditions can be
implemented more simply by writing $a' = a + \frac{B}{2}$ and
allowing $m_{a'}, m_B \in Z$ but otherwise arbitrary.  Thus we
rewrite
\begin{equation}
\label{basicP1} {\cal L}_0 = {\cal L}[f_1, a'] + {\cal L}[f_2, -a'
+ B]
\end{equation}
Now consider a specific choice where  $f_2$ is in an integer
quantum Hall state with $\sigma_{xy} = 1$. Then we can integrate
out $f_2$ to get
\begin{equation}
\label{fluxatt} {\cal L}_1= {\cal L}[f_1, a'] + \frac{1}{4\pi}
(-a' + B) d(-a' + B) + 2CS_g
\end{equation}
(We included the $CS_g$ term to keep track of the thermal Hall
effect $\kappa_{xy}$). Now we can consider two different possible
phases of $f_1$. If $f_1$ is in a trivial insulator ($\sigma_{xy}
= 0$) we can integrate it out to get just the last two terms of
${\cal L}_1$. Integrating out $a'$ we get a trivial theory. This
reproduces a trivial insulator of $\Phi$.  Next we consider
varying UV parameters to put $f_1$ in a phase where it has a
$\sigma_{xy} = -1$ (and the accompanying $\kappa_{xy}$).
Integrating it out we get
\begin{equation}
{\cal L}_{sf} = - \frac{1}{2\pi} B da' + \frac{1}{4\pi} BdB
\end{equation}
Integrating out $a'$ now Higgses $B$, and we interpret this as a
``superfluid" of $\Phi$ where the global $U_B(1)$ symmetry is
spontaneously broken.

A phase transition between the superfluid and trivial phases can
then be described as the ``integer quantum Hall transition" of
$f_1$ in Eqn. \ref{fluxatt} as it transitions from $\sigma_{xy} =
0$ to $\sigma_{xy} = -1$. A low energy model for this transition
is just a free massless Dirac fermion $\chi$ which is then coupled
to $a'$ as in Eqn. \ref{fluxatt}. We then get (denoting $a'$ by
$a$)
\begin{equation}
\label{fdualxy} {\cal L}_{dualxy} = i\bar{\chi} D_a \chi +
\frac{1}{4\pi} (-a + B) d(-a + B) + 2CS_g
\end{equation}

We also know that this same transition (between the superfluid and
the trivial phase) of $\phi$ can be described by the Wilson-Fisher
fixed point of $\phi$. It is then natural to {\em conjecture} that
Eqn. \ref{fdualxy} is dual to the Wilson-Fisher theory of $\phi$.

\bibliography{duality}

\end{document}